\documentclass{aa}  

\usepackage{bm}
\usepackage{color}
\usepackage{comment}
\usepackage[english]{babel} 
\usepackage[utf8]{inputenc}
\usepackage{lipsum}
\usepackage{graphicx}
\usepackage{float}
\usepackage{lettrine}
\usepackage{cuted}
\usepackage{graphicx}
\usepackage[table]{xcolor}
\usepackage{amsmath,tabu}
\usepackage{multirow}
\usepackage[T1]{fontenc}
\usepackage{libertine}
\usepackage{microtype}
\usepackage{mathrsfs}
\usepackage{physics}
\usepackage{makeidx}
\usepackage[nottoc]{tocbibind}
\usepackage{subcaption}
\usepackage{caption}
\usepackage{wrapfig}
\usepackage{booktabs}
\usepackage{listings}
\usepackage{comment}
\usepackage{courier}
\usepackage{gensymb}
\usepackage{hyperref}
\usepackage{amsmath}
\usepackage{graphicx}
\usepackage{hvfloat,lipsum}
\newsavebox{\myimage}
\usepackage{appendix}
\usepackage[flushleft]{threeparttable}

\usepackage{color}

\definecolor{codegreen}{rgb}{0,0.6,0}
\definecolor{codegray}{rgb}{0.5,0.5,0.5}
\definecolor{codepurple}{rgb}{0.58,0,0.82}
\definecolor{backcolour}{rgb}{0.95,0.95,0.92}
 
\lstdefinestyle{mystyle}{
    backgroundcolor=\color{backcolour},   
    commentstyle=\color{codegreen},
    keywordstyle=\color{magenta},
    numberstyle=\tiny\color{codegray},
    stringstyle=\color{codepurple},
    basicstyle=\footnotesize,
    breakatwhitespace=false,         
    breaklines=true,                 
    captionpos=b,                    
    keepspaces=true,                 
    numbers=left,                    
    numbersep=5pt,                  
    showspaces=false,                
    showstringspaces=false,
    showtabs=false,                  
    tabsize=2
}
 
\usepackage{natbib}
\bibpunct{(}{)}{;}{a}{}{,} 

\hypersetup{
    colorlinks=true,
    linkcolor=black,
    filecolor=magenta,      
    urlcolor=cyan,
    citecolor = teal
}
\usepackage[bottom]{footmisc}

\usepackage{etoolbox}

\BeforeBeginEnvironment{appendices}{\clearpage}

\usepackage{graphicx}
\usepackage{txfonts}
\begin{document}

   \title{Fast outflows in protoplanetary nebulae and young planetary nebulae observed by \textit{Herschel}/HIFI}


   \author{
          M. Lorenzo\inst{1}\thanks{currently at Centro de Astrobiología (CSIC-INTA), Torrejón de Ardoz, Madrid, Spain}  \and D. Teyssier\inst{2} \and V. Bujarrabal\inst{3}  \and P. García-Lario\inst{1} \and J. Alcolea\inst{4}   \and E. Verdugo\inst{1}   \and A. Marston\inst{1}
          }

   \institute{European Space Astronomy Centre, ESA, PO Box 78, 28691 Villanueva de la Ca\~nada, Madrid, Spain
   \and 
   Telespazio Vega UK Ltd. for ESA/ESAC, Madrid, Spain
   \and
   Observatorio Astron\'omico Nacional (IGN), Ap 112, 28803 Alcal\'a de Henares, Spain
   \and
   Observatorio Astron\'omico Nacional (OAN-IGN), Calle Alfonso XII 3, 28014 Madrid, Spain\\
             }


 
  \abstract
   {Fast outflows and their interaction with slow shells (generally known as the fossil circumstellar envelope of asymptotic giant branch stars) play an important role in the structure and kinematics of protoplanetary and planetary nebulae (pPNe, PNe). To properly study their effects within these objects, we also need to observe the intermediate-temperature gas, which is only detectable in the far-infrared (FIR) and submillimetre (submm) transitions.}
   {We study the physical conditions of the outflows presented in a number of pPNe and PNe, with a focus on their temperature and excitation states.}
   {We carried out {\it Herschel}/HIFI observations in the submm lines of $^{12}$CO in nine pPNe and nine PNe and complemented them with low-$J$ CO spectra obtained with the IRAM 30m telescope and taken from the literature. The spectral resolution of HIFI allows us to identify and measure the different nebular components in the line profiles. The comparison with large velocity gradient (LVG) model predictions was used to estimate the physical conditions of the warm gas in the nebulae, such as excitation conditions, temperature, and density.} 
   {We found high kinetic temperatures for the fast winds of pPNe, typically reaching between 75 K and 200 K. In contrast, the high-velocity gas in the sampled PNe is colder, with characteristic temperatures between 25 K and 75 K, and it is found in a lower excitation state. We interpret this correlation of the kinetic temperature and excitation state of fast outflows with the amount of time elapsed since their acceleration (probably driven by shocks) as a consequence of the cooling that occurred during the pPN phase.}
   {}
   \keywords{stars: AGB \& post-AGB - circumstellar matter - stars:mass-loss - planetary nebulae: general}

   \maketitle
%

\section{Introduction}
\label{sec:introduction}

Throughout the very short post-AGB phase, lasting less than a few thousand years, a dramatic change occurs in the morphology and dynamics of the circumstellar envelopes (CSEs) of evolved stars of intermediate mass. Starting from an intense mass-loss period that results in the formation of a dense, spherical, and slowly expanding CSE, asymptotic giant branch (AGB) stars then turn into the central stars of planetary nebulae (PNe). These PNe present complex morphological structures, often with strong axial-symmetry. These extreme kinetic and morphological changes are thought to arise from the interaction between fast, collimated (jet-like) winds and the slow, dense layers of the CSE. The transformation takes place in the intermediate stage known as the protoplanetary nebula (pPN) phase.

Collimated winds, commonly known as fast outflows, are ejected close to the central core and carry a significant fraction of nebular mass, that is, $\sim$ 0.1-0.3 M$_\odot$ \citep{bujarrabal_2001},  with velocities around 100 km s$^{-1}$. The interaction between these fast outflows and the remnant of the slowly expanding circumstellar envelope (AGB-CSE) is believed to be the main agent responsible for the change of kinematics and structure observed in pPNe, sculpting the final shape of PNe \citep{balick_2002}. Several observations of molecular lines have identified fast outflows in many pPNe and PNe; see examples in \citealp{bujarrabal_1998, bujarrabal_2001, alcolea_2001, castro_2002, sanchez-contreras_2004, fong_2006}, and references therein. In most cases, they exhibit strong emission of molecular lines with composite profiles. Typically, these profiles consist of a central core that comes from the earlier slow AGB wind and broad line wings arising from fast post-AGB outflows. In molecular line emission, these outflows show bipolar or hourglass-like geometries and dynamics in almost all well-studied cases, although the optical images of many PNe and pPNe tend to be more complex, with multipolar or point-like symmetries.

Until now, most of what we know about the structure, dynamics, and physical conditions of molecular material in nebulae has been derived  mainly from observations of low-$J$ transitions of molecular lines, especially the \mbox{$J$=2--1} and \mbox{$J$=1--0} transitions of CO. These lines primarily trace molecular gas in the cold, (typically) outer layers of the circumstellar envelope, where the expansion velocity is already constant and the temperature, as it is decreasing with the radius as $\sim$ 1/r, typically reaches between 10-20 K. There are also many studies analysing molecules with higher dipole moments, such as HCN, HCO+, and HNC, which trace denser gas than CO (e.g. \citealp{bachiller_1997, zhang_2008, milam_2009, edwards_ziurys_2013, edwards_2014, schmidt_ziurys_2016, bublitz_2019}, and references therein). However, to completely understand the shaping mechanism of the highly collimated jets in pPNe and PNe, and, consequently, their evolution, we also need to observe warmer regions (100 K $\lesssim$ T$_K$ $\lesssim$ 1000 K), which is where acceleration takes place. A proper study of intermediate-temperature gas requires observations of higher $J$ transitions of CO that are found in the far-infrared (FIR) and submillimetre (submm) ranges. 

Unfortunately, there are very few studies based on the observation of the molecular gas of pPNe and PNe in the FIR and sub-mm lines.  \citet{justtanont_2000} were the first to analyse the intermediate-temperature gas around evolved stars in the FIR. They presented ISO-LWS observations of high-$J$ transitions of CO (from \mbox{$J$=14--13} up to \mbox{$J$=37--36}) in CRL 618, CRL 2688 and NGC 7027, but the low spectral resolution of their data prevented them from identifying the nebular components responsible for the line emission. Higher spectral resolution was achieved with \textit{Herschel} in the FIR/submm observations presented in \citet{bujarrabal_2012}. By comparing their results with theoretical predictions, they found warm fast winds in CRL 618 and CRL 2688 (T$_k$ $\geq$ 100 K) and colder outflows in OH 231+4.2 and NGC 6302 (T$_k$ $\sim$ 30 K).

\textit{Herschel}, and in particular its HIFI heterodyne spectrometer, provides the necessarily high spectral resolution ($\frac{\Delta \upsilon}{\upsilon}$ < 10$^{-6}$) in the FIR and submm to determine the dynamics and physical conditions of the shocked layers of this type of nebulae and to resolve the different features of their profiles. In this paper, we present observations of intermediate-excitation CO lines in a sample of eighteen nebulae carried out with HIFI. These observations have been obtained as part of the guaranteed-time project SUbmilimetre Catalogue of Circumstellar Envelopes of StarS (SUCCESS), a submm/FIR CO survey of a large sample of AGB and post-AGB stars. In particular, this programme obtained observations of the $^{12}$CO \mbox{$J$=5--4} and \mbox{$J$=9--8} lines in 18 post-AGB stars. Of this sample of nebulae, three were also observed in the $^{12}$CO \mbox{$J$=14--13} line and five were followed up in $^{12}$CO \mbox{$J$=2--1} and \mbox{$J$=1--0} lines using the IRAM 30m telescope.

\section{Observations}
\label{sec:observations}

\subsection{Herschel/HIFI observations}
\label{sec:observations_HIFI}

Observations in the submm lines of CO were obtained in the context of the {\it Herschel} SUCCESS Guaranteed Time project \citep{teyssier_2011} executed with the HIFI high-resolution spectrometer \citep{degraauw_2010}. The post-AGB stars considered in this paper represent about a quarter of the whole sample observed by this programme. Details on how the data were acquired and processed can be found in \citet{danilovich_2015}. No particular spectral smoothing was applied to the data for their analysis, however, rebinning was applied in some of the spectra given in the paper.

\subsection{Supplementary data}
\label{sec:observations_sup}

\begin{figure}
    \begin{center}
        \setlength\abovecaptionskip{-0.01\baselineskip}
        \setlength\belowcaptionskip{-0.7\baselineskip}
        \includegraphics[width= \linewidth]{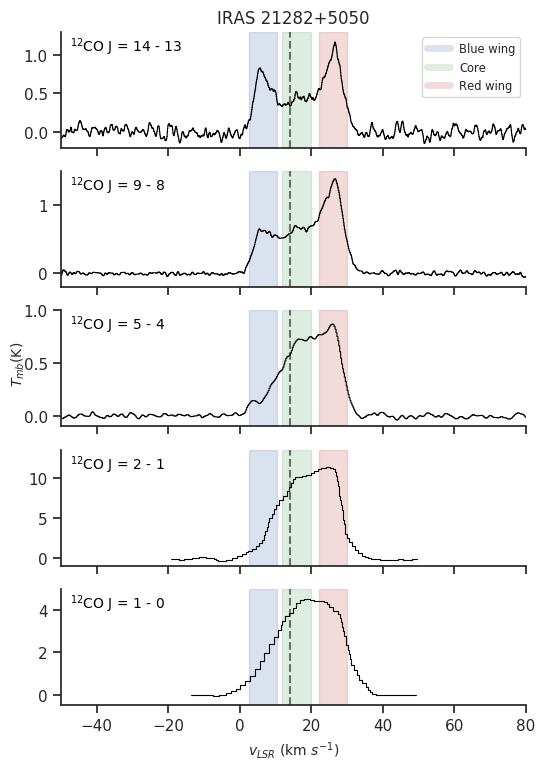}
        \caption{Observations of IRAS 21282+5050. Shadowed regions mark the LSR velocity ranges adopted to determine the characteristic intensity of each feature. The dashed vertical lines indicates the systemic velocity as tabulated in Table~\ref{tab:properties}.}\label{fig:IRAS21282_ranges}
    \end{center}
\end{figure}

\begin{figure}
    \begin{center}
        \setlength\abovecaptionskip{-0.01\baselineskip}
        \setlength\belowcaptionskip{-0.7\baselineskip}
        \includegraphics[width= \linewidth]{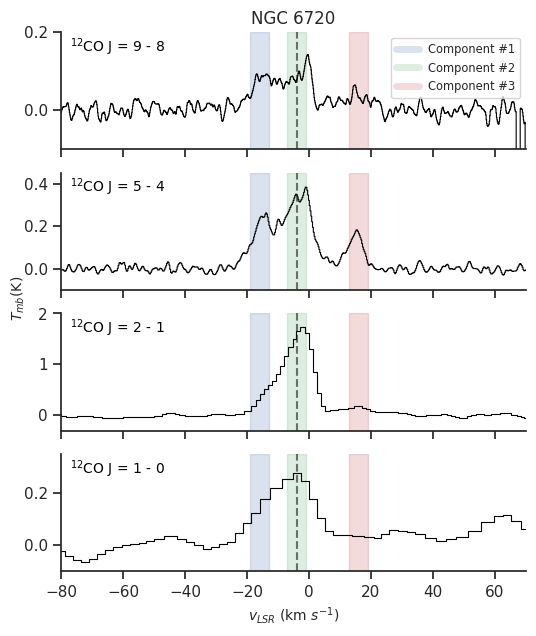}
        \caption{Observations of NGC 6720. The shadowed regions indicate the velocity ranges adopted to determine the characteristic intensity of the detected velocity components; see Section~\ref{sec:results_ranges} for further details.}\label{fig:NGC6720_ranges}
    \end{center}
\end{figure}

Our HIFI data were complemented with low-$J$ CO spectra obtained during an observing campaign at the IRAM 30m telescope, aimed at getting better calibrated data for some of the sources where the only available spectra had been taken several decades ago. These observations were conducted in December 2012 and comprised five sources from our sample of post-AGB stars. For the rest of the sources, we relied on data digitised from the literature or unpublished data available to members of our team. Table~\ref{tab:lowco} summarises the origin of the spectra used in this paper for the low-excitation CO lines. The corresponding spectra are shown in Appendix~\ref{sec:appxdata}. In the analysis described in Section~\ref{sec:results}, we distinguish among the compact and extended sources and we provide examples of the information shown together with the spectra of these sources in Figures~\ref{fig:IRAS21282_ranges} and \ref{fig:NGC6720_ranges}, respectively. For some sources, spectra are also given for additional CO transitions (e.g.~Fig.\ref{fig:NGC6537_ranges}). These were used in support to the selection of the velocity ranges applied in our analysis (Section~\ref{sec:results_ranges}).

In two cases (IRAS 21282+5050 and OH17.7-2.0), the CO data stem from interferometric observations as no single-dish counterpart could be found in the literature. The two sources concerned by these observations are either compact (OH17.7-2.0), or their interferometric data were combined with short spacings obtained at the 30m (IRAS 21282+5050). We have therefore assumed that there is no significant flux loss in the corresponding data. The conversion from surface brightness to temperature units we used were: 45 mK Jy$^{-1}$ beam$^{-1}$ (CO \mbox{$J$=1--0}) and 29 mK Jy$^{-1}$ beam$^{-1}$ (CO \mbox{$J$=2--1}) for the IRAM Plateau de Bure observations of \citet{castrocarrizo_2010}, and 1.19 K Jy$^{-1}$ beam$^{-1}$ (CO \mbox{$J$=1--0}) for the OVRO observations of \citet{sanchez-contreras_2007}.

\begin{table}[t]

\caption{Details on the origin of the low-$J$ CO spectra used in this study.}
\label{tab:lowco}

\resizebox{0.5\textwidth}{!}{%
\centering
\begin{tabular}{lcc}
\hline
\hline
Source & CO \mbox{$J$=1--0} & CO \mbox{$J$=2--1} \\ \hline
IRAS 07134+1005 &   \multicolumn{2}{c}{\citet{bujarrabal_1992}$^a$} \\
IRAS 19500--1709 &  \multicolumn{2}{c}{Bujarrabal private communication$^a$} \\
IRAS 21282+5050 &  \multicolumn{2}{c}{\citet{castrocarrizo_2010}$\mathbf{^{a,b}}$} \\
IRAS 22272+5435 &   \multicolumn{2}{c}{Bujarrabal private communication$^a$} \\
M 1--92   &  \multicolumn{2}{c}{New 30m data} \\
M 2--56   &  \multicolumn{2}{c}{Bujarrabal private communication$^a$}  \\
OH 17.7--2.0  & \citet{sanchez-contreras_2007}$^c$  & \citet{heske_1990}$^a$ \\
R Sct  &  \multicolumn{2}{c}{Bujarrabal private communication$^a$} \\
89 Her  &   \multicolumn{2}{c}{Bujarrabal private communication$^a$}  \\
NGC 6537   & \multicolumn{2}{c}{New 30m data}  \\
M 1--16   &  \multicolumn{2}{c}{New 30m data}\\
M 1--17   &  \multicolumn{2}{c}{New 30m data} \\
M 3--28  &   \multicolumn{2}{c}{New 30m data} \\
NGC 2346  & \multicolumn{2}{c}{\citet{bachiller_1989}$^a$} \\
NGC 6072  &  \multicolumn{2}{c}{\citet{cox_1992}$^d$} \\
NGC 6720  & \multicolumn{2}{c}{\citet{bachiller_1989}$^a$}\\
NGC 6781  &   \multicolumn{2}{c}{\citet{bachiller_1993}$^a$}  \\
NGC 7293  &   \citet{zack_2013}$^e$ & \citet{young_1999}$^f$\\

\hline
\end{tabular}
}
\begin{tablenotes}
      \small
      \item \textbf{Notes.} $^a$ IRAM 30 m telescope, $^b$ IRAM Plateau de Bure interferometer, \mbox{$^c$ OVRO interferometer,} $^d$ SEST telescope, $^e$ ARO 12 m telescope, \mbox{$^f$ CSO 10.4 m telescope.}
    \end{tablenotes}
\end{table}

\begin{table}[t]

\caption{Frequencies and telescope parameters for the CO transitions used in this work.}
\label{tab:telescopes}
\centering
\begin{tabular}{lccc}
\hline
\hline
Transition & \begin{tabular}[c]{@{}c@{}}Frequency\\(GHz)\end{tabular}  & Telescope & \begin{tabular}[c]{@{}c@{}}$\theta$\\('')\end{tabular} \\ \hline
$^{12}$CO \mbox{$J$=1--0} & 115.271 & IRAM & 21.4 \\
  &   & PdBI & 4.9$\times$3.9 \\
  &   & OVRO & 10.9$\times$7.1 \\
  &   & SEST & 45 \\
  &   & ARO & 55 \\
$^{12}$CO \mbox{$J$=2--1} & 230.538 & IRAM & 10.7 \\
  &   & PdBI & 2.6$\times$1.4 \\
  &   & SEST & 23 \\
  &   & CSO & 31 \\
$^{12}$CO \mbox{$J$=3--2} & 345.796 & APEX  & 17.3 \\
  &   & ARO SMT & 22 \\
$^{12}$CO \mbox{$J$=4--3} & 461.041 & ARO SMT  & 16 \\
$^{12}$CO \mbox{$J$=5--4} & 576.268 & HIFI & 36.1 \\
$^{12}$CO \mbox{$J$=6--5} & 691.473 & HIFI & 30.4 \\
  &   & ARO SMT & 11 \\
$^{12}$CO \mbox{$J$=9--8} & 1036.912 & HIFI & 20.1 \\
$^{12}$CO \mbox{$J$=10--9} & 1151.985 & HIFI & 18.2 \\
$^{12}$CO \mbox{$J$=14--13} & 1611.794 & HIFI & 12.9 \\
$^{12}$CO \mbox{$J$=16--15} & 11841.345 & HIFI & 11.5 \\
\hline
\end{tabular}

\begin{tablenotes}
      \small
      \item \textbf{Notes.} IRAM is the 30 m telescope at the Institute de Radioastronomie Millimétrique; PdBI refers to the IRAM Plateau de Bure interferometer (now known as NOEMA); OVRO refers to the millimeter interferometer of the Owens Valley Radio Observatory; SEST is the 15 m Swedish-ESO Sub-millimeter Telescope; ARO refers to the 12 m telescope of the Arizona Radio Observatory; CSO is the 10.4 m Leighton telescope of the Caltech Submilimeter Observatory; ARO SMT is the 10 m Submillimeter Telescope of the Arizona Radio Observatory; APEX refers to the Atacama Pathfinder EXperiment; and HIFI is the Heterodyne Instrument for the Far-Infrared aboard \textit{Herschel}.
    \end{tablenotes}
\end{table}

\section{Description of the sample}
\label{sec:sample}

\begin{table*}[t]

\caption{Observed sources.}
\label{tab:properties}

\resizebox{1\textwidth}{!}{%
\centering
\begin{tabular}{lcccccccccc}
\hline
\hline
Source &
\begin{tabular}[c]{@{}c@{}}$\alpha$\\ (J2000)\end{tabular} &
\begin{tabular}[c]{@{}c@{}}$\delta$\\ (J2000)\end{tabular} &
Type &
SpT &
\begin{tabular}[c]{@{}c@{}}$d$\\ (kpc)\end{tabular} &
\begin{tabular}[c]{@{}c@{}}$L$\\ (10$^3$ L$_{\odot}$)\end{tabular} &
\begin{tabular}[c]{@{}c@{}}$v_{\rm sys}$\\ (km s$^{-1}$) \end{tabular} &
\begin{tabular}[c]{@{}c@{}}$\dot{M}$ \\ (M$_{\odot}$ yr$^{-1}$) \end{tabular} &
Morph. &
Ref. \\ \hline
{\bf pPNe} \\
IRAS 07134+1005 & 07 16 10.258 & +09 59 47.99 & C & F5Iab& 3 & 13.5 & 72 & $5\times10^{-5}$ & b & 1, 2 \\

IRAS 19500--1709 & 19 52 52.701 & -17 01 50.30 & C & F2-6 & 1 & 1.5 & 25 & - & - & 1\\

IRAS 21282+5050 & 21 29 58.420 & +51 03 59.80 & C & 09.5, WC11 & 3 & 5.3 & 14 & 6$\times$10$^{-5}$ & - & 1, 2 \\

IRAS 22272+5435 & 22 29 10.374 & +54 51 06.35 & C & G5Ia & 1.7 & 8.3 & -28 & 2$\times$10$^{-5}$ & s &1\\

M 1--92 & 19 36 18.910 & +29 32 50.00 & O & B0.5IV & 2.5 & 10 & -1 & 2.2$\times$10$^{-4}$ & b & 1, 3 \\

M 2--56 & 23 56 36.400 & +70 48 18.20 & - & Be & 3 & 10 & -27 & 2.3$\times$10$^{-5}$ & b & 1\\

OH 17.7--2.0 & 08 30 31.000 & -14 28 57.00 & O & F0 & 2 & 2.9 & 61 & 1$\times$10$^{-4}$ & b & 1, 4\\

R Sct$^{\dagger}$ & 18 47 28.950 & -05 42 18.53 & O & G0-K2 & 0.4 & 4 & 56 & 3.3$\times$10$^{-7}$ & - & 1\\

89 Her$^{\dagger}$ & 17 55 25.189 & +26 02 59.97 & O & F2Ibe & 0.6 & 3.3 & -8 & 4$\times$10$^{-6}$ & b & 1, 5\\
\hline

{\bf Compact PNe} \\
NGC 6537 & 18 05 13.104 & -19 50 34.88 & O* & - & 1.6 & 2 & 8.6 & 1$\times$10$^{-7}$ & b & 6, 7, 8 \\

M 1--16 & 07 37 18.955 & -09 38 49.67 & O & - & 1.8 & 0.12 & 50 & 2$\times$10$^{-5}$ & e & 2 \\

M 1--17 & 07 40 22.206 & -11 32 29.81 & - & -& 5.8 & - & 28 & - & c & 9, 10\\ 

M 3--28 & 18 32 41.288 & -10 05 50.03 & C & - & 2.5 & - & 32 & - & q & 6, 11\\

\hline

{\bf Extended PNe} \\
NGC 2346 & 07 09 22.521 & -00 48 23.60 & - & A5V & 0.9 & - & 2 & - & b & 6 \\

NGC 6072 & 16 12 58.079 & -36 13 46.06 & - & B & 1.39 & 0.39 & 15 & - & b & 6 \\

NGC 6720 & 18 53 35.079 & +33 01 45.03 & O & hgO(H) & 0.7 & 1 & -4 & 2$\times$10$^{-4}$ & b & 6, 2 \\

NGC 6781 & 19 18 28.085 & +06 32 19.29 & C & DAO & 0.95 & 0.35 & 17 & - & e & 6 \\

NGC 7293 & 22 29 38.550 & -20 50 13.60 & C & DAO.5 & 0.22 & 0.09 & -24 & 1$\times$10$^{-4}$ & b & 6, 2 \\

\hline
\end{tabular}
}
\begin{tablenotes}
      \small
      \item $^{\dagger}$: Bipolar nebula with rotational disk
      \item *: \citet{edwards_ziurys_2013} questioned how O-rich NGC 6537 is.
      \item Morph. abbreviations: b: bipolar or hourglass-like, c:compact, e: elliptical, q: quadrupolar, s: spherical.
      \item \textbf{References:} (1): \citet{bujarrabal_2001}; (2): \citet{fong_2006}; (3): \citet{bujarrabal_1998}; (4): \citet{bujarrabal_1994}; (5): \citet{alcolea_1995}; (6): \citet{frew_2008}; (7): \citet{cuesta_1995}; (8): \citet{edwards_ziurys_2013}; (9): \citet{maciel_1984}; (10): \citet{bachiller_1991}; (11): \citet{ziurys_2020}.
    \end{tablenotes}


\end{table*}

Our sample is presented in Table \ref{tab:properties}, alongside with their observed coordinates and other parameters. It consists of 18 post-AGBs: nine protoplanetary nebulae and nine planetary nebulae. Five of the PNe are considered extended sources with respect to the size of the HIFI beams (see their values in Table \ref{tab:telescopes}). 
In this section, we present a brief description of each source and a qualitative analysis of their observed CO profiles.


\subsection{IRAS 07134+1005}
\label{sec:IRAS07134}
IRAS 07134+1005 (HD 56126) is a pPN with a double-peaked SED \citep{kwok_1989}, 
presenting a high abundance of s-process elements, a typical characteristic of post-AGBs that have experienced the third dredge-up. 
Mid-IR images of IRAS 07134+1005 \citep{meixner_1997} show an elliptical detached shell surrounding two aligned peaks, which these authors interpreted as limb-brightened peaks of a shell with an equatorial density enhancement. The profile of the CO \mbox{$J$=9--8} line obtained in our observations shows a double peak around the line core, indicative of high-velocity shocks (Fig.~\ref{fig:IRAS07134_ranges}).

\subsection{IRAS 19500--1709}
\label{sec:IRAS19500}
IRAS 19500--1709 (HD 187885) is a carbon-rich pPN. The high abundance of s-process elements in its shell and a feature at 21 $\mu$m in its spectra \citep{vanWinckel_2000} suggests that the third dredge-up has already occurred in this object. IRAS 19500--1709 has a bi-modal spectral energy distribution (SED) characteristic of pPN with detached circumstellar dust envelope \citep{hrivnak_1989}. Our observed lines reflect this composite structure. They show, in most of them, a central line core (with an expansion velocity of 11 km s$^{-1}$) and wide line wings whose expansion velocity goes up to $\sim$ 30 km s$^{-1}$. Figure \ref{fig:IRAS19500_ranges} shows that the contribution from the remnant of the slow AGB wind is dominating the profile at low $J$, but then decreases significantly in the \mbox{$J$=5--4} line and above. This suggests that the temperature of the fast outflows is high in contrast with the slow expanding AGB-CSE. In addition, we can see a clear absorption in the blue wing of \mbox{$J$=2--1} line, supporting the idea that the fast component presents higher temperatures than the slow gas and that it is found closer to the central star.
    
\subsection{IRAS 21282+5050}
\label{sec:IRAS21282}
IRAS 21282+5050 is a young, C-rich protoplanetary nebula. The CO \mbox{$J$=1--0} maps by \citet{shibata_1989} show a clumpy expanding shell whose nature was interpreted as an expanding toroidal structure with an axis along the east-west direction, perpendicular to the line of sight. The spectra of the lower-energy transitions displayed in Figure \ref{fig:IRAS21282_ranges} exhibit profiles dominated by a roughly symmetric component around the systemic velocity. However, as the energy of the transition increases, the profile seems more of a composite as well as steep-sided, revealing the three typical features: two line wings growing in intensity with increasing $J$ and a line-core decreasing with the excitation, indicative of a high temperature in the object's outflows.

\subsection{IRAS 22272+5435}
\label{sec:IRAS22272}
IRAS 22272+5435 (HD 235858) is a C-rich pPN \citep{kwok_1989, omont_1995}. Studies at optical \citep{ueta_2000} and near-infrared \citep{gledhill_2001} wavelengths have identified an axisymmetric nature (P.A. $\sim$ -35º) for the nebula and a toroidal dust shell normal to the axis. Maps of CO \mbox{$J$=1--0} \citep{fong_2006} have shown that its molecular gas expands following a spherical structure with multiple protrusions in a quadrupolar geometry. As can be seen in Fig. \ref{fig:IRAS22272_ranges}, its profile displays a parabolic-like shape as commonly observed in AGB profiles. 

\subsection{M 1--92}
\label{sec:M1-92}
M 1--92, also known as Minkowski’s Footprint, is a very well studied O-rich pPN that presents OH masers. It shows a clear hourglass-like structure consisting of a thin flat disk expanding radially and two symmetrical lobes \citep{alcolea_2007}. The CO maps show that most of its molecular gas is located in the equatorial disk and in the walls of two empty lobules, where high velocities are observed \citep{bujarrabal_1998}. Figure \ref{fig:M1-92_ranges} shows intermediate velocity features that disappear as the energy levels increase.

\subsection{M 2--56}
\label{sec:M2-56}
M 2--56 is a pPN with a bipolar appearance in the optical and an hourglass-like shape in CO maps. These maps show a dense ring probably arising from the fossil AGB envelope and two high-velocity components at 11''-15'' from the centre along the symmetry axis. It also presents two empty attached shells close to the equatorial disk, expanding along the nebula axis at low velocities \citep{castro_2002}. As we can seen in Fig. \ref{fig:M2-56_ranges}, the CO lines of M 2--56 are composite with a strong blue wing that remains as the dominant feature of the spectral profile at high values of $J$. This remarkable difference between the blue and the red wings is probably due to an intrinsic asymmetry.

\subsection{OH 17.7--2.0}
\label{sec:OH17}
OH17.7-2.0 (IRAS 18276+143) is an OH/IR pPN that presents a double-peaked SED indicative of a detached dust shell \citep{bertre_1987}. Polarised intensity images \citep{gledhill_2005} have found hourglass-like geometry with a symmetric axis at P.A. $\sim$ 23º. Horn-like features seen in its position-velocity diagram suggest the presence of outflows close to the source \citep{bains_2003}. A relative absence of OH masers is observed in its polar regions, compared with the strong emission detected near its equator (also seen in M 1--92). This fact suggests that these areas may have been excavated by fast outflows creating empty symmetric lobes of gas \citep{sanchez-contreras_2007}. Although our HIFI observations for this source are relatively noisy (Fig.~\ref{fig:OH17_ranges}), for CO \mbox{$J$=9--8}, they reveal in  a feature around 40 km s$^{-1}$ expanding at $\sim$ 10 km s$^{-1}$.

\subsection{R Sct}
\label{sec:R_SCT}
R Sct (HR 7066, HIP 92202) is a peculiar RV Tau variable whose evolutionary stage is still under debate. It presents a complex profile in all CO lines (Fig~\ref{fig:R_SCT_ranges}). In the low-$J$ transition \mbox{$J$=2--1}, we observe a narrow double peak, which is a characteristic property of rotating disks \citep{bujarrabal_2013}. A strong blue wing is seen at these low energy levels, but as $J$ increases, the emission of this feature is abated,  suggesting a low temperature for the approaching outflow. Within our sample, this source is the one exhibiting the narrowest line profile ($\sim$ 10\,km s$^{-1}$).

\subsection{89 Her}
\label{sec:89_HER}
89 Her (HR 6685) is a low-mass pPN. It is a close binary system known to present a remarkable near-infrared excess due to the emission of hot dust at $\sim$ 1000 K \citep{waters_1993, deruyter_2006} and a clear far-infrared excess suggesting the presence of large grains in its dusty envelope \citep{shenton_1995}. In the \mbox{$J$=2--1} and \mbox{$J$=1--0} CO lines presented in Fig. \ref{fig:89_HER_ranges}, we observe sharp profiles with a prominent central peak and weak wings at low velocities. At higher $J$, the structure of the profile does not change, yet their emission becomes weaker and approaches the noise level, while the wings are practically undetectable. According to its CO maps \citep{bujarrabal_2007, bujarrabal_2013}, this profile corresponds to a compact and unresolved component in the centre of the nebula which is suspected to be a rotating disk and an hourglass-like extended outflow (with a maximum velocity of $\sim$ 7 km s$^{-1}$).

\subsection{NGC 6537}
\label{sec:NGC6537}
The Red Spider nebula is one of the PNe with the highest excitation  known thus far \citep{ashley_1988}. Its spider morphology is mainly observed in the atomic emission lines, but it can also be seen in molecular hydrogen. NGC 6537 is hourglass-like and possibly multi-polar in its nebular shape \citep{cuesta_1995}. Because this PN is relatively young (its age is estimated to be $\sim$ 1600 yr, \citet{matsuura_2005}), it has been suggested that the remnant molecular gas has not yet been accelerated enough in the polar outflows \citep{edwards_2014}. In Fig. \ref{fig:NGC6537_ranges}, we observe that the strong blue wing, seen as a left peak in \mbox{$J$=2--1} and \mbox{$J$=1--0} lines, decreases with $J$, suggesting a low temperature in the approaching outflow. There is an important interstellar medium contamination between 20 and $\sim$ 45 km s$^{-1}$ in the two lowest-$J$ CO lines, which we have tried to avoid in our study. The same tendency is seen in \mbox{$J$=6--5} and \mbox{$J$=4--3} lines presented in \citet{edwards_2014}. 

Although the FIR emission of this source fills  the HIFI beam for the \mbox{$J$=5--4} line entirely and exceeds it for the other lines, we considered it as a compact PN because we pointed at its centre and not their external velocity components, in contrast with what was carried out for the extended PNe. However, the fact that the lines do not come from an area smaller than the HIFI beams prevent us from performing accurate comparisons with the other compact objects and theoretical models (see Section~\ref{sec:results}).

\subsection{M 1--16}
\label{sec:M1-16}
M 1--16 is a young, compact planetary nebula known to exhibit multiple outflows with different velocities. It presents a slow outflow (with an expansion velocity of 19 km s$^{-1}$), which exhibits an elliptical structure and high-velocity flows consisting of three bipolar jets with shock velocities of $\sim$ 300 km s$^{-1}$. These fast outflows have excavated extended bipolar cavities producing symmetric lobes of molecular gas \citep{huggins_2000}. The most remarkable characteristics seen in our observed lines presented in Fig. \ref{fig:M1-16_ranges} are the extended and strong wings, indicative of high-velocity outflows and their decreasing contribution with the increase of the excitation, leaving a peak with a broad but weak pedestal in \mbox{$J$=9--8}.
    
\subsection{M 1--17}
\label{sec:M1-17}
M 1--17 is a compact planetary nebula whose emission is weak at practically all wavelengths, except at the observed CO lines. \citet{bachiller_1991}  studied this source in \mbox{$J$=2--1} and \mbox{$J$=1--0} and identified two components in its profile: a slow wind, with a double-peaked profile extending over 34 km s$^{-1}$ and a fast wind, with a triangular-shape profile extending over 78 km s$^{-1}$. Our observations at intermediate-energy transitions (Fig.~\ref{fig:M1-17_ranges}) provide additional information about the molecular content of this nebula and reveal in particular a new component further away from the systemic velocity, expanding over less than 5 km s$^{-1}$. As the energy increases, the medium-velocity components known so far lose intensity and the component in the outskirts kicks in, indicating a higher temperature in the latter and hinting towards a possible acceleration gradient.

\subsection{M 3--28}
\label{sec:M3-28}
M 3--28 is a small PN, with a bipolar, butterfly-like morphology and a filamentary structure \citep{calvet_1978}. As shown in Fig. \ref{fig:M3-28_ranges}, the line profile is relatively broad, but its exact extent at the red end cannot be accurately assessed because of the strong interstellar medium contamination that is especially visible in the low excitation lines. Our observation at higher $J$, however, allows us to appreciate, for the first time,  the full breadth of the profile, which spans nearly 80\,km s$^{-1}$ overall. This line contamination was taken into account when selecting the velocity ranges of interest for this study (see Section~\ref{sec:results_ranges}). Apart from that, the global profile is that of a double peak on either side of a weak line core, similar to that observed in other sources, such as M 1--17 or IRAS 21282+5050. 

\subsection{NGC 2346}
\label{sec:NGC2346}
NGC 2346 is an extended bipolar PN. It has an hourglass-like geometry consisting of two open bipolar lobes and an expanding toroidal structure in its centre. This equatorial torus consists of large amounts of molecular gas and dust, and it is tilted 56$^{\circ}$ with respect to the plane of the sky \citep{su_2004}. The source is extended compared to the beams of our observations (see Fig.~\ref{fig:NGC2346_optico}) and the targeted region corresponds to an edge of this toroidal structure. Because, in terms of size, the CO \mbox{$J$=2--1} beam is smaller than the others, our profile consists of just one feature at the systemic velocity of the nebula. In contrast, in the other CO lines, we observe four features corresponding to four different velocity components (Fig~\ref{fig:NGC2346_ranges}). Their emission drastically decreases at high-energy transitions, suggesting a low temperature in all these components.

\subsection{NGC 6072}
\label{sec:NGC6072}
Near-infrared and molecular hydrogen images \citep{kwok_2010} have characterised NGC 6072 as having an elliptical, equatorial ring and three pairs of lobes, indicating a multipolar structure. This morphology is the result of fast outflows interacting with the slow winds of the AGB phase and whose orientation is changing over time. Most of our observed CO lines present a double-peaked profile, indicative of a detached molecular shell expanding at $\sim$ 15 km s$^{-1}$ (Fig.~\ref{fig:NGC6072_ranges}). However, the emission is fully extinguished in the CO \mbox{$J$=14--13} line, suggesting a low temperature in this expanding shell.

\subsection{NGC 6720}
\label{sec:NGC6720}
The Ring Nebula is an extended, O-rich planetary nebula with a bipolar morphology tilted with respect to the line of sight (\citealp{kwok_2008, odell_2013}). Observations of CO \mbox{$J$=2--1} and \mbox{$J$=1--0} lines carried out by \citet{bachiller_1989} and presented in Fig. \ref{fig:NGC6720_ranges} revealed unusually-high-velocity molecular gas. Owing, most likely, to the larger size of the CO \mbox{$J$=5--4} beam in our HIFI observations, we detected two more velocity components at roughly $\sim$ 15 km s$^{-1}$ on either side of the systemic velocity. The fact that the emission of all of these components remains visible in the CO \mbox{$J$=9--8} line suggests high-temperature molecular gas at the outskirts of this nebula.
    
\subsection{NGC 6781}
\label{sec:NGC6781}
NGC 6781 has an elliptical morphology very similar to that of the Ring nebula. High-resolution mapping of CO \mbox{$J$=2--1} and \mbox{$J$=1--0} lines by \citet{bachiller_1993} has provided a detailed kinematic structure of its envelope. A model, consisting of a thin, expanding, ellipsoidal shell, truncated at the two ends, was proposed by the authors to explain the observation. As we can see in Fig. \ref{fig:NGC6781_ranges}, our line profiles are very complex, revealing a clumpy nature for the CO shell.

\subsection{NGC 7293}
\label{sec:NGC7293}
The Helix nebula is the closest planetary nebula of our sample.  It presents an extreme bipolar morphology tilted with respect to the line of sight and a central toroidal ring. This last feature expands around the ionized nebula with arcs and filaments at high latitudes (\citealp{meaburn_2005, zeigler_2013}). The edge of one of these filaments corresponds to the position we have observed in this work, at offset (-435'', 75'') from the central star (Fig.~\ref{fig:NGC7293_optico}). As can be seen in Fig. \ref{fig:NGC7293_ranges}, the profile reveals various features that we treat as separate velocity components resolved along the line of sight. Their relative intensity remains constant, except in the CO \mbox{$J$=9--8} line, where their emission drops drastically, indicating that none of the observed components exhibits a high temperature. This observation suggests that most of the molecular gas of the PN has not recently been ejected, an assumption that is also supported by the findings of \citet{zack_ziurys_2013} and \citet{schmidt_2018}.

\section{Analysis and discussion}
\label{sec:results}

\subsection{Characteristic intensities}
\label{sec:results_ranges}

The different velocity components evidenced in our high-resolution profiles of the CO lines provide detailed information about the excitation condition in the outflows of our objects. In particular, the relative strength of the respective spectral components and their evolution with $J$ can vary significantly between the various sources and we use  this signature here as a proxy of the evolutionary state of the nebula. 

For this, we follow the method
presented in \citet{bujarrabal_2012}, where the contribution of each line component is estimated through its characteristic intensity over a selected velocity range in the relevant CO transition. 
%
%
We note beforehand that the physical meaning of the selected velocity ranges is different depending on the relative size of our objects with respect to the diameter of the instrument beams (for the record, the applicable beam sizes are collected in Table~\ref{tab:telescopes} and plotted on their optical and FIR images in Appendix~\ref{sec:appxa}). In the case of compact objects, we interpret the velocity components as those of a central line core associated with the remnant of the slow AGB wind, bracketed by two (blue and red) line wings, assumed to come from the fast outflows resulting from the post-AGB axial acceleration. For two specific sources (M 1--92 and M 1--17), we also considered additional intermediate velocity ranges, that we analyse in Section~\ref{intermediate}. In the case of sources spatially resolved in most of the probed CO line (thereafter, `extended' sources), we interpret the various features in their profiles as different velocity components probed along the line of sight crossing the nebular gas, mostly in their outskirts.

In order to define the respective velocity ranges in a systematic and objective fashion, we have taken into account the information provided in the low-$J$ transition of CO, the systemic velocity assumed for the object, and the spatial distribution of the CO emission available, and for the extended sources, in particular. 
We followed a systematic procedure consisting of the selection of the velocity range of the central core that is consistent with the systemic velocity of the source. The width of the velocity component was chosen as a compromise between a sufficient number of spectral channels compared to the noise level and line ranges being sufficiently separated to avoid too much mixing between the gas probed in the respective regions. The same width was then applied to the regions associated to the red and blue wings.
Overall, this leads to red and blue regions being relatively symmetrical on either side of the central core. For the narrowest profiles, the requirement to separate the envelope regions probed in each component is hard to fulfil, but we checked that small adjustments up and down the chosen widths do not impact the conclusions derived in this paper.
In the case of M 3--28, the strong interstellar medium contamination, which is especially visible in the lower-$J$ transitions was taken into account to adjust the position of the red wing, avoiding that it would fall too close to this contribution. 
For M 1--17 and M 1--92, additional velocity regions were considered, as explained in Section~\ref{intermediate}.

For the extended source sample, the various features are interpreted as separate velocity components along the line of sight. For this reason, the selection essentially considered the three main lines contributing to the observed profiles. The strategy regarding the choice of the velocity region width was the same as for the compact sources. In all cases, except for NGC 7293, one of the components is centred exactly on the assumed systemic velocity. In NGC 7293, the systemic velocity of -24\,km s$^{-1}$ does not match well with the blue-shifted peak observed in the higher spectral resolution profile obtained for $J$=2--1 and above. The velocity corresponding to this peak (-21.7\,km s$^{-1}$) was, therefore, assigned instead to one of the components in this source. In all cases, the choice of the other two components was guided by the other line peak intensities observed in the line profile, implying that they did not necessarily fall on either side of the component selected at the systemic velocity.

The adopted velocity ranges and their characteristic intensities are compiled, alongside their corresponding 1-$\sigma$ noise uncertainties, in Table \ref{tab:comp_results} (compact objects) and Table~\ref{tab:ext_results} (extended objects). In addition, the green, blue, and red coloured intervals drawn in Figs. \ref{fig:IRAS07134_ranges} to \ref{fig:NGC7293_ranges} denote the selected regions.

\begin{table*}
\caption{Characteristic intensities of the three representative features identified in the CO profiles of the compact objects, and their corresponding coordinates in the pseudo cc-diagram presented in Figs. \ref{fig:ccdiagram_dots} and \ref{fig:ccdiagram_arrows}.}
\label{tab:comp_results}
\resizebox*{!}{0.95\textheight}{%
\centering
\begin{tabular}{lllcccccc}
\hline \hline
Source &
$^{12}$CO line &
\begin{tabular}[c]{@{}c@{}}Observation\\ ID\end{tabular} &
\begin{tabular}[c]{@{}c@{}}Central core\\ average T$_{mb}$(K)\end{tabular} &
\begin{tabular}[c]{@{}c@{}}Blue wing\\ av. T$_{mb}$(K)\end{tabular} &
\begin{tabular}[c]{@{}c@{}}Red wing\\ av. T$_{mb}$(K)\end{tabular} &
\begin{tabular}[c]{@{}c@{}}Noise\\ r.m.s. (K)\end{tabular} &
\begin{tabular}[c]{@{}c@{}}Line ratios\\ Central core\end{tabular} & 
\begin{tabular}[c]{@{}c@{}}Line ratios\\ Averaged line wings\end{tabular}\\
\hline
\multicolumn{2}{c}{LSR velocity ranges (km s$^{-1}$)} & & \multicolumn{1}{c}{69.5 -- 74.5} & \multicolumn{1}{c}{61.5 -- 66.5} & \multicolumn{1}{c}{77.5 -- 82.5} & \multicolumn{1}{c}{} & \multicolumn{1}{c}{} & \multicolumn{1}{c}{}\\
IRAS 07134+1005 
 & \mbox{$J$=1--0} & & 1.32 & 0.54 & 1.05 & 0.05 & R$_{54/21}$ = 0.13 & R$_{54/21}$ = 0.15 \\
 & \mbox{$J$=2--1} & & 2.05 & 0.48 & 1.22 & 0.04 & & \\
 & \mbox{$J$=5--4} & 1342244408 & 0.263 & 0.090 & 0.134 & 0.004 & R$_{98/54}$ = 1.07 & R$_{98/54}$ = 2.05\\
 & \mbox{$J$=9--8} & 1342229927 & 0.281 & 0.236 & 0.200 & 0.004 & & \\
\hline
\multicolumn{2}{c}{LSR velocity ranges (km s$^{-1}$)} & & \multicolumn{1}{c}{19.5 -- 29.5} & \multicolumn{1}{c}{1 -- 11} & \multicolumn{1}{c}{40 -- 50} & \multicolumn{1}{c}{} & \multicolumn{1}{c}{} & \multicolumn{1}{c}{} \\
IRAS 19500--1709 
 & \mbox{$J$=1--0} & & 0.83 & 0.15 & 0.06 & 0.01 & &  \\
 & \mbox{$J$=2--1} & & 2.06 & 0.62 & 0.35 & 0.01 & R$_{54/21}$ = 0.05 & R$_{54/21}$ = 0.11 \\
 & \mbox{$J$=5--4} & 1342216805 & 0.104 & 0.068 & 0.036 & 0.003 & &  \\
 & \mbox{$J$=9--8} & 1342216811 & 0.224 & 0.162 & 0.076 & 0.003 & R$_{98/54}$ = 2.16 & R$_{98/54}$ = 2.24 \\
 & \mbox{$J$=14--13} &  
1342231774 & 0.13 & 0.09 & 0.05 & 0.01 & & \\ 
\hline
\multicolumn{2}{c}{LSR velocity ranges (km s$^{-1}$)} & & \multicolumn{1}{c}{11.9 -- 19.9} & \multicolumn{1}{c}{2.5 -- 10.5} & \multicolumn{1}{c}{22 -- 30} & \multicolumn{1}{c}{} & \multicolumn{1}{c}{} & \multicolumn{1}{c}{} \\
IRAS 21282+5050
 & \mbox{$J$=1--0} & & 1.21 & 0.52 & 1.05 & 0.01  & & \\
 & \mbox{$J$=2--1} & & 2.30 & 0.78 & 1.99 & 0.02  & R$_{54/21}$ = 0.28 & R$_{54/21}$ = 0.30 \\
 & \mbox{$J$=5--4} &    
1342200921 & 0.653 & 0.199 & 0.704 & 0.004  & & \\
 & \mbox{$J$=9--8} &    
1342210667 & 0.626 & 0.522 & 1.054 & 0.004 & R$_{98/54}$ = 0.96 & R$_{98/54}$ = 2.06 \\
 & \mbox{$J$=14--13} & 1342220506 & 0.42 & 0.57 & 0.78 & 0.01 & & \\
\hline
\multicolumn{2}{c}{LSR velocity ranges (km s$^{-1}$)} & & \multicolumn{1}{c}{-30.5 -- -25.5} & \multicolumn{1}{c}{-37.3 -- -32.3} & \multicolumn{1}{c}{-23.2 -- -18.2} & \multicolumn{1}{c}{} & \multicolumn{1}{c}{} & \multicolumn{1}{c}{} \\
IRAS 22272+5435 
 & \mbox{$J$=1--0} & & 3.14 & 2.03 & 1.69 & 0.02  & R$_{54/21}$ = 0.034 & R$_{54/21}$ = 0.03\\
 & \mbox{$J$=2--1} & & 8.76 & 3.69 & 3.45 & 0.02  & & \\
 & \mbox{$J$=5--4} & 1342200922 & 0.294 & 0.120 & 0.115 & 0.004 & R$_{98/54}$ = 1.52 & R$_{98/54}$ = 2.12  \\
 & \mbox{$J$=9--8} & 1342213360 & 0.448 & 0.253 & 0.245 & 0.004 & &\\
\hline
\multicolumn{2}{c}{LSR velocity ranges (km s$^{-1}$)} & & \multicolumn{1}{c}{-7.5 -- 7.5} & \multicolumn{1}{c}{-41.5 -- -26.5} & \multicolumn{1}{c}{26.5 -- 41.5} & \multicolumn{1}{c}{} & \multicolumn{1}{c}{} & \multicolumn{1}{c}{} \\
M 1--92
 & \mbox{$J$=1--0} & & 0.60 & 0.16 & 0.20 & 0.01  & R$_{54/21}$ = 0.059 & R$_{54/21}$ = 0.08 \\
 & \mbox{$J$=2--1} & & 1.48 & 0.29 & 0.55 & 0.02  & & \\
 & \mbox{$J$=5--4} & 1342210076 & 0.087 & 0.032 & 0.030 & 0.003 & R$_{98/54}$ = 0.51 & R$_{98/54}$ = 1.64 \\
 & \mbox{$J$=9--8} &    
1342220500 & 0.045 & 0.045 & 0.055 & 0.002 & & \\
\hline
\multicolumn{2}{c}{LSR velocity ranges (km s$^{-1}$)} & & \multicolumn{1}{c}{-29 -- -13} & \multicolumn{1}{c}{-52 -- -36} & \multicolumn{1}{c}{5 -- 21} & \multicolumn{1}{c}{} & \multicolumn{1}{c}{} & \multicolumn{1}{c}{} \\
M 2--56 
 & \mbox{$J$=1--0} & & 0.181 & 0.152 & 0.058 & 0.003  & R$_{54/21}$ = 0.07 & -  \\
 & \mbox{$J$=2--1} & & 0.86 & 0.82 & 0.28 & 0.01  & & \\
 & \mbox{$J$=5--4} & 1342201534 & 0.058 & 0.067 & $\lesssim$ 0.005 & 0.003 & R$_{98/54}$ $\lesssim$ 0.08 & - \\
 & \mbox{$J$=9--8} & 1342213359 & $\lesssim$ 0.005 & $\lesssim$ 0.005 & $\lesssim$ 0.005 & 0.002 & & \\
\hline
\multicolumn{2}{c}{LSR velocity ranges (km s$^{-1}$)} & & \multicolumn{1}{c}{59.5 -- 64.5} & \multicolumn{1}{c}{49.5 -- 54.5} & \multicolumn{1}{c}{69.5 -- 74.5} & \multicolumn{1}{c}{} & \multicolumn{1}{c}{} & \multicolumn{1}{c}{} \\
OH 17.7--2.0 
 & \mbox{$J$=1--0} & & 1.48 & 1.13 & 0.84 & 0.07  & R$_{54/21}$  $\lesssim$ 0.01 & -\\
 & \mbox{$J$=2--1} & & 0.98 & 0.48 & 0.44 & 0.05  & & \\
 & \mbox{$J$=5--4} &    
1342218429 & $\lesssim$ 0.009 & $\lesssim$ 0.009 & $\lesssim$ 0.009 & 0.005 & - & - \\
 & \mbox{$J$=9--8} & 1342228622 & 0.064 & $\lesssim$ 0.008 & $\lesssim$ 0.008 & 0.004 & & \\
\hline
\multicolumn{2}{c}{LSR velocity ranges (km s$^{-1}$)} & & \multicolumn{1}{c}{54.5 -- 56.5} & \multicolumn{1}{c}{51 -- 53} & \multicolumn{1}{c}{57.2 -- 59.2} & \multicolumn{1}{c}{} & \multicolumn{1}{c}{} & \multicolumn{1}{c}{} \\
R Sct 
 & \mbox{$J$=1--0} & & 0.91 & 0.57 & 0.91 & 0.03  & R$_{54/21}$ = 0.05 & - \\
 & \mbox{$J$=2--1} & & 1.48 & 0.53 & 1.36 & 0.02  & & \\
 & \mbox{$J$=5--4} &    
1342230387 & 0.069 & $\lesssim$ 0.012 & 0.044 & 0.006 & R$_{98/54}$ $\lesssim$ 0.16 & - \\
 & \mbox{$J$=9--8} & 1342229922 & $\lesssim$ 0.011 & $\lesssim$ 0.011 & $\lesssim$ 0.011 & 0.005 & & \\
\hline
\multicolumn{2}{c}{LSR velocity ranges (km s$^{-1}$)} & & \multicolumn{1}{c}{-10 -- -6} & \multicolumn{1}{c}{-15.5 -- -11.5} & \multicolumn{1}{c}{-5 -- -1} & \multicolumn{1}{c}{} & \multicolumn{1}{c}{} & \multicolumn{1}{c}{} \\
89 Her 
 & \mbox{$J$=1--0} & & 0.62 & 0.2 & 0.18 & 0.01  & R$_{54/21}$ = 0.03 & - \\
 & \mbox{$J$=2--1} & & 1.36 & 0.44 & 0.30 & 0.01  & & \\
 & \mbox{$J$=5--4} & 1342214335 & 0.043 & $\lesssim$ 0.010 & $\lesssim$ 0.010 & 0.005 & R$_{98/54}$ = 0.71 & - \\
 & \mbox{$J$=9--8} & 1342218624 & 0.030 & $\lesssim$ 0.009 & $\lesssim$ 0.009 & 0.005 & & \\
\hline
\multicolumn{2}{c}{LSR velocity ranges (km s$^{-1}$)} & & \multicolumn{1}{c}{5.5 -- 10.5} & \multicolumn{1}{c}{-0.5 -- 4.5} & \multicolumn{1}{c}{11.5 -- 16.5} & \multicolumn{1}{c}{} & \multicolumn{1}{c}{} & \multicolumn{1}{c}{} \\
NGC 6537 
 & \mbox{$J$=1--0} & & 1.72 & 1.04 & 1.09 & 0.03  & &  \\
 & \mbox{$J$=2--1} & & 3.48 & 2.51 & 1.88 & 0.05  &  R$_{54/21}$ = 0.14 & R$_{54/21}$ = 0.12  \\
 & \mbox{$J$=5--4} &    
1342217693 & 0.471 & 0.211 & 0.296 & 0.004  &  &  \\
 & \mbox{$J$=9--8} &    
1342215913 & 0.917 & 0.286 & 0.608 & 0.004 &  R$_{98/54}$ = 1.95 & R$_{98/54}$ = 1.7 \\
 & \mbox{$J$=14--13} &  
1342229785 & 0.94 & 0.20 & 0.62 & 0.01 & & \\
\hline
\multicolumn{2}{c}{LSR velocity ranges (km s$^{-1}$)} & & \multicolumn{1}{c}{45.8 -- 54.2} & \multicolumn{1}{c}{35.8 -- 44.2} & \multicolumn{1}{c}{55.8 -- 64.2} & \multicolumn{1}{c}{} & \multicolumn{1}{c}{} & \multicolumn{1}{c}{} \\
M 1--16 
 & \mbox{$J$=1--0} & & 1.58 & 1.43 & 1.61 & 0.02  & R$_{54/21}$ = 0.07 & R$_{54/21}$ = 0.06 \\
 & \mbox{$J$=2--1} & & 3.91 & 3.24 & 3.74 & 0.03  & & \\
 & \mbox{$J$=5--4} &    
1342220515 & 0.271 & 0.237 & 0.210 & 0.003 & R$_{98/54}$ = 0.65 & R$_{98/54}$ = 0.40 \\
 & \mbox{$J$=9--8} & 1342220495 & 0.176 & 0.083 & 0.097 & 0.003 & & \\
\hline
\multicolumn{2}{c}{LSR velocity ranges (km s$^{-1}$)} & & \multicolumn{1}{c}{25 -- 31} & \multicolumn{1}{c}{3.5 -- 9.5} & \multicolumn{1}{c}{47 -- 53} & \multicolumn{1}{c}{} & \multicolumn{1}{c}{} & \multicolumn{1}{c}{} \\
M 1--17 
 & \mbox{$J$=1--0} & & 1.16 & 0.09 & 0.13 & 0.03  & R$_{54/21}$ = 0.06 & R$_{54/21}$ = 0.13 \\
 & \mbox{$J$=2--1} & & 1.93 & 0.54 & 0.37 & 0.04  & & \\
 & \mbox{$J$=5--4} & 1342218900 & 0.113 & 0.058 & 0.056 & 0.005 & R$_{98/54}$ = 1.02 & R$_{98/54}$ = 0.80 \\
 & \mbox{$J$=9--8} & 1342220494 & 0.115 & 0.053 & 0.038 & 0.004 & & \\
\hline
\multicolumn{2}{c}{LSR velocity ranges (km s$^{-1}$)} & & \multicolumn{1}{c}{29.5 -- 34.5} & \multicolumn{1}{c}{15.5 -- 20.5} & \multicolumn{1}{c}{41.5 -- 46.5} & \multicolumn{1}{c}{} & \multicolumn{1}{c}{} & \multicolumn{1}{c}{} \\
M 3--28 
 & \mbox{$J$=1--0} & & 1.93 & 2.99 & 3.09 & 0.02  & R$_{54/21}$ = 0.11 & R$_{54/21}$ = 0.10\\
 & \mbox{$J$=2--1} & & 2.61 & 4.22 & 4.90 & 0.04  & & \\
 & \mbox{$J$=5--4} & 1342218430 & 0.289 & 0.412 & 0.474 & 0.006 & R$_{98/54}$ = 0.47 & R$_{98/54}$ = 0.48\\
 & \mbox{$J$=9--8} & 1342218214 & 0.136 & 0.174 & 0.257 & 0.004 &  & \\
\hline 
\end{tabular}
}
\end{table*}

\begin{table*}
\caption{Characteristic intensities of the identified velocity components and the coordinates of the CO profiles of extended sources in the pseudo cc-diagram shown in Fig. \ref{fig:ccdiagram_dots}.}
\label{tab:ext_results}
\resizebox{\textwidth}{!}{%
\centering
\begin{tabular}{lllccccccc}

\hline \hline
Source &
$^{12}$CO line &
\begin{tabular}[c]{@{}c@{}}Observation\\ ID\end{tabular} &
\begin{tabular}[c]{@{}c@{}}Component \#1\\ average T$_{mb}$(K)\end{tabular} &
\begin{tabular}[c]{@{}c@{}}Component \#2\\ av. T$_{mb}$(K)\end{tabular} &
\begin{tabular}[c]{@{}c@{}}Component \#3\\ av. T$_{mb}$(K)\end{tabular} &
\begin{tabular}[c]{@{}c@{}}Noise\\ r.m.s. (K)\end{tabular} &
\begin{tabular}[c]{@{}c@{}}Averaged line ratio\\of all components\end{tabular}  \\

\hline
\multicolumn{2}{c}{LSR velocity ranges (km s$^{-1}$)} & & \multicolumn{1}{c}{12 -- 17} & \multicolumn{1}{c}{-0.5 -- 4.5} & \multicolumn{1}{c}{21.5 -- 26.5} & \multicolumn{1}{c}{} & \multicolumn{1}{c}{}\\
NGC 2346 
 & \mbox{$J$=1--0} & & 0.76 & 0.12 & 0.17 & 0.04 & R$_{54/21}$ = 0.27 \\
 & \mbox{$J$=2--1} & & 1.75 & 0.07 & 0.09 & 0.02 & \\
 & \mbox{$J$=5--4} & 1342218901 & 0.400 & 0.069 & 0.142 & 0.003 & R$_{98/54}$ $\lesssim$ 0.06\\
 & \mbox{$J$=9--8} & 1342229928 & 0.061 & $\lesssim$ 0.007 & $\lesssim$ 0.007 & 0.004 &  \\
\hline
\multicolumn{2}{c}{LSR velocity ranges (km s$^{-1}$)} & & \multicolumn{1}{c}{11 -- 19} & \multicolumn{1}{c}{-6 -- 2} & \multicolumn{1}{c}{25 -- 33} & \multicolumn{1}{c}{} & \multicolumn{1}{c}{} \\
NGC 6072 
 & \mbox{$J$=1--0} & & 0.39 & 0.31 & 0.35 & 0.04 & \\
 & \mbox{$J$=2--1} & & 0.61 & 0.29 & 0.82 & 0.02 & R$_{54/21}$ = 0.87\\
 & \mbox{$J$=5--4} & 1342213733 & 0.859 & 0.154 & 0.980 & 0.003 & \\
 & \mbox{$J$=9--8} & 1342216328 & 0.150 & $\lesssim$ 0.006 & 0.293 & 0.003 & R$_{98/54}$ = 0.24 \\
 & \mbox{$J$=14--13} & 1342229786 & $\lesssim$ 0.02 & $\lesssim$ 0.02 & $\lesssim$ 0.02 & 0.01 &  \\
\hline
\multicolumn{2}{c}{LSR velocity ranges (km s$^{-1}$)} & & \multicolumn{1}{c}{-7 -- -1} & \multicolumn{1}{c}{-19 -- -13} & \multicolumn{1}{c}{13 -- 19} & \multicolumn{1}{c}{} & \multicolumn{1}{c}{} \\
NGC 6720 
 & \mbox{$J$=1--0} & & 0.18 & 0.30 & $\lesssim$ 0.04 & 0.02 & R$_{54/21}$ = 0.26\\
 & \mbox{$J$=2--1} & & 0.36 & 1.57 & $\lesssim$ 0.06 & 0.03 \\
 & \mbox{$J$=5--4} & 1342210075 & 0.207 & 0.323 & 0.127 & 0.004 & R$_{98/54}$ = 0.33 \\
 & \mbox{$J$=9--8} & 1342220499 & 0.076 & 0.084 & $\lesssim$ 0.007 & 0.003 &  \\
\hline
\multicolumn{2}{c}{LSR velocity ranges (km s$^{-1}$)} & & \multicolumn{1}{c}{13 -- 21} & \multicolumn{1}{c}{-4.4 -- 3.6} & \multicolumn{1}{c}{25.8 -- 33.8} & \multicolumn{1}{c}{} & \multicolumn{1}{c}{} \\
NGC 6781 
 & \mbox{$J$=1--0} & & 0.46 & 0.90 & 0.19 & 0.06 & R$_{54/21}$ = 0.36\\
 & \mbox{$J$=2--1} & & 1.43 & 2.60 & $\lesssim$ 0.14 & 0.07 \\
 & \mbox{$J$=5--4} & 1342230381 & 0.435 & 0.631 & 0.332 & 0.003 & R$_{98/54}$ = 0.12 \\
 & \mbox{$J$=9--8} & 1342229920 & 0.049 & 0.098 & $\lesssim$ 0.006 & 0.003 &  \\
\hline
\multicolumn{2}{c}{LSR velocity ranges (km s$^{-1}$)} & & \multicolumn{1}{c}{-18.5 -- -15.5} & \multicolumn{1}{c}{-23.2 -- -20.2} & \multicolumn{1}{c}{-15 -- -12} & \multicolumn{1}{c}{} & \multicolumn{1}{c}{} \\
NGC 7293 
 & \mbox{$J$=1--0} & & 1.99 & 1.45 & 1.75 & 0.11 & R$_{54/21}$ = 0.16\\
 & \mbox{$J$=2--1} & & 1.65 & 0.97 & 1.07 & 0.01 \\
 & \mbox{$J$=5--4} &    
1342210082 & 0.265 & 0.131 & 0.251 & 0.004 & R$_{98/54}$ $\lesssim$ 0.06 \\
 & \mbox{$J$=9--8} & 1342220502 & $\lesssim$ 0.007 & $\lesssim$ 0.007 & $\lesssim$ 0.007 & 0.004 &  \\
\hline
\end{tabular}
}
\end{table*}

\subsection{Pseudo colour-colour diagrams}
\label{sec:results_ccdiagram}

\begin{figure}[!ht]
    \begin{center}
        \setlength\abovecaptionskip{-0.01\baselineskip}
        \setlength\belowcaptionskip{-0.7\baselineskip}
        \includegraphics[width= 1\linewidth]{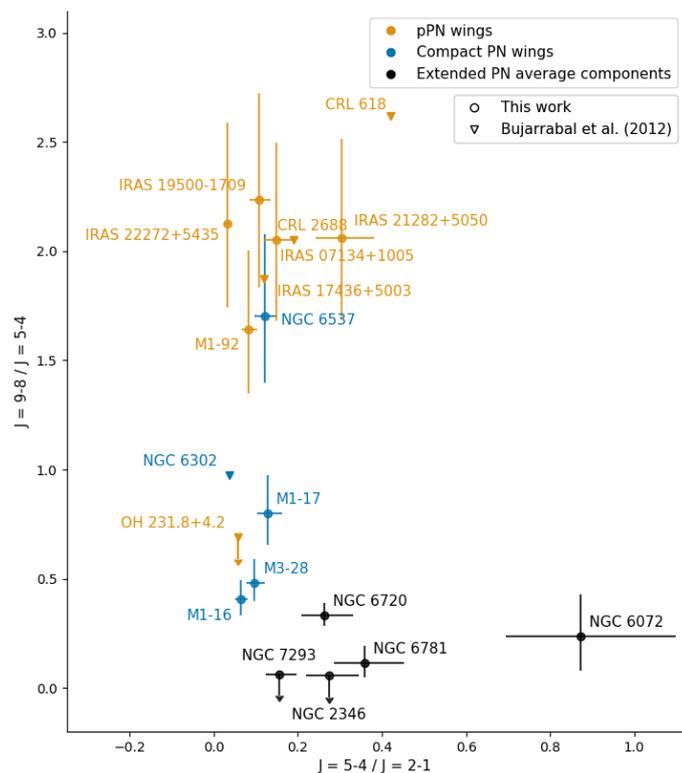}
        \caption{Pseudo colour-colour diagram comparing the \mbox{$J$=5--4}/\mbox{$J$=2--1} and \mbox{$J$=9--8}/\mbox{$J$=5--4} intensity ratios of the wings for the compact sources and the line profiles for the extended PNe. Although they are based on different line ratios, some sources from \citet{bujarrabal_2012} are also included here (see text for details).}\label{fig:ccdiagram_dots}
    \end{center}
\end{figure}

\begin{figure}
    \begin{center}
        \setlength\abovecaptionskip{-0.01\baselineskip}
        \setlength\belowcaptionskip{-0.7\baselineskip}
        \includegraphics[width=1\linewidth]{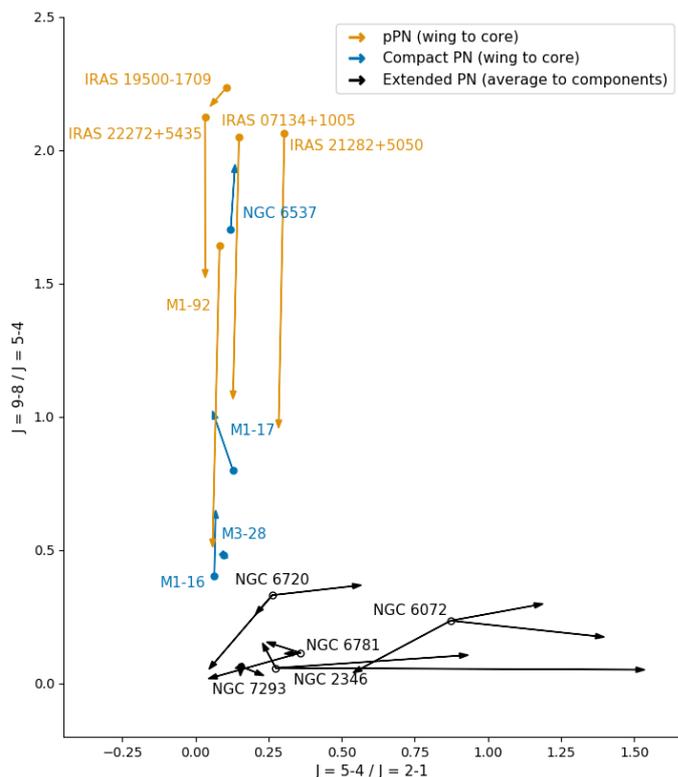}
        \caption{Pseudo colour-colour diagram comparing the \mbox{$J$=5--4}/\mbox{$J$=2--1} and \mbox{$J$=9--8}/\mbox{$J$=5--4} intensity ratios of the wings and cores of the compact sources (joined by arrows pointing towards the line cores) and the identified velocity components of the extended PNe (connecting ratios of each component to its average).}\label{fig:ccdiagram_arrows}
    \end{center}
\end{figure}

The relative contribution of the respective low-to-intermediate excitation CO lines can be explored by mean of the corresponding line ratio, distinctively formed for the various velocity intervals defined in the previous section. In this study, we define R$_{54/21}$ and R$_{98/54}$ as the respective ratio of the \mbox{$J$=5--4} / \mbox{$J$=2--1} and \mbox{$J$=9--8} / \mbox{$J$=5--4} characteristic intensities derived previously.  Their uncertainties were calculated using standard error propagation, assuming absolute calibration errors of 20\% for \mbox{$J$=2--1}, and 15\% for \mbox{$J$=5--4} and \mbox{$J$=9--8} (HIFI Handbook\footnote{\url{https://www.cosmos.esa.int/web/herschel/legacy-documentation-hifi}}). As explained in Section~\ref{sec:results_ranges}, the physical meaning of the chosen velocity interval is different for the compact and extended sources in our sample. This distinction also has implications in the analysis of the ratios discussed here. For compact sources, we distinguish the ratios computed over the line core intervals and those computed over the line wings intervals -- for the latter, we take an average of both the red and blue wings. For extended sources, the ratios are computed for each separate velocity component, as well as based on the averaged intensities over all considered components. These respective ratios are compiled in Table \ref{tab:comp_results} (compact sources) and Table \ref{tab:ext_results} (extended sources).

We then plot the R$_{98/54}$ ratio against the R$_{54/21}$ one, which we call in the following pseudo colour-colour diagrams (thereafter cc-diagram). Although different CO transitions were probed by \citet{bujarrabal_2012}, we decided to also represent in this diagram the ratios obtained for five sources (CRL 2688, CRL 618, IRAS 07134+1005, OH 231.8+4.2 and NGC 6302) by replacing the \mbox{$J$=5--4} line by the \mbox{$J$=6--5} one, and the \mbox{$J$=9--8} line by the \mbox{$J$=10--9} one. While not strictly comparable to our own ratios, we added those points to check the general trends in the pseudo colour-colour space.

Figure~\ref{fig:ccdiagram_dots} shows this diagram for the wings of the compact sources, and the intensity averages of the extended objects. Three different groups emerge from this plot: extended planetary nebulae, compact planetary nebulae and protoplanetary nebulae. Extended planetary nebulae (in black) all show low values of the average R$_{98/54}$ ratio (typically below unity), while their R$_{54/21}$ ratio spreads over an interval range from 0.16 to 0.9. Compact planetary nebulae (in blue), however, seem to concentrate low values of R$_{54/21}$ (all below 0.2), while their R$_{98/54}$ ratio is typically between 0.5 and 1, with the exception of NGC 6537 (see discussion later in this paper). Protoplanetary nebulae (in orange) also show a relatively low R$_{54/21}$ ratio (around 0.1 on average), but their R$_{98/54}$ ratio, on the other hand, takes values that are systematically in excess of unity.


Although they are not strictly based on the same metrics, the sources from \citet{bujarrabal_2012} follow this general classification, but with a certain deviation. On the one hand, CRL618 shows a noticeably higher R$_{65/21}$ ratio (i.e. around 0.4), while OH231.8+4.2 only offers an upper limit to the R$_{109/65}$ ratio located well below 1. Still, the younger objects are mainly located in the upper left corner of the diagram, while the planetary nebulae tend to be in the lower area.
 
The overall trends observed in Fig. \ref{fig:ccdiagram_dots} suggest that R$_{98/54}$ (and the equivalent ratio formed from the \mbox{$J$=10--9} over \mbox{$J$=6--5} ratio for the sources in \citealp{bujarrabal_2012}) is sensitive to the evolutionary stage of the nebulae. This manifests as a higher excitation state in the high-velocity gas of pPNe than in the shocked gas probed in the PNe. NGC 6537 stands as a noticeable exception in that respect. Although considered a compact planetary nebula in our sample, the optical and FIR maps (Fig.~\ref{fig:NGC6537_optico}) indicates that the source is partially resolved within the HIFI beam for the \mbox{$J$=5--4} transition. Moreover, if the different features we see are interpreted as separate velocity components, their observational characteristics differ from those in the rest of the extended object sample, as the spectra were collected very close to the source photocentre (while most lines-of-sight in the extended source sample pointed towards offset regions of the nebula). As such, NGC 6537 is not directly comparable to either of the compact or extended PN source samples and this fact could explain its outlying location on the diagram. 

We also note that the wings of 89 Her, R Sct, OH 17.7--2.0 and M 2--56 are not detected at high-$J$ (they are not featured in Fig.\ref{fig:ccdiagram_dots}, but their ratios in the line core are shown in Fig.~\ref{fig:ccdiagram_arrows}). As we mentioned in Sect.~\ref{sec:sample}, 89 Her and R Sct are considered bipolar nebulae with a rotational disk. It is believed that this type of nebula concentrates all its gas in the disk and central region and, as such, they are considered compact objects suffering from severe dilution in the beam of the telescope. 

A comparison of the line ratios as measured in the line wings and line core of the compact objects is shown in Fig. \ref{fig:ccdiagram_arrows}. Each pair of ratios is joined with an arrow pointing toward the line cores, allowing us to trace possible trends between the relative excitation of the outflows with regard to that of the cores. We note that the arrows of the pPNe sub-sample have greater amplitude and point downwards, indicating that the temperature of their fast outflows must be higher than in the slow winds and suggesting, therefore, a more recent gas acceleration. On the contrary, the difference between the ratios in the line wings and the line cores of the compact PNe sub-sample is smaller and most of their arrows point upwards, indicative of cooler and older gas in the fast flows. This trend is similar to that reported in \citet{bujarrabal_2012} for the objects of their sample represented in our pseudo colour-colour diagram.

Figure \ref{fig:ccdiagram_arrows} also includes a visualisation of the distribution of the line ratios for the various components of the extended PNe sub-samples. In this case, the ratio in each component is connected to that of the averaged ratios shown in Fig. \ref{fig:ccdiagram_dots}. At most, this should indicate whether the individual velocity components in each object do or do not share certain conditions of excitation or evolution. Overall, there is no clear trend derived from them. In the case of NGC 6781, all three components seem to behave similarly, while NGC 2346, for example, exhibits a larger spread of physical conditions in the line ratios.

\subsection{Comparison with theoretical predictions}
\label{sec:results_LVG}

We compare in this section the distribution of our line ratios with theoretical predictions in order to estimate the physical conditions in the fast outflows of our source sample. To this aim, we have performed simplified calculation based on the large velocity gradient (LVG) approximation, for a grid of density, $n_{\rm tot}$, and kinetic temperature, $T_{\rm k}$, as well as different optical depth regimes. This assumption is justified by the kinematics at play in pPNe and many PNe. We also assume that no strong radiative interactions take place between distant points of the envelope exhibiting very different physical conditions. The code is similar to the one used by \citet{bujarrabal_2012}; we have updated the calculation of the collisional transition probabilities, including, in particular, a more accurate interpolation between rates for different temperatures and allowing for calculations at a higher number of levels.

\begin{figure}
    \begin{center}
        \setlength\abovecaptionskip{-0.01\baselineskip}
        \setlength\belowcaptionskip{-0.7\baselineskip}
        \includegraphics[width=1\linewidth]{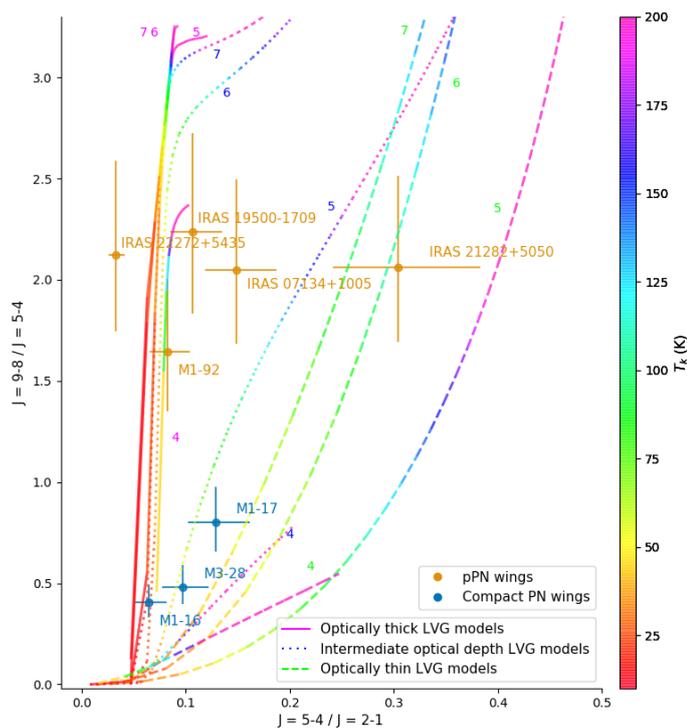}
        \caption{Pseudo colour-colour diagram comparing the \mbox{$J$=5--4}/\mbox{$J$=2--1} and \mbox{$J$=9--8}/\mbox{$J$=5--4} intensity ratios of pPNe line wings, compact PNe line wings and the LVG models for different optical depths (solid, dotted and dashed lines), densities (number indicates log$_{10}$($n_{\rm tot}$)) and temperatures (coded as shown in the colour scale). See text for further details.}\label{fig:ccdiagram_LVG}
    \end{center}
\end{figure}
 
Our computations are based on canonical envelope properties assuming the following parameters: a distance to the central star of $R$ = 6$\times$10$^{16}$ cm, an expansion velocity of $V_{\rm exp}$ = 15 km s$^{-1}$ and a relative abundance of $X(^{12}$CO) = 2$\times$ 10$^{-4}$ (for more details, see Sect. 3.1. of \citealp{bujarrabal_2012}). Opacity effects are taken into account in the calculations under the usual assumptions of the LVG approximation. Although such an assumption implies a constant value for $R$ $X/V_{\rm exp}$ (where $n_{tot}R$ $X/V_{\rm exp}$ represents the column density opacity under the LVG approximation), this approach offers an empirical set of models which is sufficient to probe the statistical behaviour of our line ratios in this grid. Similarly, because we want to locate all our objects in the same set of model space, we do not include any ad hoc modelling of the large-scale structure of the nebula. 

We also assume that the lines come from a region significantly smaller than the telescope spatial resolution, which applies to our compact source sample. For comparison with our extended nebulae, we corrected our line ratio calculations, taking into account the ratios of the applicable telescope beams (Table~\ref{tab:telescopes}).

Figure \ref{fig:ccdiagram_LVG} shows our predicted line ratios for several optical depths, densities, and temperature regimes within the ranges relevant to our data. The theoretical results for different temperatures are presented using a colour coding where the adopted scale ranges from 10\,K to 200\,K. As anticipated in Section~\ref{sec:results_ccdiagram}, the $^{12}$CO \mbox{$J$=9--8} / \mbox{$J$=5--4} ratio appears as a good proxy for the kinetic temperature, especially in the optically thick regime. The slope of this relationship is also related to the density. Conversely, the $^{12}$CO \mbox{$J$=5--4} / \mbox{$J$=2--1}  ratio appears to be inversely proportional to both optical depth and density. Unfortunately, predictions from the various model grids overlap in the pseudo colour-colour diagram, leading to some degeneracy in the derivation of the physical conditions from our measurements.
Figure \ref{fig:ccdiagram_LVG} also shows the intensity ratios of the line wings for the compact sources.


From the comparison of our pPN observational results with the LVG predictions, we observe that high density optically thick models predict too low temperatures (< 25\,K) for the outflows of this kind of objects, while the optically thin approximations assign densities excessively high (larger than 10$^{7}$ cm$^{-3}$). Consequently, we used the model grid based on an intermediate optical depth and optically thick models of 10$^{4}$ cm$^{-3}$, for which the fast outflows of protoplanetary nebulae are compatible with densities between 10$^{4}$ and 10$^{6}$ cm$^{-3}$ and temperatures in the range of 75--200\,K. Doing the same exercise for the compact PNe, we derive lower kinetic temperatures in the 25--75\,K range and densities in the 10$^{4}$--10$^{6}$ cm$^{-3}$ range.

We interpret this result as a relation between the temperature of the high-velocity gas found in the nebulae and the time elapsed since its acceleration due to shocks. The heated and accelerated gas reaches high temperatures during the ejection of very fast outflows, but it cools in a short period of time. This is consistent with the results already reported by \citet{bujarrabal_2012}. The correlation between the density of the warm regions and the evolutionary stage of the nebulae, however, is not so clear. 

\subsection{Intermediate velocity ranges}
\label{intermediate}

Aside from the main spectral components analysed above, we note the existence of intermediate velocity features in some of our sources, particularly in M 1--92 and M 1--17. Following the same methods described in the previous sections, we derived the characteristic intensities and respective $R_{54/21}$ and $R_{98/54}$ ratios of these intermediate line regions for the two mentioned nebulae. Their values are collected in Table \ref{tab:interm_results}, alongside\ the LSR velocity ranges adopted (also marked in Figs. \ref{fig:M1-92_ranges} and \ref{fig:M1-17_ranges}). The substantial difference between the ratios of these regions and those estimated in the line wings is indicative of a cooler gas component of intermediate velocity in the nebulae that, in the case of \mbox{M 1--92}, could correspond with the lobe walls identified in \citet{alcolea_2007} and \citet{bujarrabal_1998}. 

\begin{table*}
\caption{Characteristic intensities of two intermediate velocity ranges identified in the CO profiles of the pPNe M 1--17 and M 1--92, and their derived ratios}
\label{tab:interm_results}
\resizebox*{!}{0.18\textheight}{%
\centering
\begin{tabular}{llccccc}
\hline \hline
Source &
$^{12}$CO line &
\begin{tabular}[c]{@{}c@{}}Blue interm. wing\\av. T$_{mb}$(K)\end{tabular} &
\begin{tabular}[c]{@{}c@{}}Red interm. wing\\av. T$_{mb}$(K)\end{tabular} &
\begin{tabular}[c]{@{}c@{}}Noise\\ r.m.s. (K)\end{tabular} &
\begin{tabular}[c]{@{}c@{}}Line ratios\\ Blue interm. line wings\end{tabular} &
\begin{tabular}[c]{@{}c@{}}Line ratios\\ Red interm. line wings\end{tabular}\\
\hline
\multicolumn{2}{c}{LSR velocity ranges (km s$^{-1}$)} & \multicolumn{1}{c}{-25 -- -10.5} & \multicolumn{1}{c}{10.5 -- 25} & \multicolumn{1}{c}{} & \multicolumn{1}{c}{} & \multicolumn{1}{c}{}\\
M 1--92 
 & \mbox{$J$=1--0} & 0.36 & 0.43 & 0.01 & R$_{54/21}$ = 0.05 & R$_{54/21}$ = 0.04 \\
 & \mbox{$J$=2--1} & 0.88 & 1.18 & 0.02 & & \\
 & \mbox{$J$=5--4} & 0.048 & 0.052 & 0.002 & R$_{98/54}$ = 0.55 & R$_{98/54}$ = 0.68 \\
 & \mbox{$J$=9--8} & 0.026 & 0.035 & 0.002 & & \\
\hline
\multicolumn{2}{c}{LSR velocity ranges (km s$^{-1}$)} & \multicolumn{1}{c}{11.5 -- 23.5} & \multicolumn{1}{c}{32 -- 44} & \multicolumn{1}{c}{} & \multicolumn{1}{c}{} & \multicolumn{1}{c}{}\\
M 1--17  
 & \mbox{$J$=1--0} & 1.06 & 1.11 & 0.03 & R$_{54/21}$ = 0.06 & R$_{54/21}$ = 0.08 \\
 & \mbox{$J$=2--1} & 1.86 & 1.89 & 0.04 & & \\
 & \mbox{$J$=5--4} & 0.112 & 0.144 & 0.005 & R$_{98/54}$ = 1.19 & R$_{98/54}$ = 0.94\\
 & \mbox{$J$=9--8} & 0.133 & 0.135 & 0.004 & & \\
\hline
\end{tabular}
}
\end{table*}

\section{Conclusions}
\label{sec:conclusions}
We used the high-resolution spectrometer HIFI on board \textit{Herschel} to survey a sample of nine protoplanetary nebulae (pPNe) and nine planetary nebulae (PNe) in intermediate-excitation molecular lines of $^{12}$CO (\mbox{$J$=5--4}, \mbox{$J$=9--8}, and \mbox{$J$=14--13} for a subset of those). We complemented these data with observations of the low-$J$ transitions of CO using IRAM 30m telescope and literature (Table \ref{tab:lowco}). The selected lines are shown in Figs. \ref{fig:IRAS07134_ranges} to \ref{fig:NGC7293_ranges} and described briefly in Sect. \ref{sec:sample}.

As in \citet{bujarrabal_2012}, we carried out a simplified analysis of the CO lines in terms of the emission of three representative profile components. The physical significance behind these regions depends on the relative size of the objects with regard to the dimension of HIFI beams. Five of the PNe in our sample were considered extended sources and we interpreted their three main profile features as three different velocity components. On the other hand, the representative profile features of the compact nebulae were analysed as a line core (known to come from the remnant AGB shell and rotating disks in two cases) and two line wings (associated with the fast outflows). 

In Tables \ref{tab:comp_results} and \ref{tab:ext_results}, we compiled the averaged intensities of the characteristic components found in compact and extended sources respectively, together with their errors. These tables indicate the intensity ratios \mbox{$J$=5--4}/\mbox{$J$=2--1} and \mbox{$J$=9--8}/\mbox{$J$=5--4} derived for the line core and line wings of compact objects, and for the entire line profile of extended nebulae. These ratios are represented in Figs. \ref{fig:ccdiagram_dots} and \ref{fig:ccdiagram_arrows} as so-called pseudo colour-colour diagrams. In these diagrams, our sample is clearly divided with regard to their evolutionary stage and to their relative extension. While the wings of pPNe are located at the top left of the diagram and are significantly more excited than their cores, the wings of PNe are found at its bottom left and are less excited than their cores. This observation suggests a correlation between the evolutionary stage of nebulae and the excitation state of their fast outflows. With respect to the apparent spatial extension, compact objects remain at the left of the pseudo cc-diagram, while the extended sources are found at its bottom right.

We performed radiative transfer calculations based on the LVG
approximation to derive the physical properties of the nebular gas. We found that the kinetic temperatures of the fast gas presented in pPNe tend to be higher (with typical values between 75-200\,K) than those found in PNe (with characteristic temperatures between 25 and 75\,K). This result indicates a relation between the temperature of fast outflows and the time elapsed since their ejection. It is interpreted as the effect of cooling in the heated and accelerated gas during the pPN phase, when the shocks propagate across the nebula. This conclusion is strengthened by the significantly high excitation state found for the high-velocity line wings in M 1--92 (much higher than for the intermediate-velocity wings); the  high-velocity emission is indeed known to come from the tips of the nebula lobes, which are directly impinged by the post-AGB jets.

\begin{acknowledgements}

We thank the anonymous referee for their useful and constructive comments on the paper. We also thank Prof. Hans Olofsson for his thorough review of this manuscript. HIFI  has  been  designed  and  built  by  a  consortium  of  institutes  and  university  departments  from  across  Europe,  Canada  and  the  USA under  the  leadership  of  SRON  Netherlands  Institute  for  Space  Research, Groningen,  The  Netherlands,  and  with  major  contributions  from  Germany, France  and  the  USA.  Consortium  members  are:  Canada:  CSA,  U.Waterloo; France:  CESR,  LAB,  LERMA,  IRAM;  Germany:  KOSMA,  MPIfR,  MPS; Ireland:   NUI   Maynooth;   Italy:   ASI,   IFSI-INAF,   Osservatorio   Astrofisico di  Arcetri-INAF;  The  Netherlands:  SRON,  TUD;  Poland:  CAMK,  CBK; Spain:  Observatorio Astron\'omico  Nacional  (IGN),  Centro  de  Astrobiolog\'ia (CSIC-INTA);  Sweden:  Chalmers  University  of  Technology  –  MC2,  RSS  \& GARD;  Onsala Space Observatory; Swedish National Space Board, Stockholm University -- Stockholm   Observatory;   Switzerland:   ETH   Zurich,   FHNW; USA:  Caltech,  JPL,  NHSC. This paper is based on obser-vations  carried  out  at  the  IRAM-30  m  single-dish  telescope.  IRAM  is  sup-ported  by  INSU/CNRS  (France),  MPG  (Germany)  and  IGN  (Spain). JA and VB are partly funded by the Spanish MICIIN grant AYA2016-78994-P.

\end{acknowledgements}


\bibliography{sample}

\begin{thebibliography}{65}
\expandafter\ifx\csname natexlab\endcsname\relax\def\natexlab#1{#1}\fi

\bibitem[{{Alcolea} \& {Bujarrabal}(1995)}]{alcolea_1995}
{Alcolea}, J. \& {Bujarrabal}, V. 1995, \aap, 303, L21

\bibitem[{{Alcolea} {et~al.}(2001){Alcolea}, {Bujarrabal}, {S{\'a}nchez
  Contreras}, {Neri}, \& {Zweigle}}]{alcolea_2001}
{Alcolea}, J., {Bujarrabal}, V., {S{\'a}nchez Contreras}, C., {Neri}, R., \&
  {Zweigle}, J. 2001, \aap, 373, 932

\bibitem[{{Alcolea} {et~al.}(2007){Alcolea}, {Neri}, \&
  {Bujarrabal}}]{alcolea_2007}
{Alcolea}, J., {Neri}, R., \& {Bujarrabal}, V. 2007, \aap, 468, L41

\bibitem[{{Ashley} \& {Hyland}(1988)}]{ashley_1988}
{Ashley}, M. C.~B. \& {Hyland}, A.~R. 1988, \apj, 331, 532

\bibitem[{{Bachiller} {et~al.}(1989){Bachiller}, {Bujarrabal},
  {Martin-Pintado}, \& {Gomez-Gonzalez}}]{bachiller_1989}
{Bachiller}, R., {Bujarrabal}, V., {Martin-Pintado}, J., \& {Gomez-Gonzalez},
  J. 1989, \aap, 218, 252

\bibitem[{{Bachiller} {et~al.}(1997){Bachiller}, {Forveille}, {Huggins}, \&
  {Cox}}]{bachiller_1997}
{Bachiller}, R., {Forveille}, T., {Huggins}, P.~J., \& {Cox}, P. 1997, \aap,
  324, 1123

\bibitem[{{Bachiller} {et~al.}(1991){Bachiller}, {Huggins}, {Cox}, \&
  {Forveille}}]{bachiller_1991}
{Bachiller}, R., {Huggins}, P.~J., {Cox}, P., \& {Forveille}, T. 1991, \aap,
  247, 525

\bibitem[{{Bachiller} {et~al.}(1993){Bachiller}, {Huggins}, {Cox}, \&
  {Forveille}}]{bachiller_1993}
{Bachiller}, R., {Huggins}, P.~J., {Cox}, P., \& {Forveille}, T. 1993, \aap,
  267, 177

\bibitem[{{Bains} {et~al.}(2003){Bains}, {Gledhill}, {Yates}, \&
  {Richards}}]{bains_2003}
{Bains}, I., {Gledhill}, T.~M., {Yates}, J.~A., \& {Richards}, A.~M.~S. 2003,
  \mnras, 338, 287

\bibitem[{{Balick} \& {Frank}(2002)}]{balick_2002}
{Balick}, B. \& {Frank}, A. 2002, \araa, 40, 439

\bibitem[{{Bublitz} {et~al.}(2019){Bublitz}, {Kastner},
  {Santander-Garc{\'\i}a}, {Bujarrabal}, {Alcolea}, \& {Montez}}]{bublitz_2019}
{Bublitz}, J., {Kastner}, J.~H., {Santander-Garc{\'\i}a}, M., {et~al.} 2019,
  \aap, 625, A101

\bibitem[{{Bujarrabal} {et~al.}(1998){Bujarrabal}, {Alcolea}, \&
  {Neri}}]{bujarrabal_1998}
{Bujarrabal}, V., {Alcolea}, J., \& {Neri}, R. 1998, \apj, 504, 915

\bibitem[{{Bujarrabal} {et~al.}(1992){Bujarrabal}, {Alcolea}, \&
  {Planesas}}]{bujarrabal_1992}
{Bujarrabal}, V., {Alcolea}, J., \& {Planesas}, P. 1992, \aap, 257, 701

\bibitem[{{Bujarrabal} {et~al.}(2012){Bujarrabal}, {Alcolea}, {Soria-Ruiz},
  {Planesas}, {Teyssier}, {Cernicharo}, {Decin}, {Dominik}, {Justtanont}, {de
  Koter}, {Marston}, {Melnick}, {Menten}, {Neufeld}, {Olofsson}, {Schmidt},
  {Sch{\"o}ier}, {Szczerba}, \& {Waters}}]{bujarrabal_2012}
{Bujarrabal}, V., {Alcolea}, J., {Soria-Ruiz}, R., {et~al.} 2012, \aap, 537, A8

\bibitem[{{Bujarrabal} {et~al.}(2013){Bujarrabal}, {Alcolea}, {Van Winckel},
  {Santand er-Garc{\'\i}a}, \& {Castro-Carrizo}}]{bujarrabal_2013}
{Bujarrabal}, V., {Alcolea}, J., {Van Winckel}, H., {Santand er-Garc{\'\i}a},
  M., \& {Castro-Carrizo}, A. 2013, \aap, 557, A104

\bibitem[{{Bujarrabal} {et~al.}(2001){Bujarrabal}, {Castro-Carrizo}, {Alcolea},
  \& {S{\'a}nchez Contreras}}]{bujarrabal_2001}
{Bujarrabal}, V., {Castro-Carrizo}, A., {Alcolea}, J., \& {S{\'a}nchez
  Contreras}, C. 2001, \aap, 377, 868

\bibitem[{{Bujarrabal} {et~al.}(1994){Bujarrabal}, {Fuente}, \&
  {Omont}}]{bujarrabal_1994}
{Bujarrabal}, V., {Fuente}, A., \& {Omont}, A. 1994, \aap, 285, 247

\bibitem[{{Bujarrabal} {et~al.}(2007){Bujarrabal}, {van Winckel}, {Neri},
  {Alcolea}, {Castro-Carrizo}, \& {Deroo}}]{bujarrabal_2007}
{Bujarrabal}, V., {van Winckel}, H., {Neri}, R., {et~al.} 2007, \aap, 468, L45

\bibitem[{{Calvet} \& {Cohen}(1978)}]{calvet_1978}
{Calvet}, N. \& {Cohen}, M. 1978, \mnras, 182, 687

\bibitem[{{Castro-Carrizo} {et~al.}(2002){Castro-Carrizo}, {Bujarrabal},
  {S{\'a}nchez Contreras}, {Alcolea}, \& {Neri}}]{castro_2002}
{Castro-Carrizo}, A., {Bujarrabal}, V., {S{\'a}nchez Contreras}, C., {Alcolea},
  J., \& {Neri}, R. 2002, \aap, 386, 633

\bibitem[{{Castro-Carrizo} {et~al.}(2010){Castro-Carrizo}, {Quintana-Lacaci},
  {Neri}, {Bujarrabal}, {Sch{\"o}ier}, {Winters}, {Olofsson}, {Lindqvist},
  {Alcolea}, {Lucas}, \& {Grewing}}]{castrocarrizo_2010}
{Castro-Carrizo}, A., {Quintana-Lacaci}, G., {Neri}, R., {et~al.} 2010, \aap,
  523, A59

\bibitem[{{Cox} {et~al.}(1992){Cox}, {Omont}, {Huggins}, {Bachiller}, \&
  {Forveille}}]{cox_1992}
{Cox}, P., {Omont}, A., {Huggins}, P.~J., {Bachiller}, R., \& {Forveille}, T.
  1992, \aap, 266, 420

\bibitem[{{Cuesta} {et~al.}(1995){Cuesta}, {Phillips}, \&
  {Mampaso}}]{cuesta_1995}
{Cuesta}, L., {Phillips}, J.~P., \& {Mampaso}, A. 1995, \aap, 304, 475

\bibitem[{{Danilovich} {et~al.}(2015){Danilovich}, {Teyssier}, {Justtanont},
  {Olofsson}, {Cerrigone}, {Bujarrabal}, {Alcolea}, {Cernicharo},
  {Castro-Carrizo}, {Garc{\'\i}a-Lario}, \& {Marston}}]{danilovich_2015}
{Danilovich}, T., {Teyssier}, D., {Justtanont}, K., {et~al.} 2015, \aap, 581,
  A60

\bibitem[{{De Beck} {et~al.}(2010){De Beck}, {Decin}, {de Koter}, {Justtanont},
  {Verhoelst}, {Kemper}, \& {Menten}}]{debeck_2010}
{De Beck}, E., {Decin}, L., {de Koter}, A., {et~al.} 2010, \aap, 523, A18

\bibitem[{{de Graauw} {et~al.}(2010){de Graauw}, {Helmich}, {Phillips},
  {Stutzki}, {Caux}, {Whyborn}, {Dieleman}, {Roelfsema}, {Aarts}, {Assendorp},
  {Bachiller}, {Baechtold}, {Barcia}, {Beintema}, {Belitsky}, {Benz}, {Bieber},
  {Boogert}, {Borys}, {Bumble}, {Ca{\"\i}s}, {Caris}, {Cerulli-Irelli},
  {Chattopadhyay}, {Cherednichenko}, {Ciechanowicz}, {Coeur-Joly}, {Comito},
  {Cros}, {de Jonge}, {de Lange}, {Delforges}, {Delorme}, {den Boggende},
  {Desbat}, {Diez-Gonz{\'a}lez}, {di Giorgio}, {Dubbeldam}, {Edwards},
  {Eggens}, {Erickson}, {Evers}, {Fich}, {Finn}, {Franke}, {Gaier}, {Gal},
  {Gao}, {Gallego}, {Gauffre}, {Gill}, {Glenz}, {Golstein}, {Goulooze},
  {Gunsing}, {G{\"u}sten}, {Hartogh}, {Hatch}, {Higgins}, {Honingh}, {Huisman},
  {Jackson}, {Jacobs}, {Jacobs}, {Jarchow}, {Javadi}, {Jellema}, {Justen},
  {Karpov}, {Kasemann}, {Kawamura}, {Keizer}, {Kester}, {Klapwijk}, {Klein},
  {Kollberg}, {Kooi}, {Kooiman}, {Kopf}, {Krause}, {Krieg}, {Kramer},
  {Kruizenga}, {Kuhn}, {Laauwen}, {Lai}, {Larsson}, {Leduc}, {Leinz}, {Lin},
  {Liseau}, {Liu}, {Loose}, {L{\'o}pez-Fernandez}, {Lord}, {Luinge}, {Marston},
  {Mart{\'\i}n-Pintado}, {Maestrini}, {Maiwald}, {McCoey}, {Mehdi}, {Megej},
  {Melchior}, {Meinsma}, {Merkel}, {Michalska}, {Monstein}, {Moratschke},
  {Morris}, {Muller}, {Murphy}, {Naber}, {Natale}, {Nowosielski}, {Nuzzolo},
  {Olberg}, {Olbrich}, {Orfei}, {Orleanski}, {Ossenkopf}, {Peacock}, {Pearson},
  {Peron}, {Phillip-May}, {Piazzo}, {Planesas}, {Rataj}, {Ravera}, {Risacher},
  {Salez}, {Samoska}, {Saraceno}, {Schieder}, {Schlecht}, {Schl{\"o}der},
  {Schm{\"u}lling}, {Schultz}, {Schuster}, {Siebertz}, {Smit}, {Szczerba},
  {Shipman}, {Steinmetz}, {Stern}, {Stokroos}, {Teipen}, {Teyssier}, {Tils},
  {Trappe}, {van Baaren}, {van Leeuwen}, {van de Stadt}, {Visser}, {Wildeman},
  {Wafelbakker}, {Ward}, {Wesselius}, {Wild}, {Wulff}, {Wunsch}, {Tielens},
  {Zaal}, {Zirath}, {Zmuidzinas}, \& {Zwart}}]{degraauw_2010}
{de Graauw}, T., {Helmich}, F.~P., {Phillips}, T.~G., {et~al.} 2010, \aap, 518,
  L6

\bibitem[{{de Ruyter} {et~al.}(2006){de Ruyter}, {van Winckel}, {Maas}, {Lloyd
  Evans}, {Waters}, \& {Dejonghe}}]{deruyter_2006}
{de Ruyter}, S., {van Winckel}, H., {Maas}, T., {et~al.} 2006, \aap, 448, 641

\bibitem[{{Edwards} {et~al.}(2014){Edwards}, {Cox}, \& {Ziurys}}]{edwards_2014}
{Edwards}, J.~L., {Cox}, E.~G., \& {Ziurys}, L.~M. 2014, \apj, 791, 79

\bibitem[{{Edwards} \& {Ziurys}(2013)}]{edwards_ziurys_2013}
{Edwards}, J.~L. \& {Ziurys}, L.~M. 2013, \apjl, 770, L5

\bibitem[{{Fong} {et~al.}(2006){Fong}, {Meixner}, {Sutton}, {Zalucha}, \&
  {Welch}}]{fong_2006}
{Fong}, D., {Meixner}, M., {Sutton}, E.~C., {Zalucha}, A., \& {Welch}, W.~J.
  2006, \apj, 652, 1626

\bibitem[{Frew(2008)}]{frew_2008}
Frew, D.~J. 2008, PhD thesis, Department of Physics, Macquarie University, NSW
  2109, Australia

\bibitem[{{Gledhill}(2005)}]{gledhill_2005}
{Gledhill}, T.~M. 2005, \mnras, 356, 883

\bibitem[{{Gledhill} {et~al.}(2001){Gledhill}, {Chrysostomou}, {Hough}, \&
  {Yates}}]{gledhill_2001}
{Gledhill}, T.~M., {Chrysostomou}, A., {Hough}, J.~H., \& {Yates}, J.~A. 2001,
  \mnras, 322, 321

\bibitem[{{Heske} {et~al.}(1990){Heske}, {Forveille}, {Omont}, {van der Veen},
  \& {Habing}}]{heske_1990}
{Heske}, A., {Forveille}, T., {Omont}, A., {van der Veen}, W.~E.~C.~J., \&
  {Habing}, H.~J. 1990, \aap, 239, 173

\bibitem[{{Hrivnak} {et~al.}(1989){Hrivnak}, {Kwok}, \& {Volk}}]{hrivnak_1989}
{Hrivnak}, B.~J., {Kwok}, S., \& {Volk}, K.~M. 1989, \apj, 346, 265

\bibitem[{{Huggins} {et~al.}(2000){Huggins}, {Forveille}, {Bachiller}, \&
  {Cox}}]{huggins_2000}
{Huggins}, P.~J., {Forveille}, T., {Bachiller}, R., \& {Cox}, P. 2000, \apj,
  544, 889

\bibitem[{{Justtanont} {et~al.}(2000){Justtanont}, {Barlow}, {Tielens},
  {Hollenbach}, {Latter}, {Liu}, {Sylvester}, {Cox}, \&
  {Rieu}}]{justtanont_2000}
{Justtanont}, K., {Barlow}, M.~J., {Tielens}, A.~G.~G.~M., {et~al.} 2000, \aap,
  360, 1117

\bibitem[{{Kwok} {et~al.}(2010){Kwok}, {Chong}, {Hsia}, {Zhang}, \&
  {Koning}}]{kwok_2010}
{Kwok}, S., {Chong}, S.-N., {Hsia}, C.-H., {Zhang}, Y., \& {Koning}, N. 2010,
  \apj, 708, 93

\bibitem[{{Kwok} {et~al.}(2008){Kwok}, {Chong}, {Koning}, {Hua}, \&
  {Yan}}]{kwok_2008}
{Kwok}, S., {Chong}, S.-N., {Koning}, N., {Hua}, T., \& {Yan}, C.-H. 2008,
  \apj, 689, 219

\bibitem[{{Kwok} {et~al.}(1989){Kwok}, {Volk}, \& {Hrivnak}}]{kwok_1989}
{Kwok}, S., {Volk}, K.~M., \& {Hrivnak}, B.~J. 1989, \apjl, 345, L51

\bibitem[{{Le Bertre}(1987)}]{bertre_1987}
{Le Bertre}, T. 1987, \aap, 180, 160

\bibitem[{{Maciel}(1984)}]{maciel_1984}
{Maciel}, W.~J. 1984, \aaps, 55, 253

\bibitem[{{Matsuura} {et~al.}(2005){Matsuura}, {Zijlstra}, {Gray}, {Molster},
  \& {Waters}}]{matsuura_2005}
{Matsuura}, M., {Zijlstra}, A.~A., {Gray}, M.~D., {Molster}, F.~J., \&
  {Waters}, L.~B.~F.~M. 2005, \mnras, 363, 628

\bibitem[{{Meaburn} {et~al.}(2005){Meaburn}, {Boumis}, {L{\'o}pez}, {Harman},
  {Bryce}, {Redman}, \& {Mavromatakis}}]{meaburn_2005}
{Meaburn}, J., {Boumis}, P., {L{\'o}pez}, J.~A., {et~al.} 2005, \mnras, 360,
  963

\bibitem[{{Meixner} {et~al.}(1997){Meixner}, {Skinner}, {Graham}, {Keto},
  {Jernigan}, \& {Arens}}]{meixner_1997}
{Meixner}, M., {Skinner}, C.~J., {Graham}, J.~R., {et~al.} 1997, Astronomy Data
  Image Library

\bibitem[{{Milam} {et~al.}(2009){Milam}, {Woolf}, \& {Ziurys}}]{milam_2009}
{Milam}, S.~N., {Woolf}, N.~J., \& {Ziurys}, L.~M. 2009, \apj, 690, 837

\bibitem[{{O'Dell} {et~al.}(2013){O'Dell}, {Ferland}, {Henney}, \&
  {Peimbert}}]{odell_2013}
{O'Dell}, C.~R., {Ferland}, G.~J., {Henney}, W.~J., \& {Peimbert}, M. 2013,
  \aj, 145, 170

\bibitem[{{Omont} {et~al.}(1995){Omont}, {Moseley}, {Cox}, {Glaccum}, {Casey},
  {Forveille}, {Chan}, {Szczerba}, {Loewenstein}, {Harvey}, \&
  {Kwok}}]{omont_1995}
{Omont}, A., {Moseley}, S.~H., {Cox}, P., {et~al.} 1995, \apj, 454, 819

\bibitem[{{S{\'a}nchez Contreras} {et~al.}(2004){S{\'a}nchez Contreras},
  {Bujarrabal}, {Castro-Carrizo}, {Alcolea}, \&
  {Sargent}}]{sanchez-contreras_2004}
{S{\'a}nchez Contreras}, C., {Bujarrabal}, V., {Castro-Carrizo}, A., {Alcolea},
  J., \& {Sargent}, A. 2004, \apj, 617, 1142

\bibitem[{{S{\'a}nchez Contreras} {et~al.}(2007){S{\'a}nchez Contreras}, {Le
  Mignant}, {Sahai}, {Gil de Paz}, \& {Morris}}]{sanchez-contreras_2007}
{S{\'a}nchez Contreras}, C., {Le Mignant}, D., {Sahai}, R., {Gil de Paz}, A.,
  \& {Morris}, M. 2007, \apj, 656, 1150

\bibitem[{{Schmidt} {et~al.}(2018){Schmidt}, {Zack}, \&
  {Ziurys}}]{schmidt_2018}
{Schmidt}, D.~R., {Zack}, L.~N., \& {Ziurys}, L.~M. 2018, \apjl, 864, L31

\bibitem[{{Schmidt} \& {Ziurys}(2016)}]{schmidt_ziurys_2016}
{Schmidt}, D.~R. \& {Ziurys}, L.~M. 2016, \apj, 817, 175

\bibitem[{{Shenton} {et~al.}(1995){Shenton}, {Evans}, \&
  {Williams}}]{shenton_1995}
{Shenton}, M., {Evans}, A., \& {Williams}, P.~M. 1995, \mnras, 273, 906

\bibitem[{{Shibata} {et~al.}(1989){Shibata}, {Tamura}, {Deguchi}, {Hirano},
  {Kameya}, \& {Kasuga}}]{shibata_1989}
{Shibata}, K.~M., {Tamura}, S., {Deguchi}, S., {et~al.} 1989, \apjl, 345, L55

\bibitem[{{Su} {et~al.}(2004){Su}, {Kelly}, {Latter}, {Misselt}, {Frank},
  {Volk}, {Engelbracht}, {Gordon}, {Hines}, {Morrison}, {Muzerolle}, {Rieke},
  {Stansberry}, \& {Young}}]{su_2004}
{Su}, K.~Y.~L., {Kelly}, D.~M., {Latter}, W.~B., {et~al.} 2004, \apjs, 154, 302

\bibitem[{{Teyssier} {et~al.}(2011){Teyssier}, {Alcolea}, {Bujarrabal},
  {Castro-Carrizo}, {Cernicharo}, {Garcia-Lario}, {Marston}, {Olofsson},
  {Risacher}, {Schoeier}, \& {Verdugo}}]{teyssier_2011}
{Teyssier}, D., {Alcolea}, J., {Bujarrabal}, V., {et~al.} 2011, 280, 353

\bibitem[{{Ueta} {et~al.}(2000){Ueta}, {Meixner}, \& {Bobrowsky}}]{ueta_2000}
{Ueta}, T., {Meixner}, M., \& {Bobrowsky}, M. 2000, \apj, 528, 861

\bibitem[{{Van Winckel} \& {Reyniers}(2000)}]{vanWinckel_2000}
{Van Winckel}, H. \& {Reyniers}, M. 2000, \aap, 354, 135

\bibitem[{{Waters} {et~al.}(1993){Waters}, {Waelkens}, {Mayor}, \&
  {Trams}}]{waters_1993}
{Waters}, L.~B.~F.~M., {Waelkens}, C., {Mayor}, M., \& {Trams}, N.~R. 1993,
  \aap, 269, 242

\bibitem[{{Young} {et~al.}(1999){Young}, {Cox}, {Huggins}, {Forveille}, \&
  {Bachiller}}]{young_1999}
{Young}, K., {Cox}, P., {Huggins}, P.~J., {Forveille}, T., \& {Bachiller}, R.
  1999, \apj, 522, 387

\bibitem[{{Zack} \& {Ziurys}(2013{\natexlab{a}})}]{zack_2013}
{Zack}, L.~N. \& {Ziurys}, L.~M. 2013{\natexlab{a}}, \apj, 765, 112

\bibitem[{{Zack} \& {Ziurys}(2013{\natexlab{b}})}]{zack_ziurys_2013}
{Zack}, L.~N. \& {Ziurys}, L.~M. 2013{\natexlab{b}}, \apj, 765, 112

\bibitem[{{Zeigler} {et~al.}(2013){Zeigler}, {Zack}, {Woolf}, \&
  {Ziurys}}]{zeigler_2013}
{Zeigler}, N.~R., {Zack}, L.~N., {Woolf}, N.~J., \& {Ziurys}, L.~M. 2013, \apj,
  778, 16

\bibitem[{{Zhang} {et~al.}(2008){Zhang}, {Kwok}, \&
  {Dinh-V-Trung}}]{zhang_2008}
{Zhang}, Y., {Kwok}, S., \& {Dinh-V-Trung}. 2008, \apj, 678, 328

\bibitem[{{Ziurys} {et~al.}(2020){Ziurys}, {Schmidt}, \& {Woolf}}]{ziurys_2020}
{Ziurys}, L.~M., {Schmidt}, D.~R., \& {Woolf}, N.~J. 2020, \apjl, 900, L31

\end{thebibliography}

\newpage
\appendix

\section{CO spectra of the source sample}
\label{sec:appxdata}
 In this section, we compile all the CO transition spectra used in this work, together with the adopted systemic velocity and velocity ranges considered in the analysis. See the captions in Figures~\ref{fig:IRAS21282_ranges} and \ref{fig:NGC6720_ranges} for further details.

\begin{figure}
    \begin{center}
        \setlength\abovecaptionskip{-0.01\baselineskip}
        \setlength\belowcaptionskip{-0.7\baselineskip}
        \includegraphics[width= \linewidth]{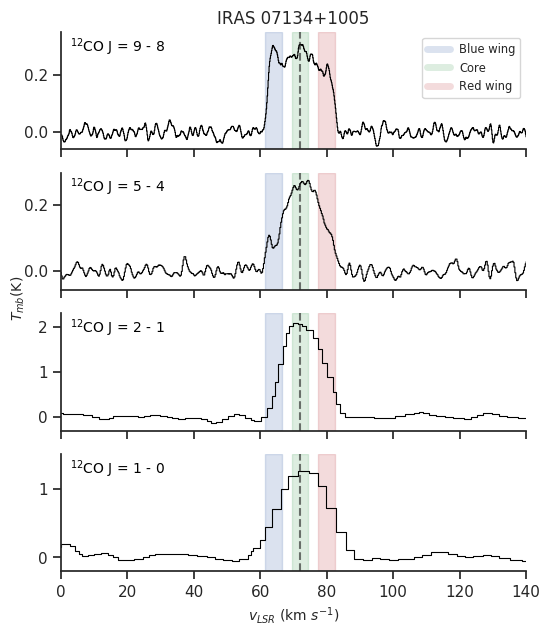}
        \caption{Observations of IRAS 07134+1005. Shadowed regions mark the LSR velocity ranges adopted for each feature and the dashed vertical line indicates the systemic velocity.}\label{fig:IRAS07134_ranges}
    \end{center}
\end{figure}

\begin{figure}
    \begin{center}
        \setlength\abovecaptionskip{-0.01\baselineskip}
        \setlength\belowcaptionskip{-0.7\baselineskip}
        \includegraphics[width= \linewidth]{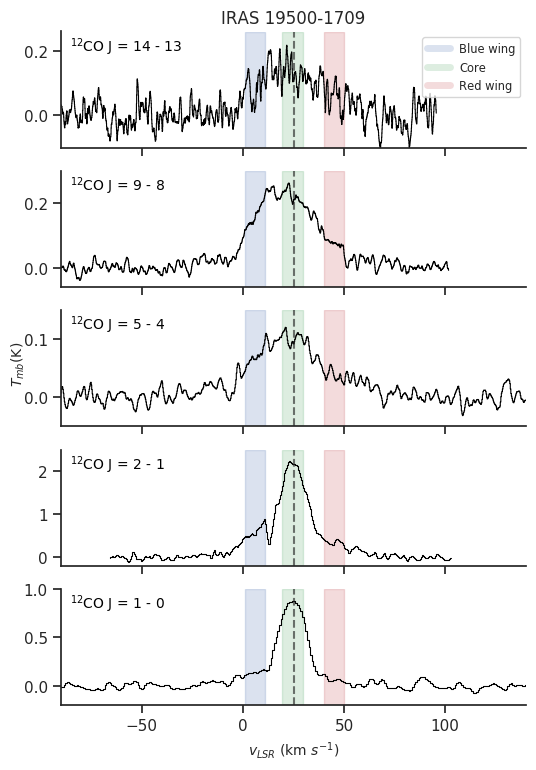}
        \caption{Observations of IRAS 19500--1709. Shadowed regions mark the LSR velocity ranges adopted for each feature and the dashed vertical line indicates the systemic velocity.}\label{fig:IRAS19500_ranges}
    \end{center}
\end{figure}

\begin{figure}
    \begin{center}
        \setlength\abovecaptionskip{-0.01\baselineskip}
        \setlength\belowcaptionskip{-0.7\baselineskip}
        \includegraphics[width= \linewidth]{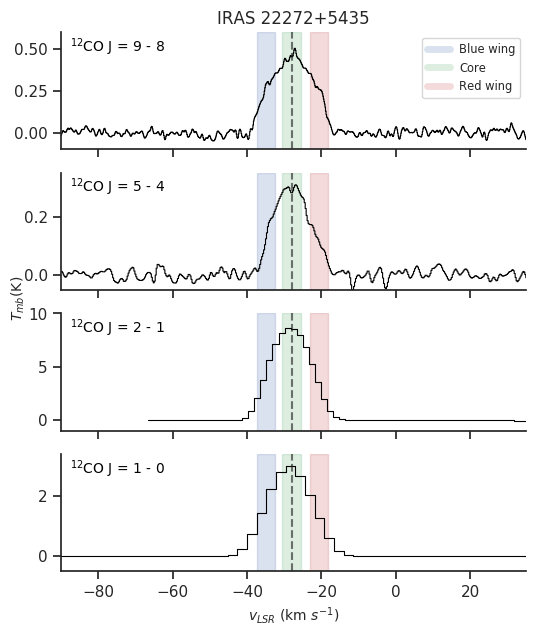}
        \caption{Observations of IRAS 22272+5435. Shadowed regions mark the LSR velocity ranges adopted for each feature and the dashed vertical line indicates the systemic velocity.}\label{fig:IRAS22272_ranges}
    \end{center}
\end{figure}

\begin{figure}
    \begin{center}
        \setlength\abovecaptionskip{-0.01\baselineskip}
        \setlength\belowcaptionskip{-0.7\baselineskip}
        \includegraphics[width= \linewidth]{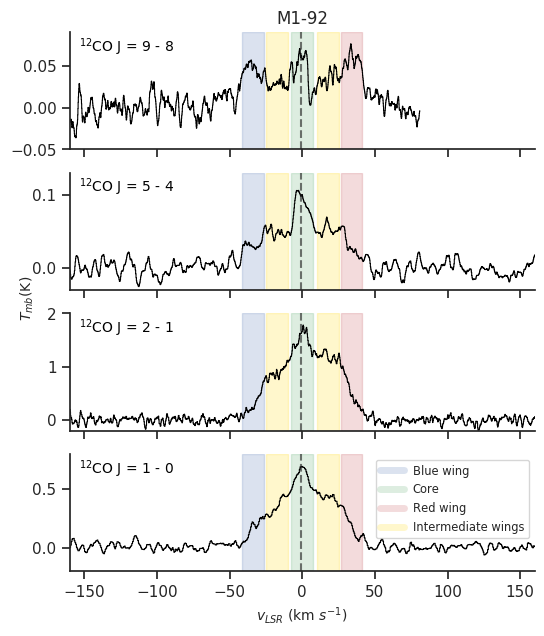}
        \caption{Observations of M 1--92. Shadowed regions mark the LSR velocity ranges adopted for each feature and the dashed vertical line indicates the systemic velocity. See Section~\ref{intermediate} for the reference to the intermediate wings.}\label{fig:M1-92_ranges}
    \end{center}
\end{figure}

\begin{figure}
    \begin{center}
        \setlength\abovecaptionskip{-0.01\baselineskip}
        \setlength\belowcaptionskip{-0.7\baselineskip}
        \includegraphics[width= \linewidth]{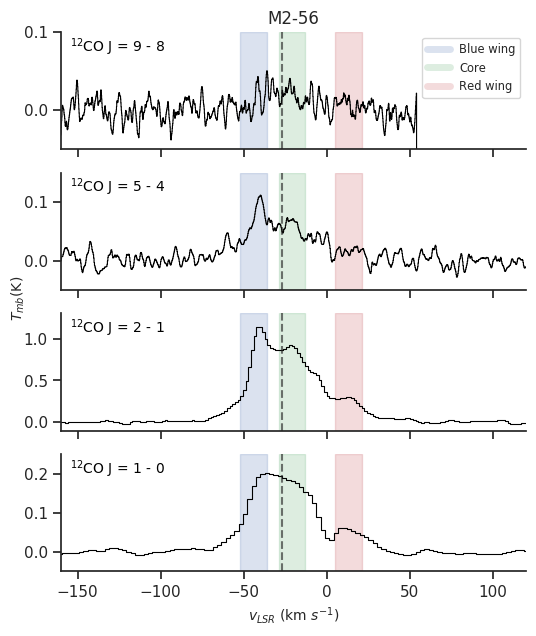}
        \caption{Observations of M 2--56. Shadowed regions mark the LSR velocity ranges adopted for each feature and the dashed vertical line indicates the systemic velocity.}\label{fig:M2-56_ranges}
    \end{center}
\end{figure}

\begin{figure}
    \begin{center}
        \setlength\abovecaptionskip{-0.01\baselineskip}
        \setlength\belowcaptionskip{-0.7\baselineskip}
        \includegraphics[width= \linewidth]{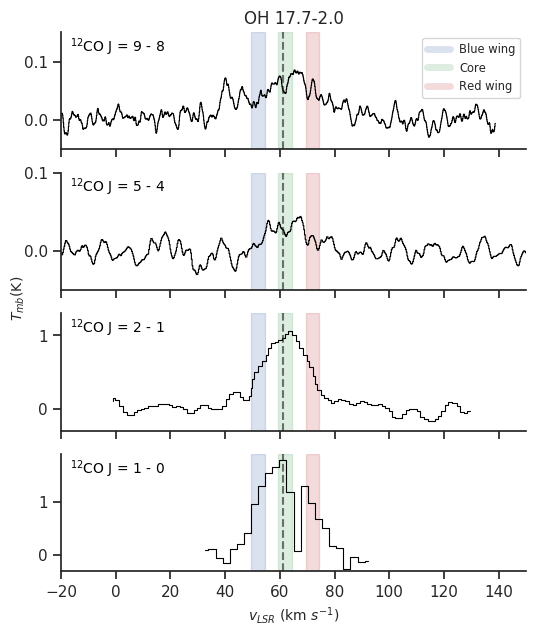}
        \caption{Observations of OH 17.7--2.0. Shadowed regions mark the LSR velocity ranges adopted for each feature and the dashed vertical line indicates the systemic velocity.}\label{fig:OH17_ranges}
    \end{center}
\end{figure}

\begin{figure}
    \begin{center}
        \setlength\abovecaptionskip{-0.01\baselineskip}
        \setlength\belowcaptionskip{-0.7\baselineskip}
        \includegraphics[width= \linewidth]{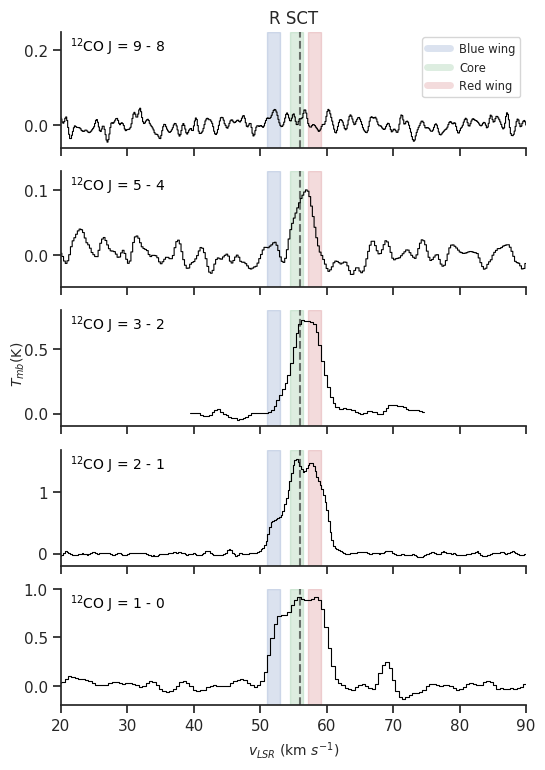}
        \caption{Observations of R Sct. Shadowed regions mark the LSR velocity ranges adopted for each feature and the dashed vertical line indicates the systemic velocity. The CO \mbox{$J$=3--2} line is taken from \citet{debeck_2010}. It is shown here only as an illustration of the line profile evolution between the low- and intermediate-$J$ CO lines, but it will not be used in our quantitative study.}\label{fig:R_SCT_ranges}
    \end{center}
\end{figure}

\begin{figure}
    \begin{center}
        \setlength\abovecaptionskip{-0.01\baselineskip}
        \setlength\belowcaptionskip{-0.7\baselineskip}
        \includegraphics[width= \linewidth]{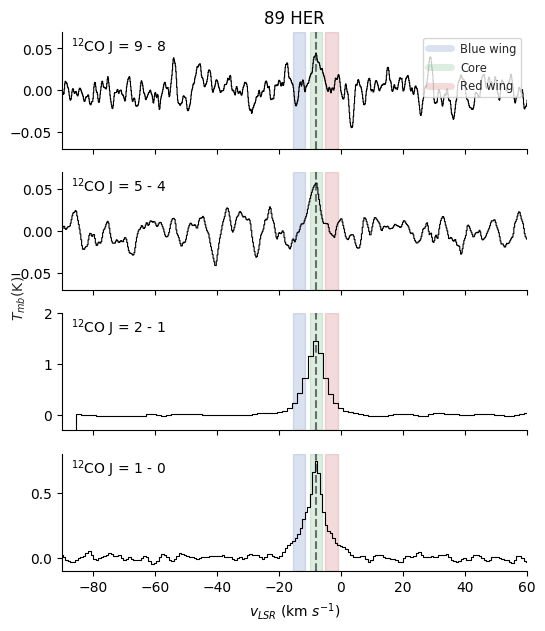}
        \caption{Observations of 89 Her. Shadowed regions mark the LSR velocity ranges adopted for each feature and the dashed vertical line indicates the systemic velocity.}\label{fig:89_HER_ranges}
    \end{center}
\end{figure}

\begin{figure}
    \begin{center}
        \setlength\abovecaptionskip{-0.01\baselineskip}
        \setlength\belowcaptionskip{-0.7\baselineskip}
        \includegraphics[width= \linewidth]{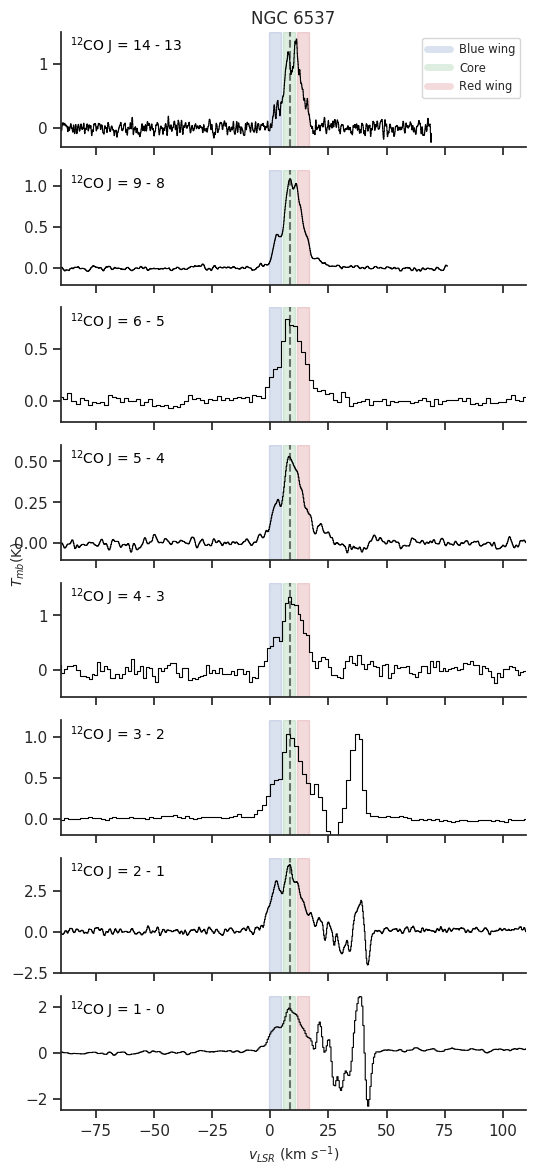}
        \caption{Observations of NGC 6537. Shadowed regions mark the LSR velocity ranges adopted for each feature and the dashed vertical line indicates the systemic velocity. The CO \mbox{$J$=3--2}, \mbox{$J$=4--3} and \mbox{$J$=6--5} transitions are taken from \citet{edwards_2014}. They are shown here just as an illustration of the line profile evolution between the low- and intermediate-$J$ CO lines, but they are used in our quantitative study.}\label{fig:NGC6537_ranges}
    \end{center}
\end{figure}

\begin{figure}
    \begin{center}
        \setlength\abovecaptionskip{-0.01\baselineskip}
        \setlength\belowcaptionskip{-0.7\baselineskip}
        \includegraphics[width= \linewidth]{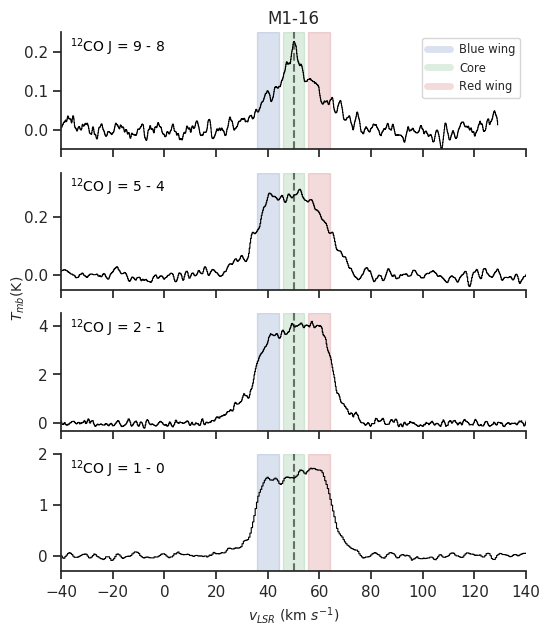}
        \caption{Observations of M 1--16. Shadowed regions mark the LSR velocity ranges adopted for each feature and the dashed vertical line indicates the systemic velocity.}\label{fig:M1-16_ranges}
    \end{center}
\end{figure}

\begin{figure}
    \begin{center}
        \setlength\abovecaptionskip{-0.01\baselineskip}
        \setlength\belowcaptionskip{-0.7\baselineskip}
        \includegraphics[width= \linewidth]{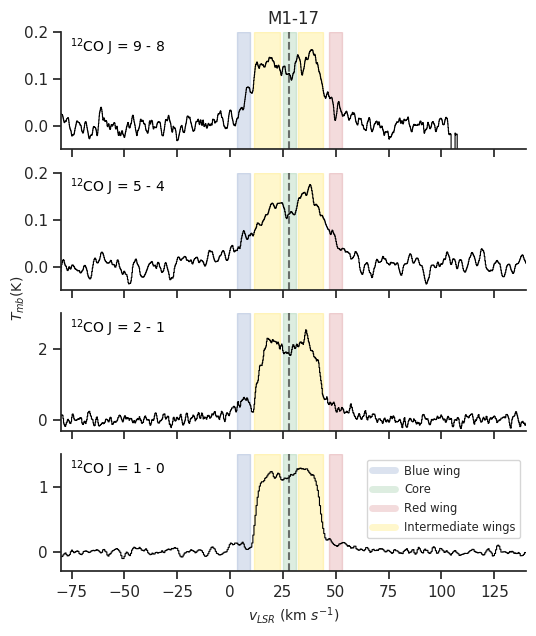}
        \caption{ Observations of M 1--17. Shadowed regions mark the LSR velocity ranges adopted for each feature and the dashed vertical line indicates the systemic velocity. See Section~\ref{intermediate} for the reference to the intermediate wings.} \label{fig:M1-17_ranges}
    \end{center}
\end{figure}

\begin{figure}
    \begin{center}
        \setlength\abovecaptionskip{-0.01\baselineskip}
        \setlength\belowcaptionskip{-0.7\baselineskip}
        \includegraphics[width= \linewidth]{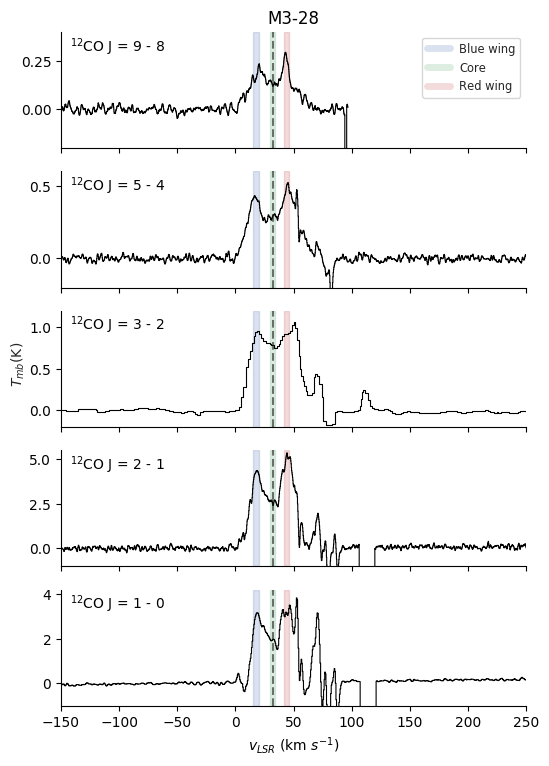}
        \caption{Observations of M 3--28. Shadowed regions mark the LSR velocity ranges adopted for each feature and the dashed vertical line indicates the systemic velocity. The CO \mbox{$J$=3--2}, transitions is taken from \citet{schmidt_ziurys_2016}. It is shown here just as an illustration of the line profile evolution between the low- and intermediate-$J$ CO lines, but is not used in our quantitative study.}\label{fig:M3-28_ranges}
    \end{center}
\end{figure}


\begin{figure}
    \begin{center}
        \setlength\abovecaptionskip{-0.01\baselineskip}
        \setlength\belowcaptionskip{-0.7\baselineskip}
        \includegraphics[width= \linewidth]{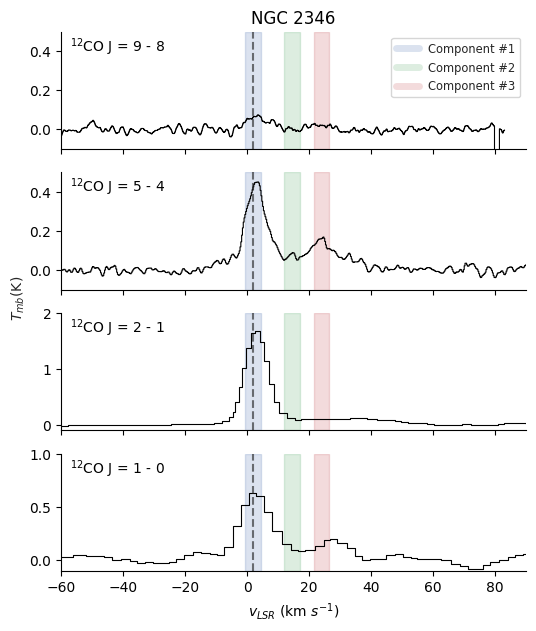}
        \caption{Observations of NGC 2346. Shadowed regions indicate the velocity ranges adopted for the detected velocity components and the dashed line marks the systemic velocity.}\label{fig:NGC2346_ranges}
    \end{center}
\end{figure}

\begin{figure}
    \begin{center}
        \setlength\abovecaptionskip{-0.01\baselineskip}
        \setlength\belowcaptionskip{-0.7\baselineskip}
        \includegraphics[width= \linewidth]{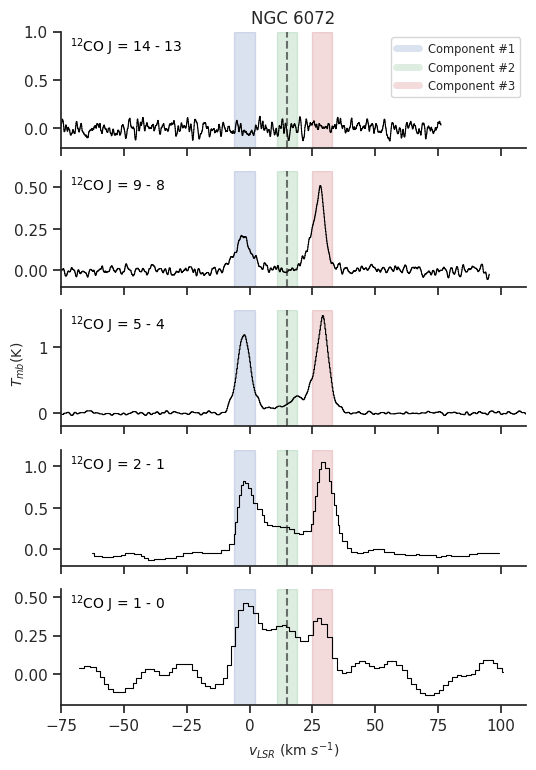}
        \caption{Observations of NGC 6072. Shadowed regions indicate the velocity ranges adopted for the detected velocity components and the dashed line marks the systemic velocity.}\label{fig:NGC6072_ranges}
    \end{center}
\end{figure}

\begin{figure}
    \begin{center}
        \setlength\abovecaptionskip{-0.01\baselineskip}
        \setlength\belowcaptionskip{-0.7\baselineskip}
        \includegraphics[width= \linewidth]{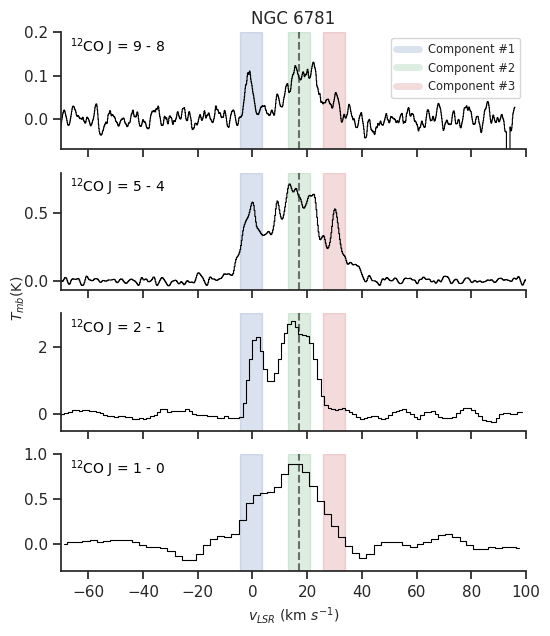}
        \caption{Observations of NGC 6781. Shadowed regions indicate the velocity ranges adopted for the detected velocity components and the dashed line marks the systemic velocity.}\label{fig:NGC6781_ranges}
    \end{center}
\end{figure}

\begin{figure}
    \begin{center}
        \setlength\abovecaptionskip{-0.01\baselineskip}
        \setlength\belowcaptionskip{-0.7\baselineskip}
        \includegraphics[width= \linewidth]{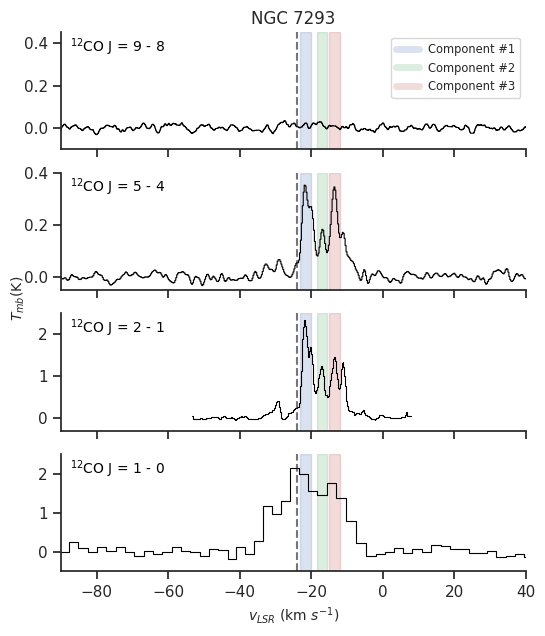}
        \caption{Observations of NGC 7293. Shadowed regions indicate the velocity ranges adopted for the detected velocity components and the dashed line marks the systemic velocity.}\label{fig:NGC7293_ranges}
    \end{center}
\end{figure}

\section{Optical and FIR images of the sample}
\label{sec:appxa}

In this section, we present the optical and FIR images (when available) of our targets. The \textit{Herschel/HIFI} beams of the lines used in this work are overlayed on these maps red contours, in order of size from largest to smallest: $^{12}$CO \mbox{$J$=5--4}, $^{12}$CO \mbox{$J$=9--8} and $^{12}$CO \mbox{$J$=14--13}.

\onecolumn

\begin{figure*}
\label{fig:map1}
 \begin{minipage}{0.45\textwidth}
     \centering
     \includegraphics[width=\linewidth]{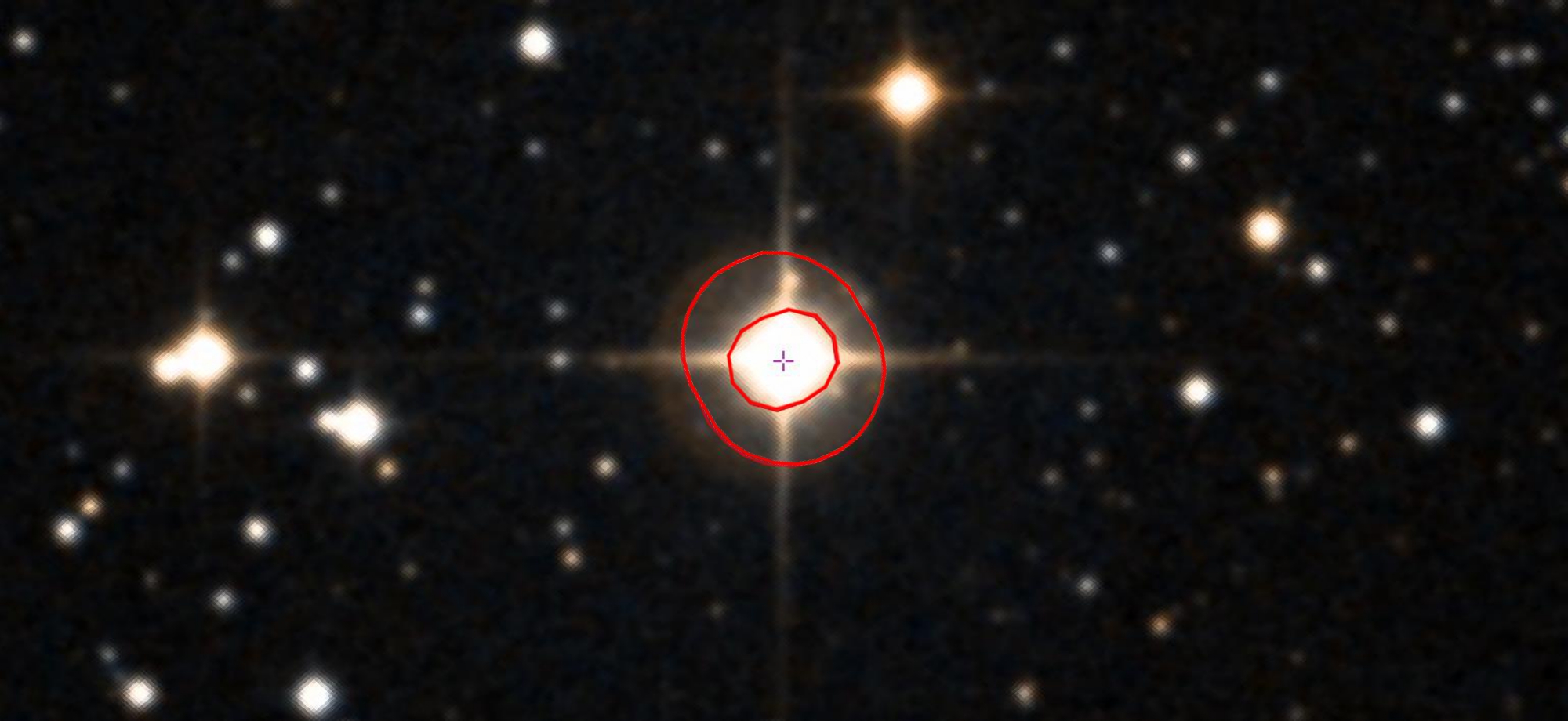}
     \caption{Image of IRAS 07134+1005 obtained from DSS2 colour.}\label{fig:IRAS07134_optico}
   \end{minipage}\hfill
   \begin{minipage}{0.45\textwidth}
     \centering
     \includegraphics[width=\linewidth]{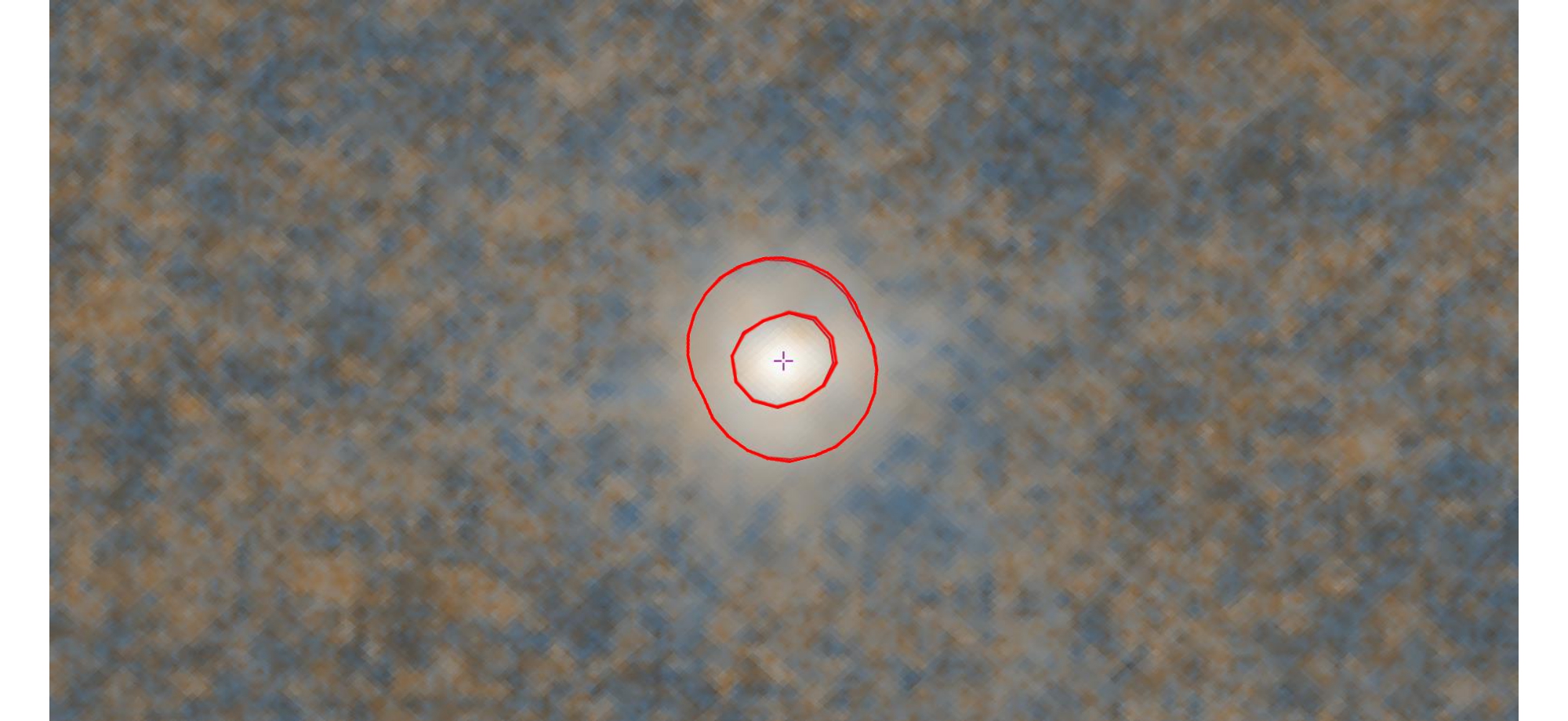}
     \caption{FIR image of IRAS 07134+1005 obtained by Herschel PACS RGB 70,160 microns.}\label{fig:IRAS07134_FIR}
   \end{minipage}
   
\begin{minipage}{0.45\textwidth}
     \centering
     \includegraphics[width=\linewidth]{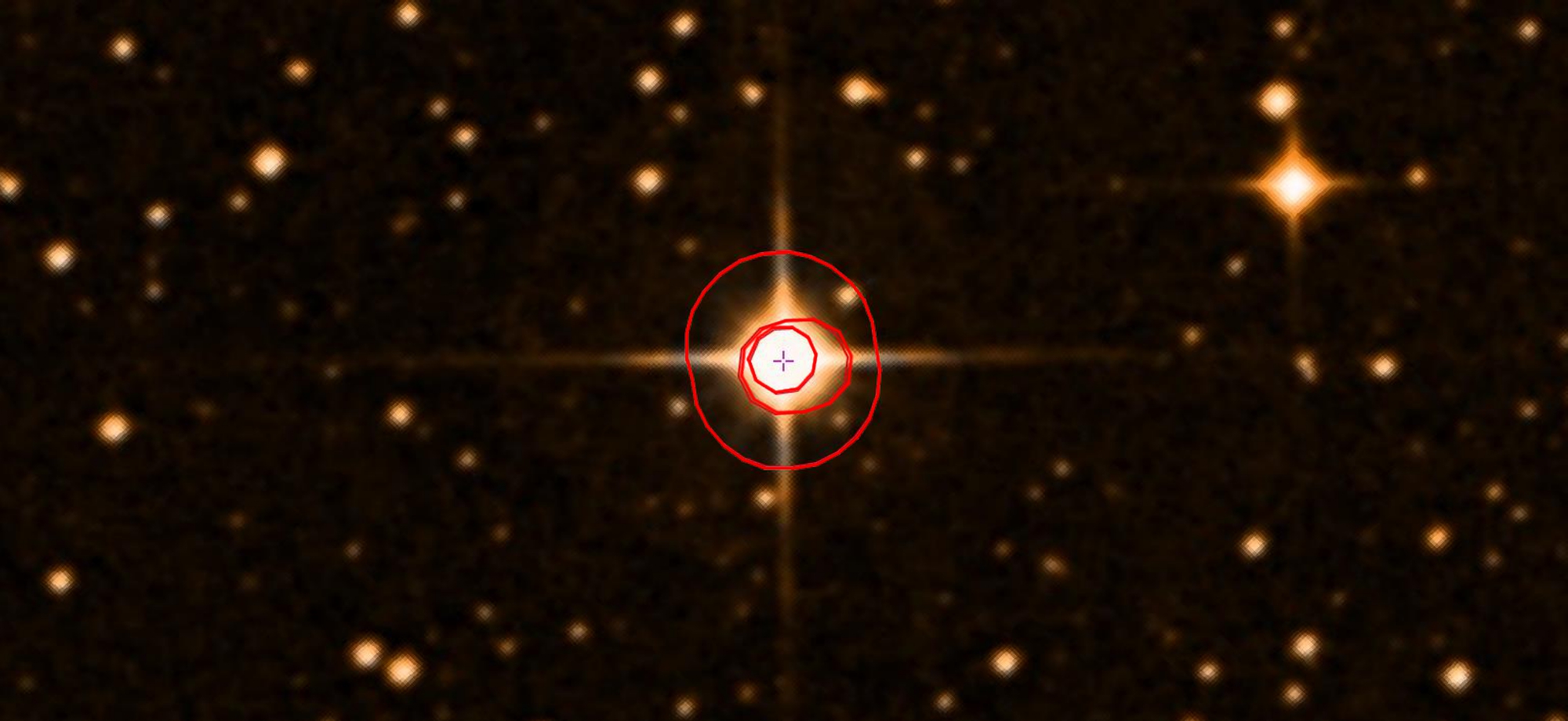}
     \caption{Image of IRAS 19500--1709 obtained from DSS2 colour.}\label{fig:IRAS19500_optico}
   \end{minipage}\hfill
   \begin{minipage}{0.45\textwidth}
     \centering
     \includegraphics[width=\linewidth]{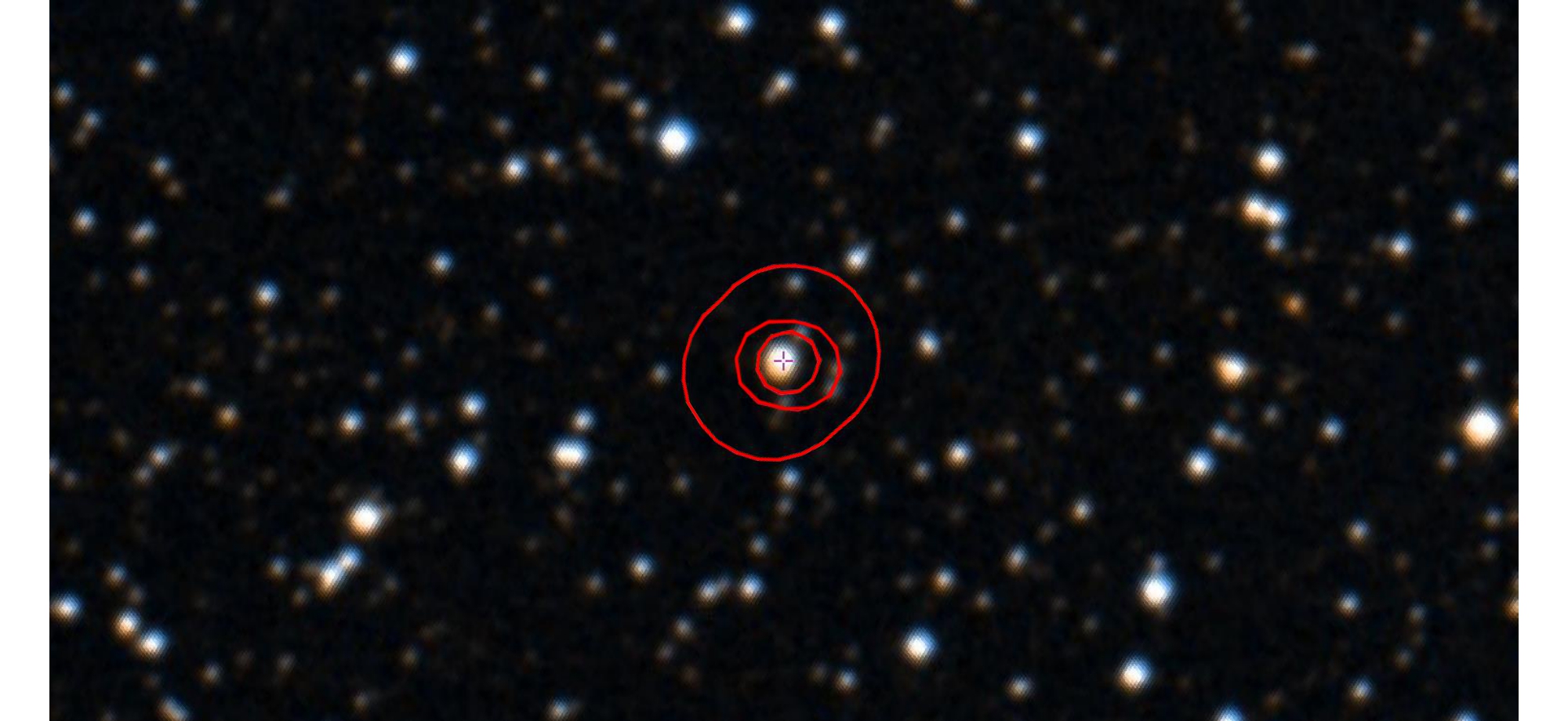}
     \caption{Image of IRAS 21282+5050 obtained from DSS2 colour.}\label{fig:IRAS21282_optico}
   \end{minipage}
   
\begin{minipage}{0.45\textwidth}
     \centering
     \includegraphics[width=\linewidth]{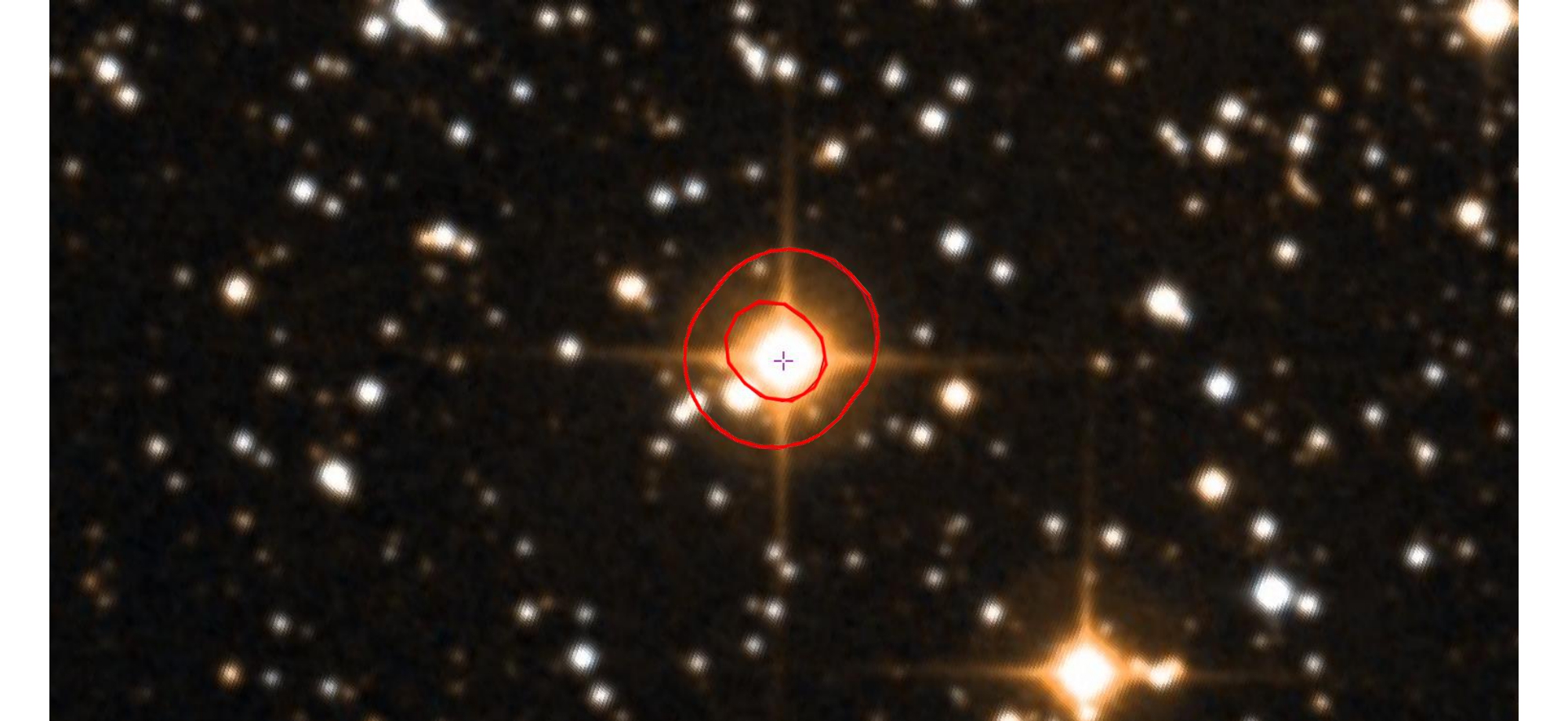}
     \caption{Image of IRAS 22272+5435 obtained from DSS2 colour.}\label{fig:IRAS22272_optico}
   \end{minipage}\hfill
   \begin{minipage}{0.45\textwidth}
   \centering
   \includegraphics[width=\linewidth]{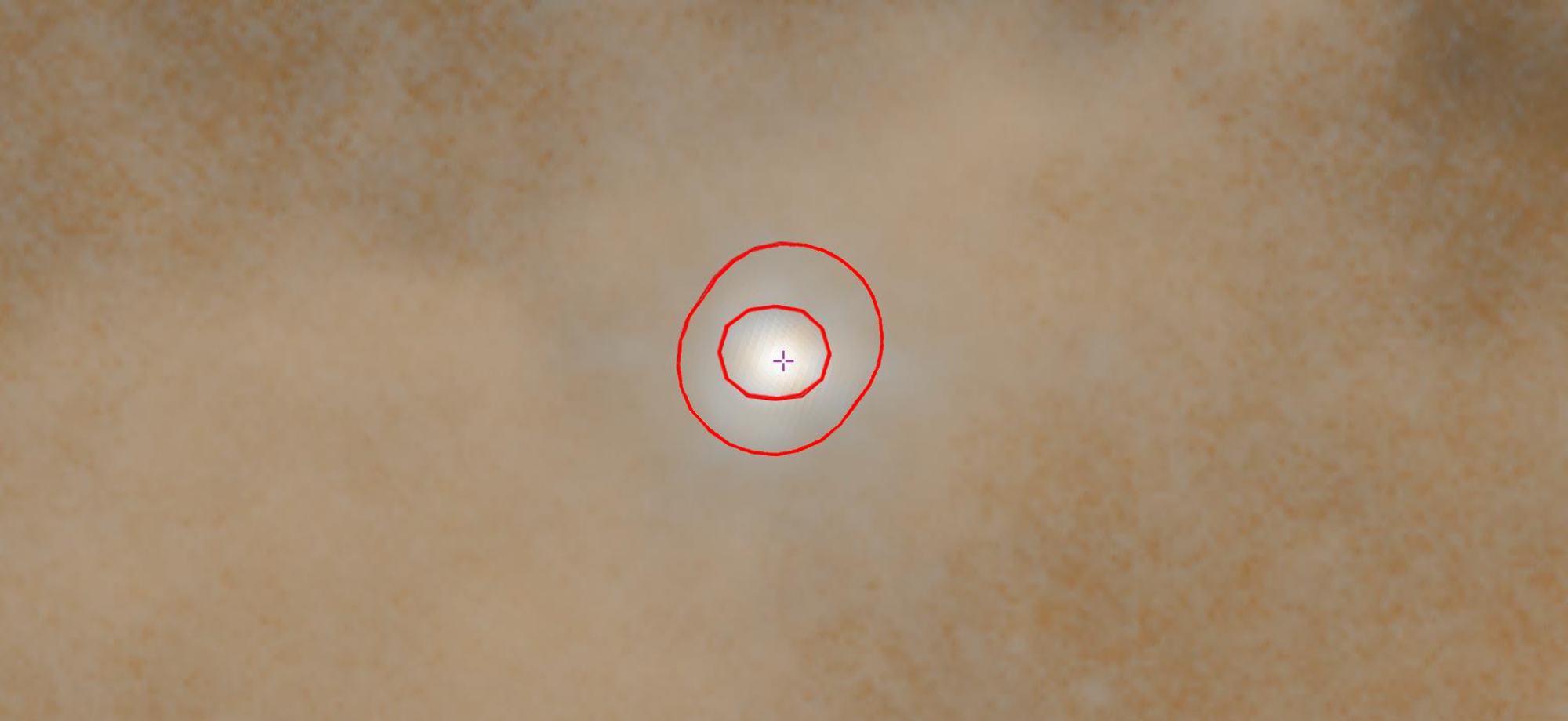}
     \caption{FIR image of IRAS 22272+5435 obtained by Herschel PACS RGB 70,160 microns.}\label{fig:IRAS22272_FIR}
   \end{minipage}
\begin{minipage}{0.45\textwidth}
     \centering
     \includegraphics[width=\linewidth]{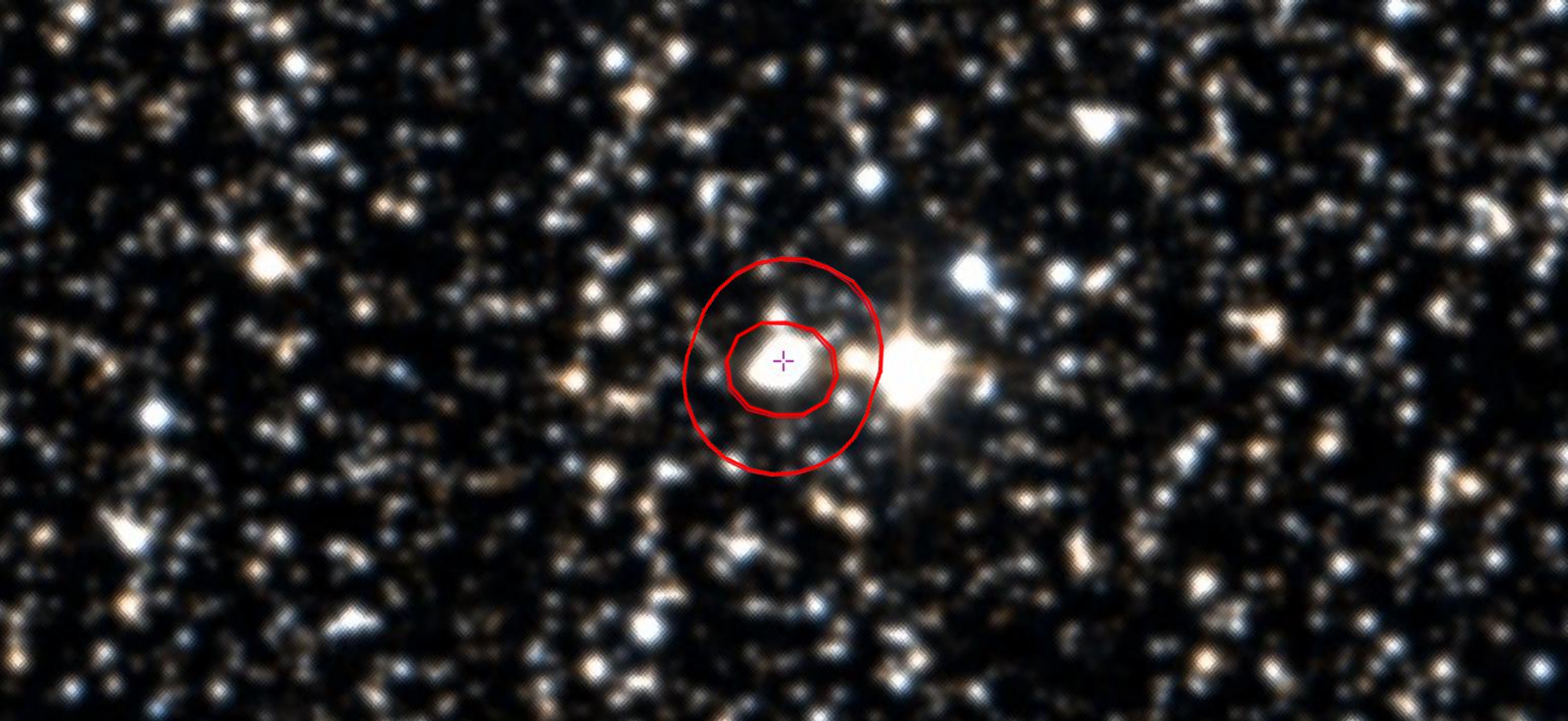}
     \caption{Image of M 1--92 obtained from DSS2 colour.}\label{fig:M1-92_optico}
   \end{minipage}\hfill
   \begin{minipage}{0.45\textwidth}
     \centering
     \includegraphics[width=\linewidth]{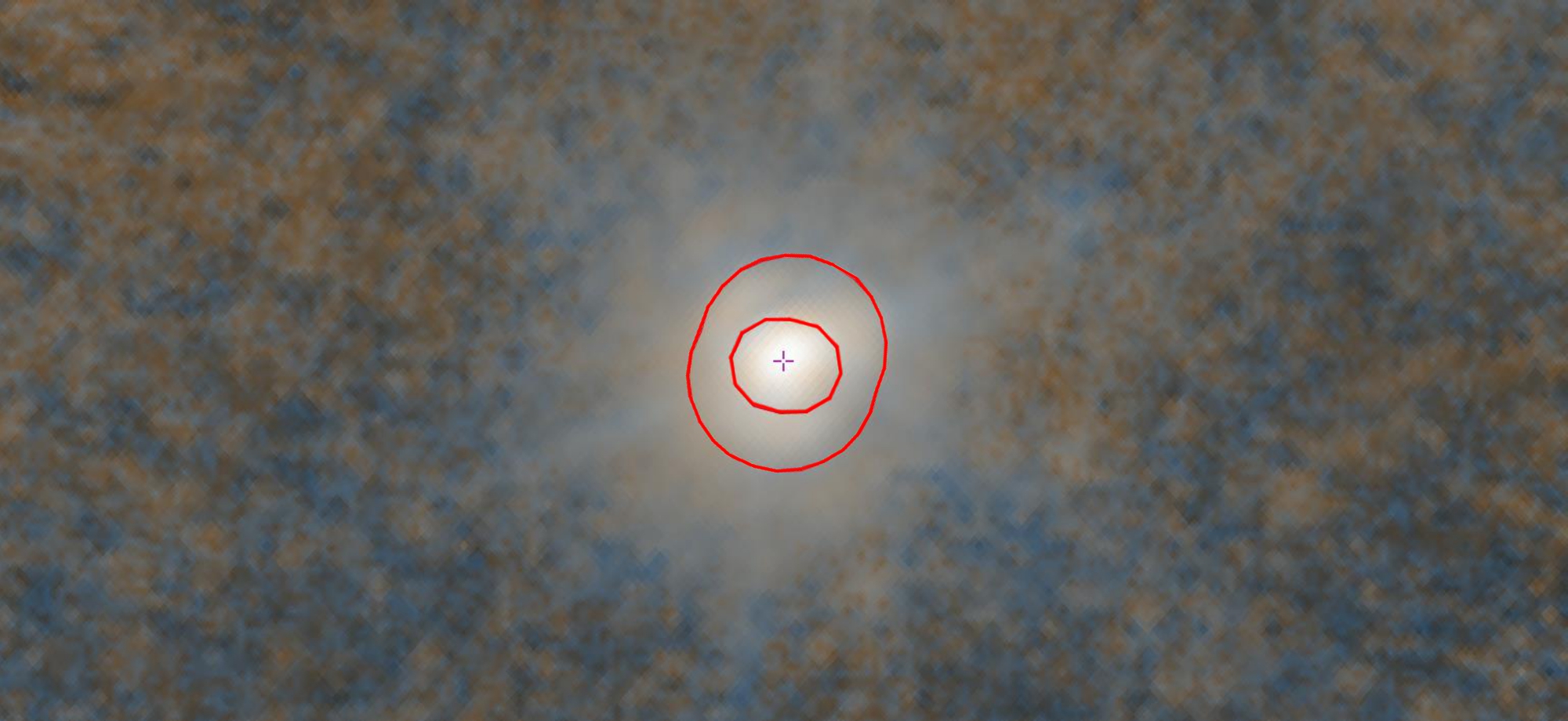}
     \caption{FIR image of M 1--92 obtained by Herschel PACS RGB 70,160 microns.}\label{fig:M1-92_FIR}
   \end{minipage}
  
\end{figure*}

\newpage

\begin{figure*}
\label{fig:map2}
 \begin{minipage}{0.45\textwidth}
     \centering
     \includegraphics[width=\linewidth]{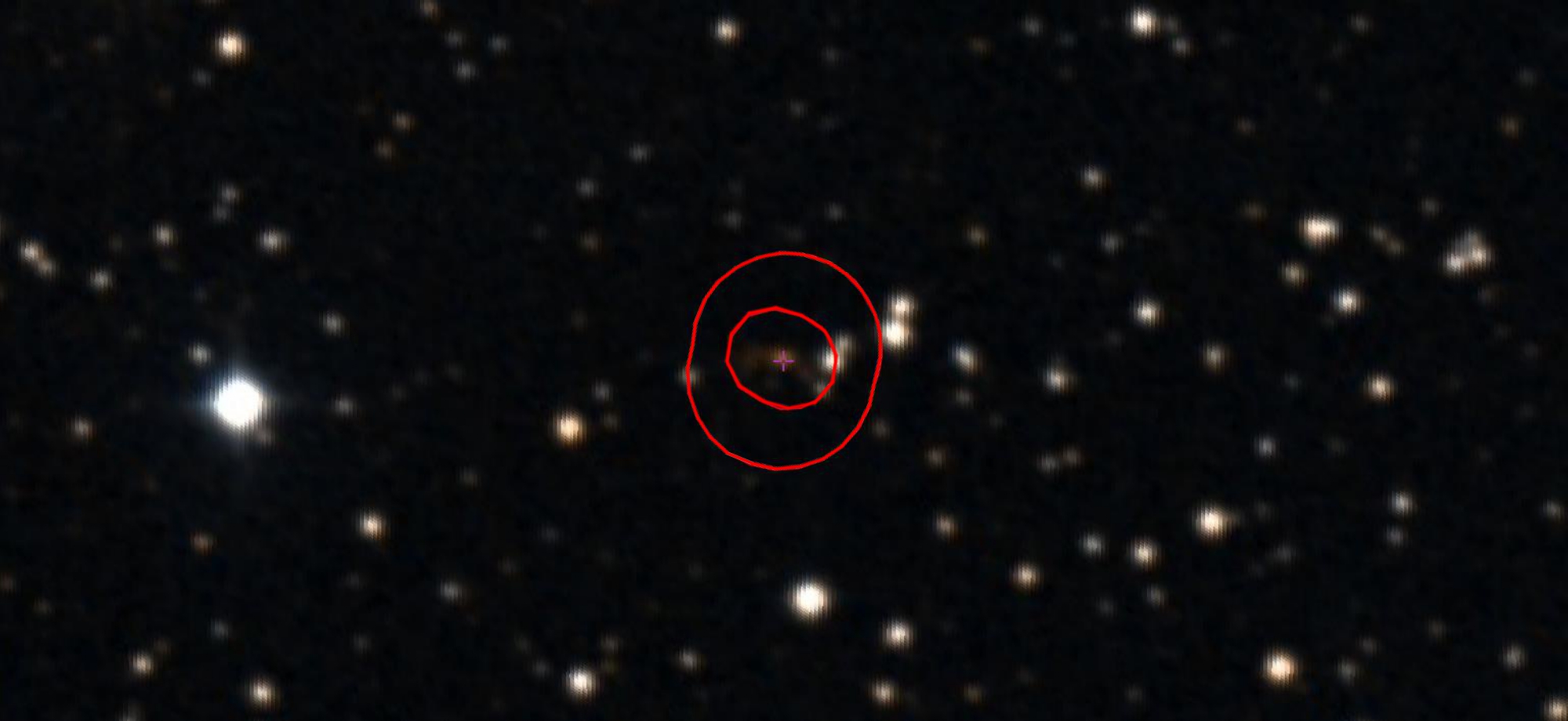}
     \caption{Image of M 2--56 obtained from DSS2 colour.}\label{fig:M2-56_optico}
   \end{minipage}\hfill
   \begin{minipage}{0.45\textwidth}
     \centering
     \includegraphics[width=\linewidth]{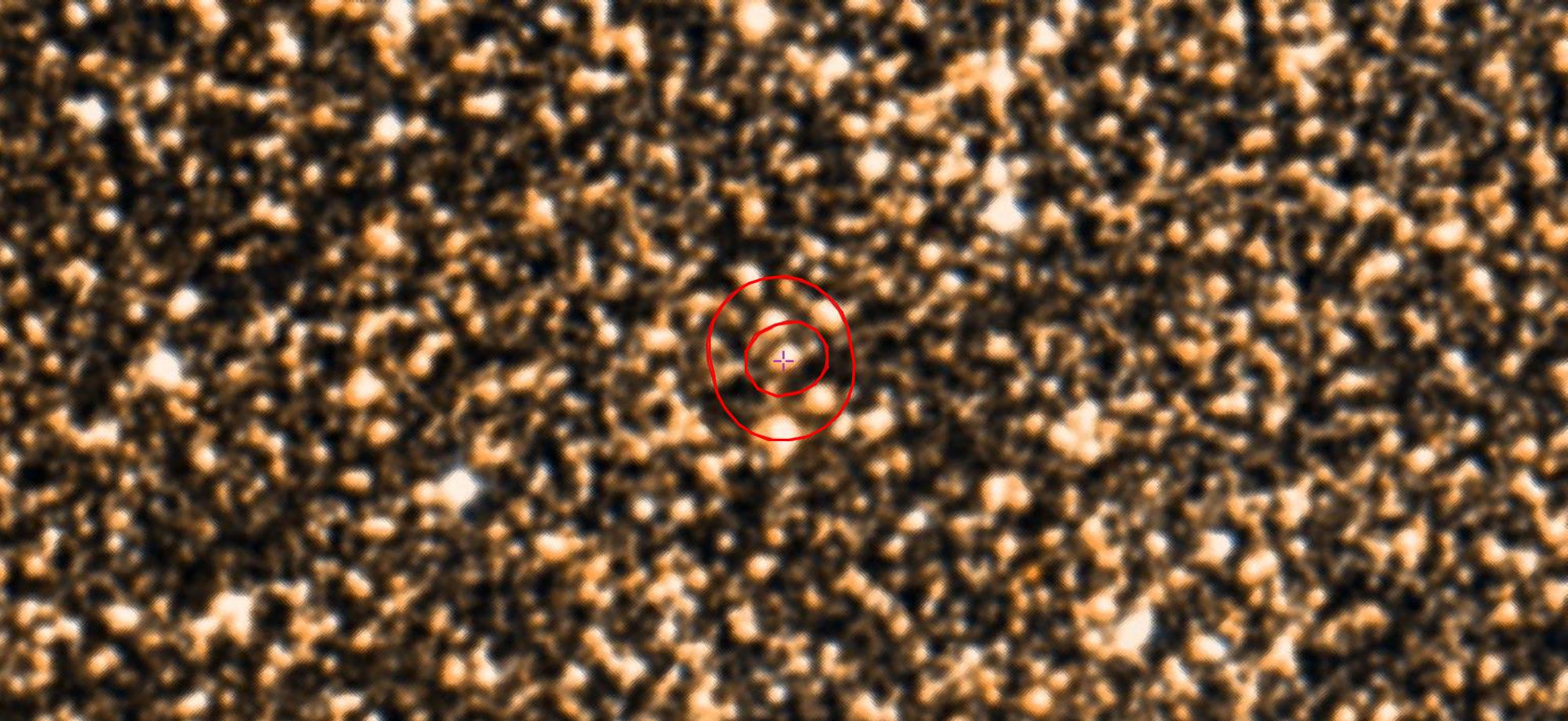}
     \caption{Image of OH 17.7--2.0 obtained from DSS2 colour.}\label{fig:OH177_20_optico}
   \end{minipage}
   
\begin{minipage}{0.45\textwidth}
     \centering
     \includegraphics[width=\linewidth]{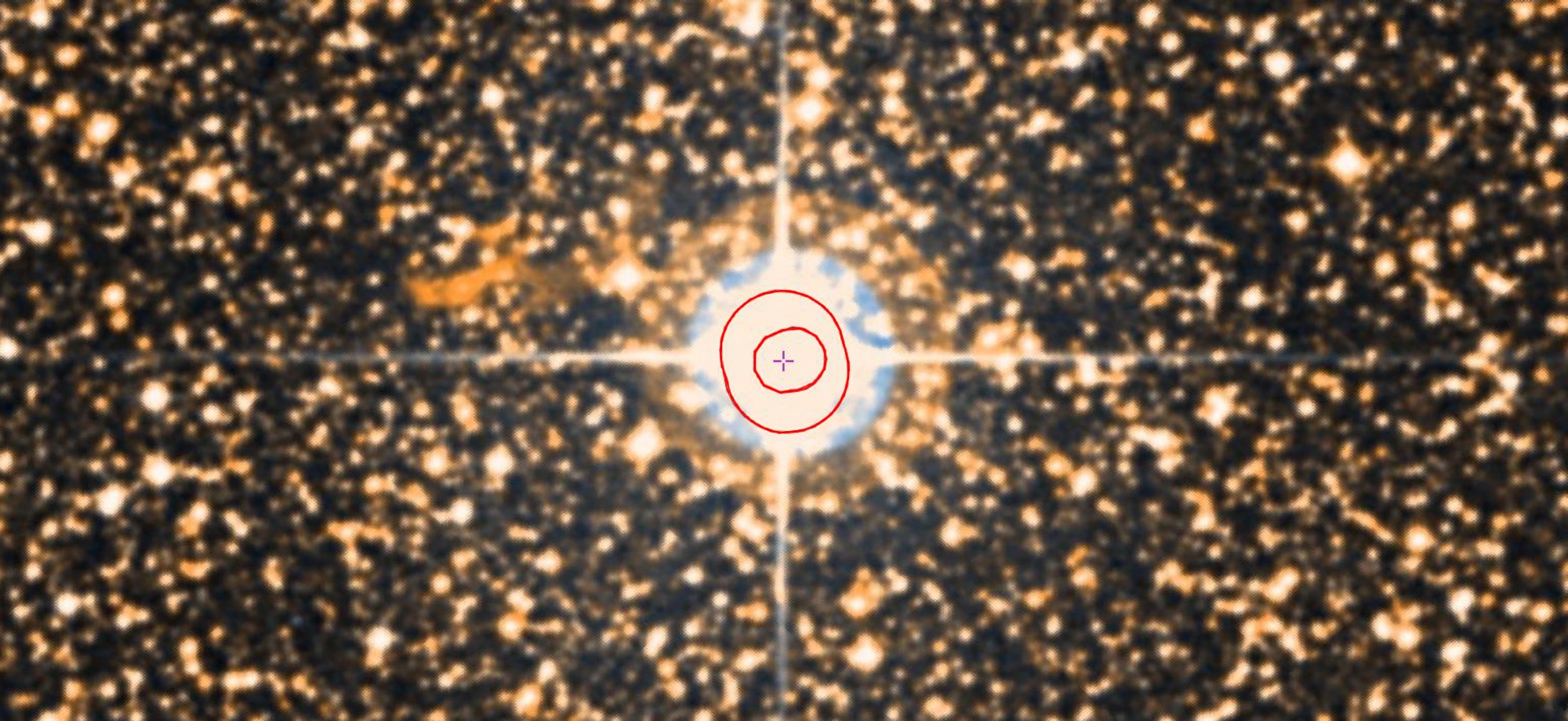}
     \caption{Image of R Sct obtained from DSS2 colour.}\label{fig:R_Sct_optico}
   \end{minipage}\hfill
   \begin{minipage}{0.45\textwidth}
     \centering
     \includegraphics[width=\linewidth]{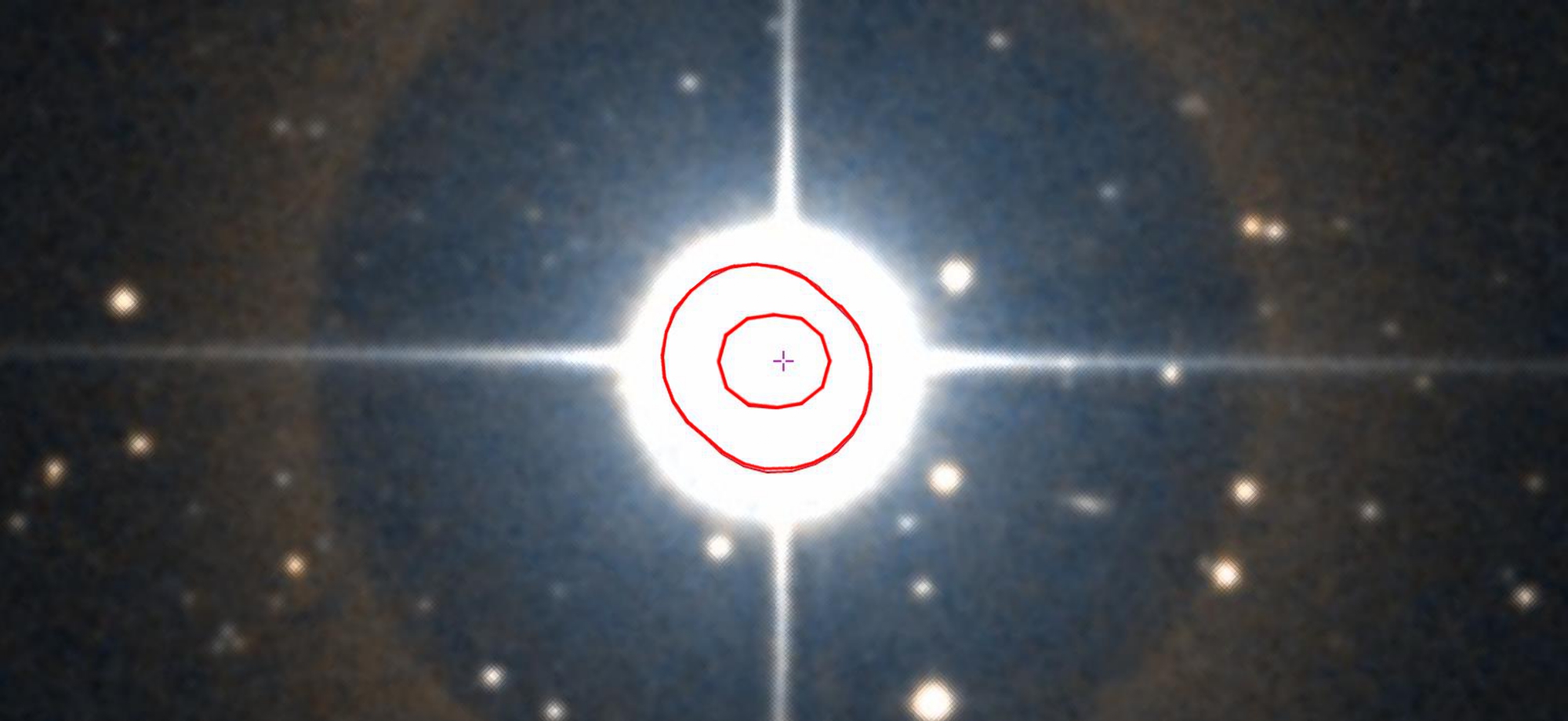}
     \caption{Image of 89 Her obtained from DSS2 colour.}\label{fig:89_Her_optico}
   \end{minipage}
   
\begin{minipage}{0.45\textwidth}
     \centering
     \includegraphics[width=\linewidth]{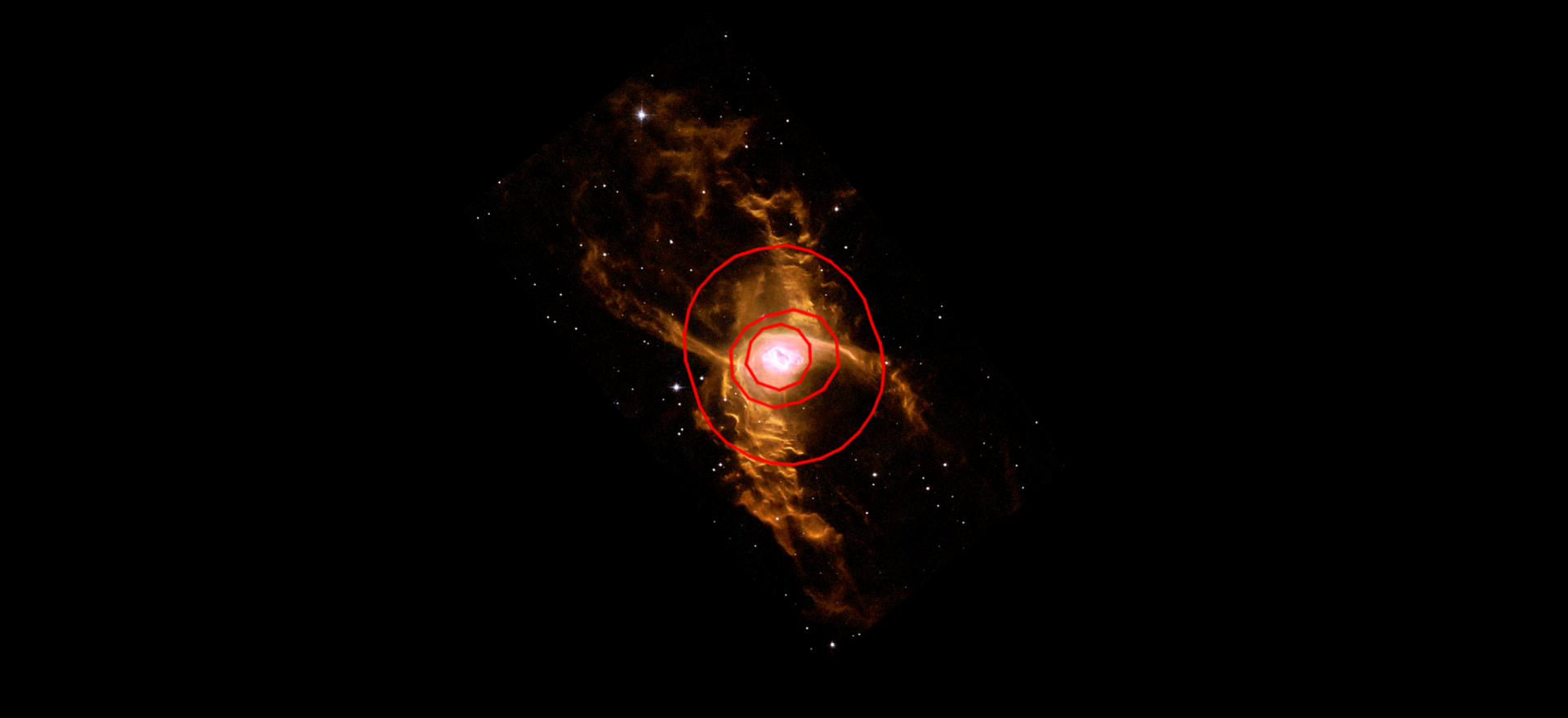}
     \caption{NASA/ESA \textit{Hubble Space Telescope} image of NGC 6537, a composite of [N II] 6584 $\AA$ (red), H$\alpha$ 6563 $\AA$ (green), and [O III] 5007 $\AA$
(blue) emission.}\label{fig:NGC6537_optico}
   \end{minipage}\hfill
   \begin{minipage}{0.45\textwidth}
     \centering
     \includegraphics[width=\linewidth]{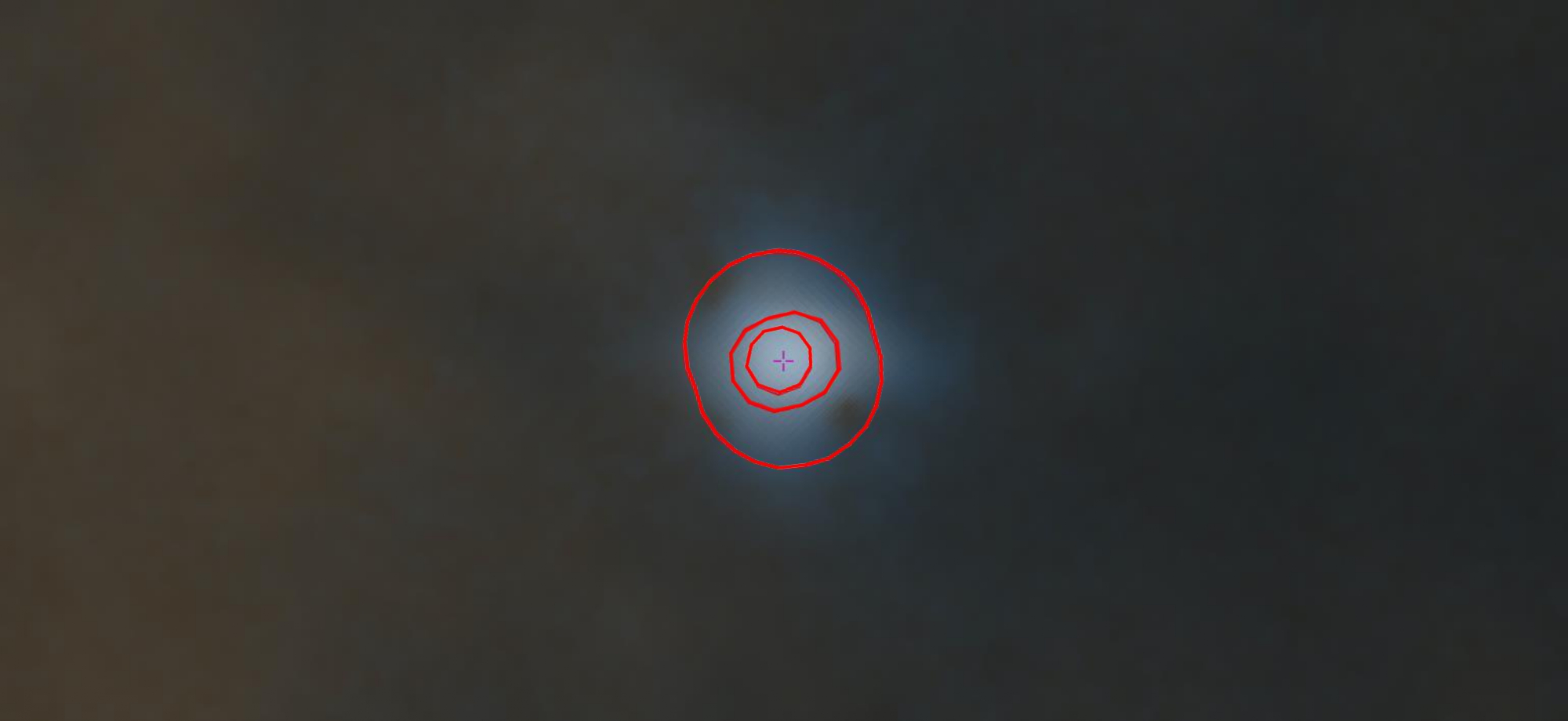}
     \caption{FIR image of NGC 6537 obtained by Herschel PACS RGB 70,160 microns.}\label{fig:NGC6537_FIR}
   \end{minipage}
   
\begin{minipage}{0.45\textwidth}
     \centering
     \includegraphics[width=\linewidth]{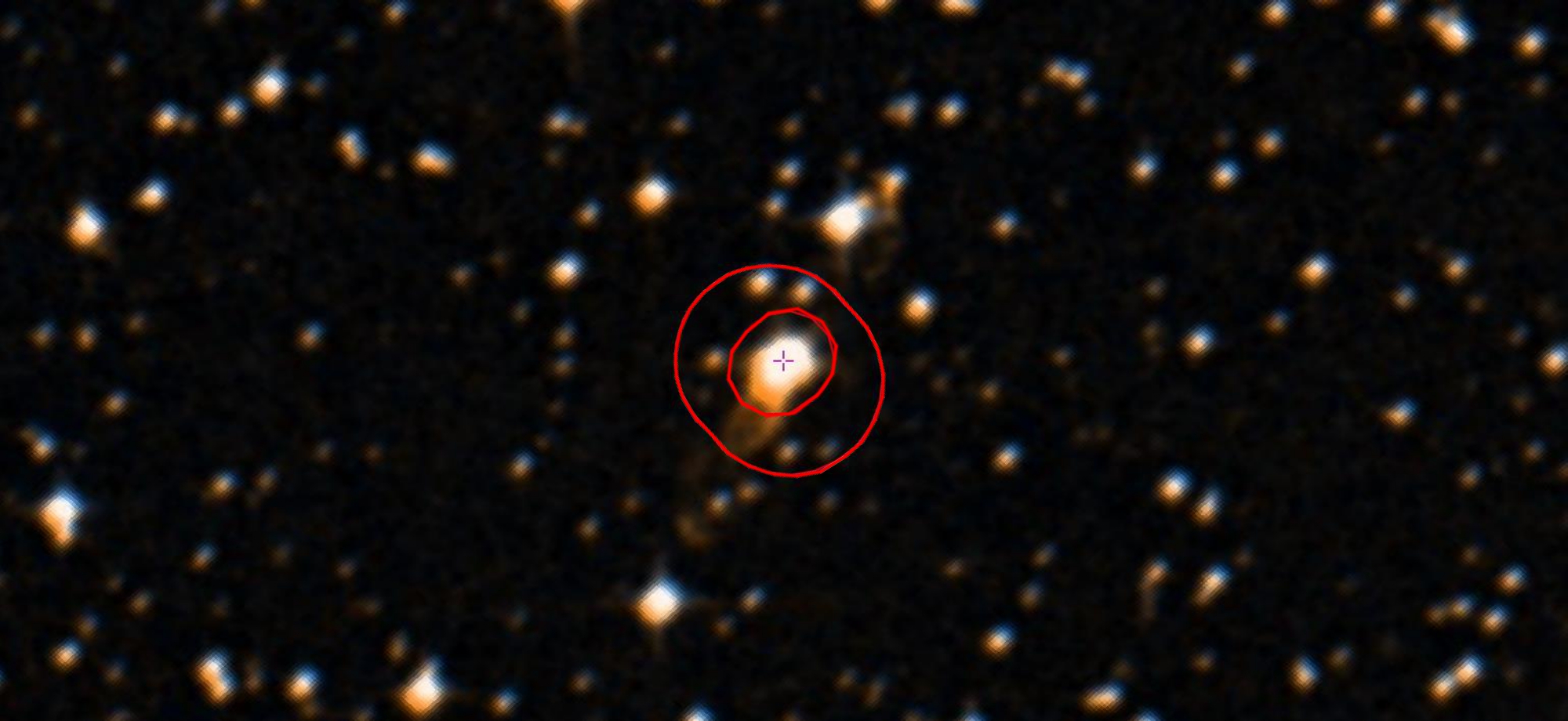}
     \caption{Image of M 1--16 obtained from DSS2 colour.}\label{fig:M1-16_optico}
   \end{minipage}\hfill
   \begin{minipage}{0.45\textwidth}
     \centering
     \includegraphics[width=\linewidth]{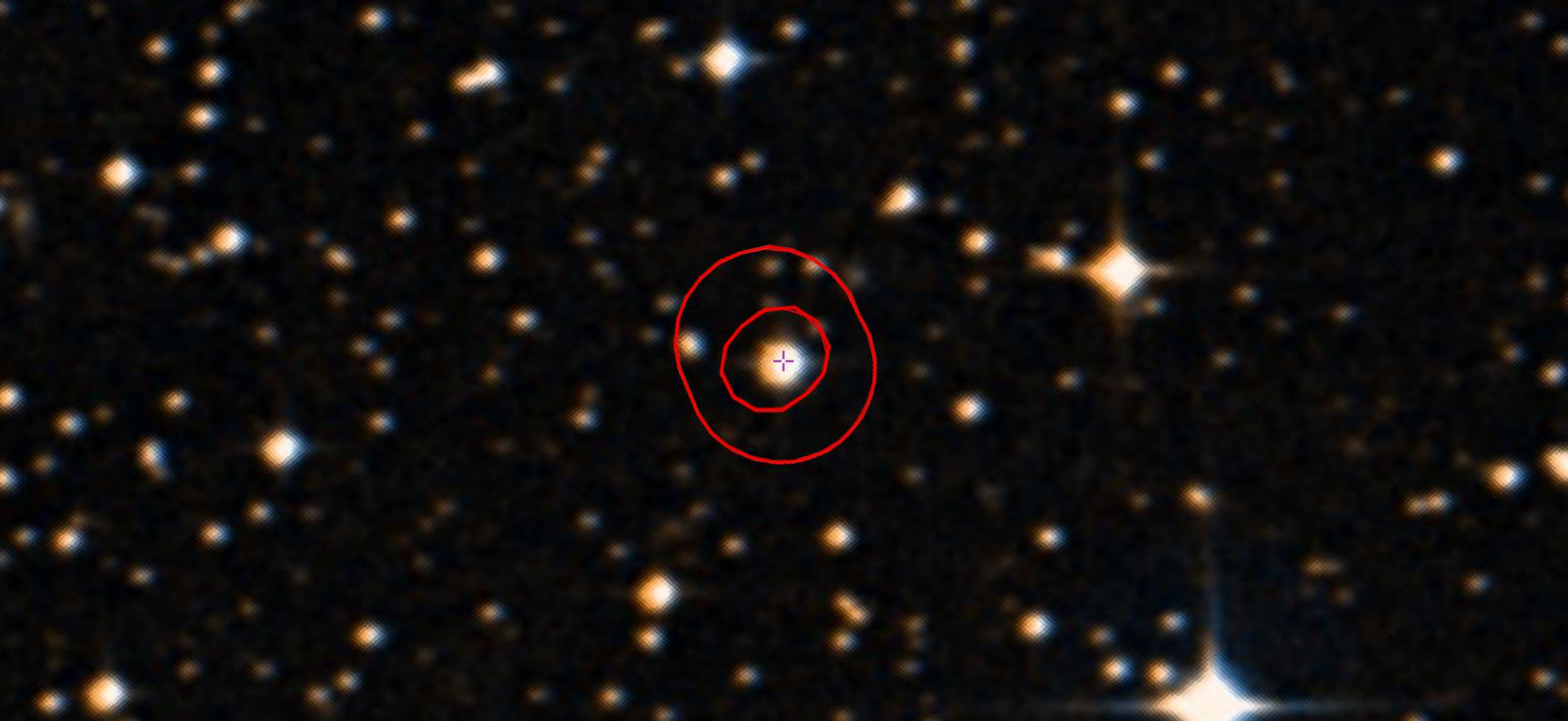}
     \caption{Image of M 1--17 obtained from DSS2 colour.}\label{fig:M1-17_optico}
   \end{minipage}
   
 \begin{minipage}{0.45\textwidth}
     \centering
     \includegraphics[width=\linewidth]{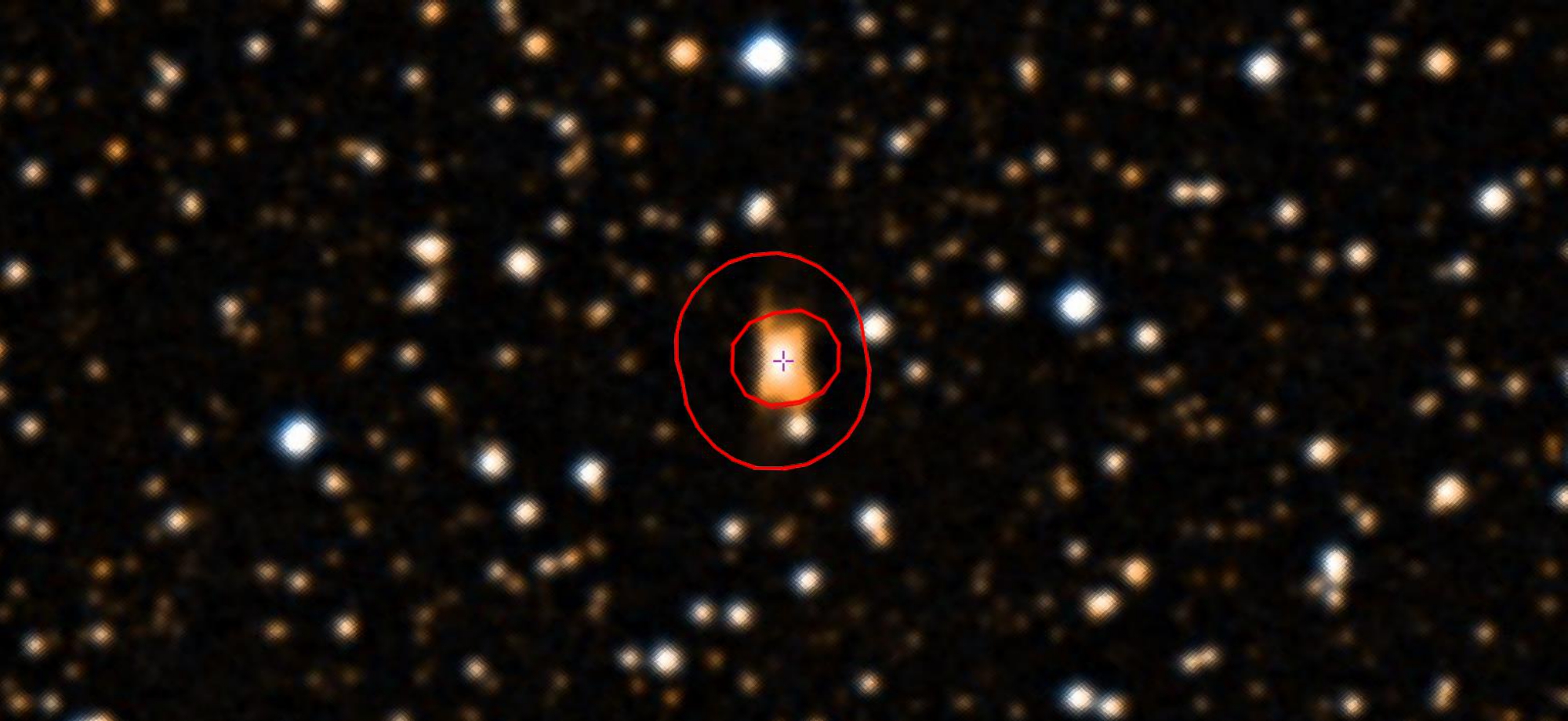}
     \caption{Image of M 3--28 obtained from DSS2 colour.}\label{fig:M3-28_optico}
   \end{minipage}\hfill
   \begin{minipage}{0.45\textwidth}
     \centering
     \includegraphics[width=\linewidth]{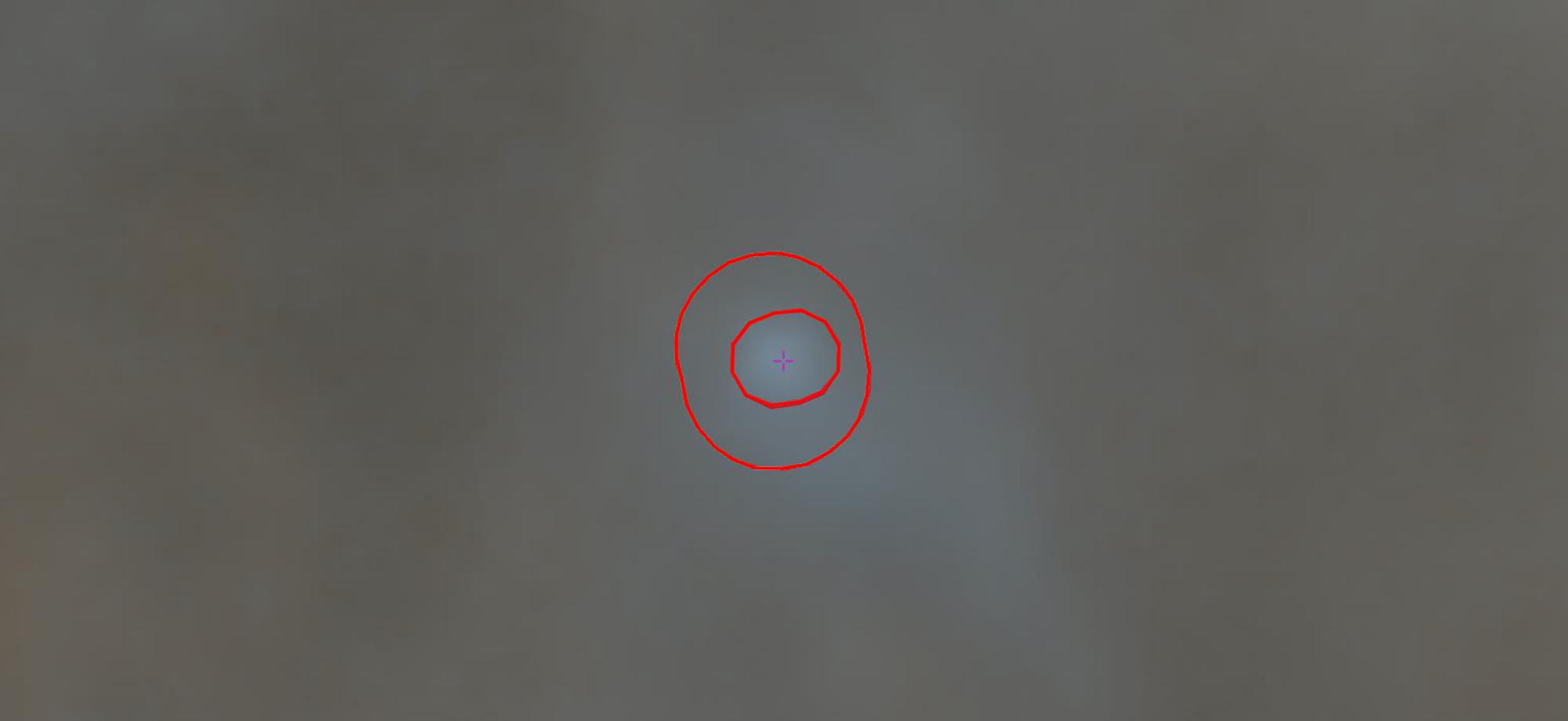}
     \caption{FIR image of M 3--28 obtained by Herschel PACS RGB 70,160 microns.}\label{fig:M3-28_FIR}
   \end{minipage}
\end{figure*}

\newpage

\begin{figure*}
\label{fig:map3}
 \begin{minipage}{0.45\textwidth}
     \centering
     \includegraphics[width=\linewidth]{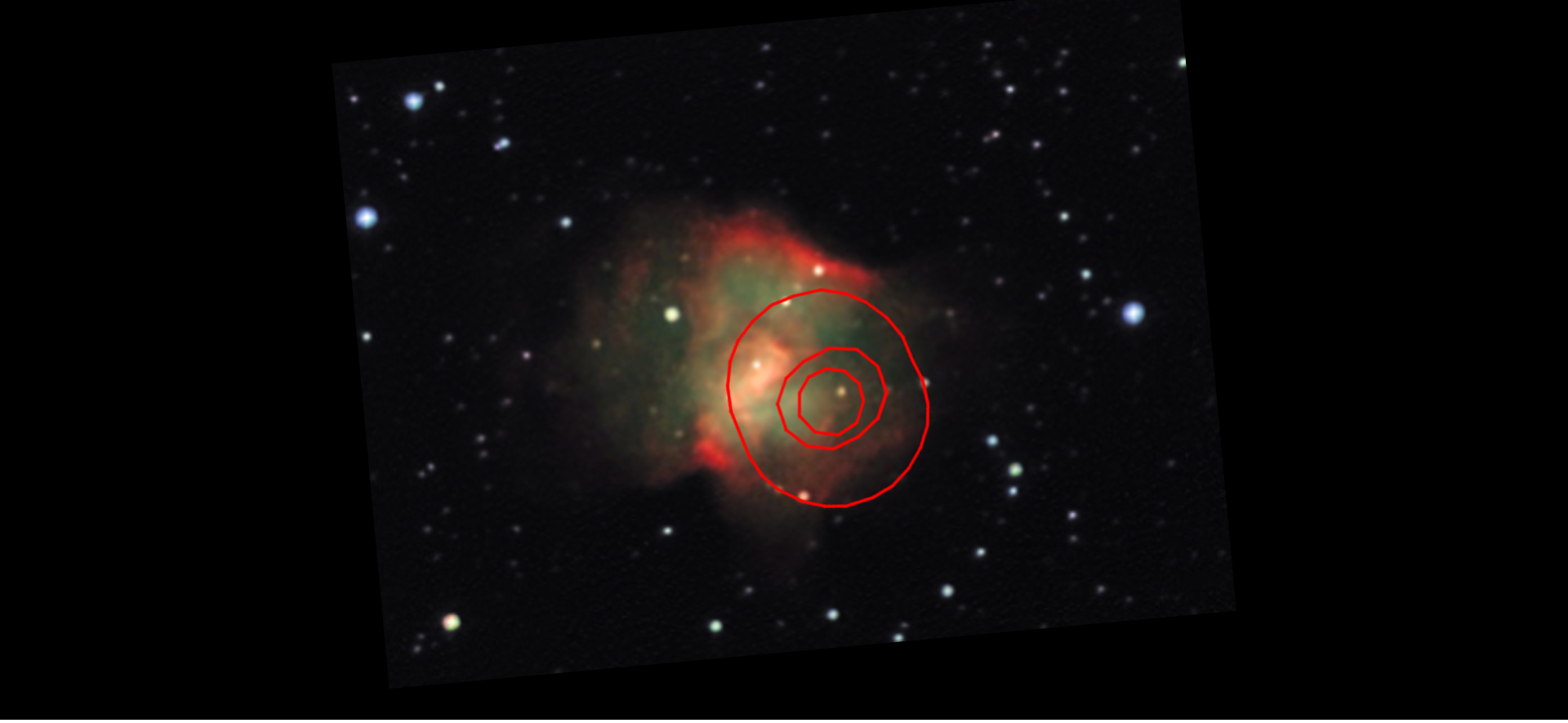}
     \caption{Composite image of NGC 6072, H$\alpha$ (red), [OIII] (green), B filters (blue), produced by Ganymed Telescope.}\label{fig:NGC6072_optico}
   \end{minipage}\hfill
   \begin{minipage}{0.45\textwidth}
     \centering
     \includegraphics[width=\linewidth]{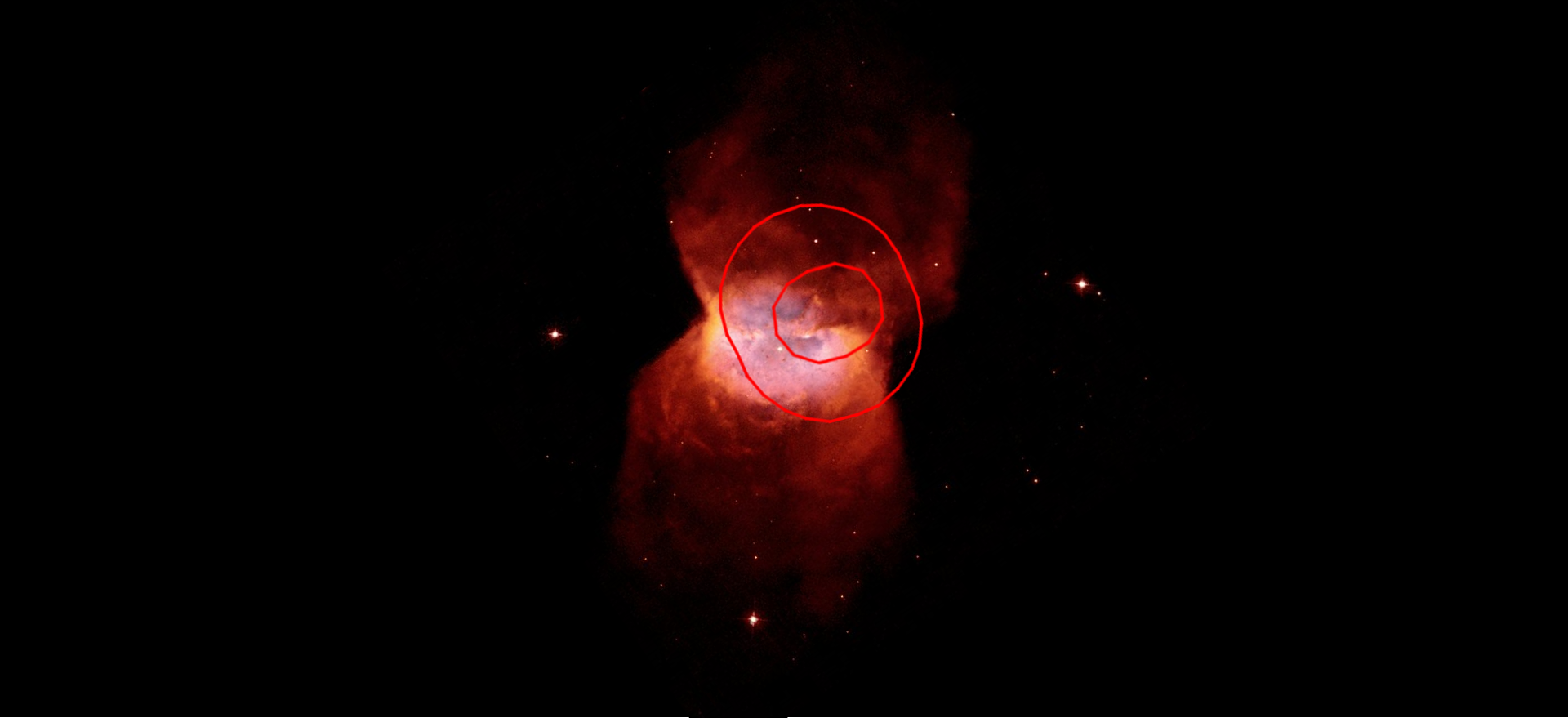}
     \caption{NASA/ESA \textit{Hubble Space Telescope} image of NGC 2346, a composite of R band 6750 $\AA$ (red), H$\alpha$ 6560 $\AA$ (green), [SII] 6730 $\AA$ (blue).}\label{fig:NGC2346_optico}
   \end{minipage}
   
 \begin{minipage}{0.45\textwidth}
     \centering
     \includegraphics[width=\linewidth]{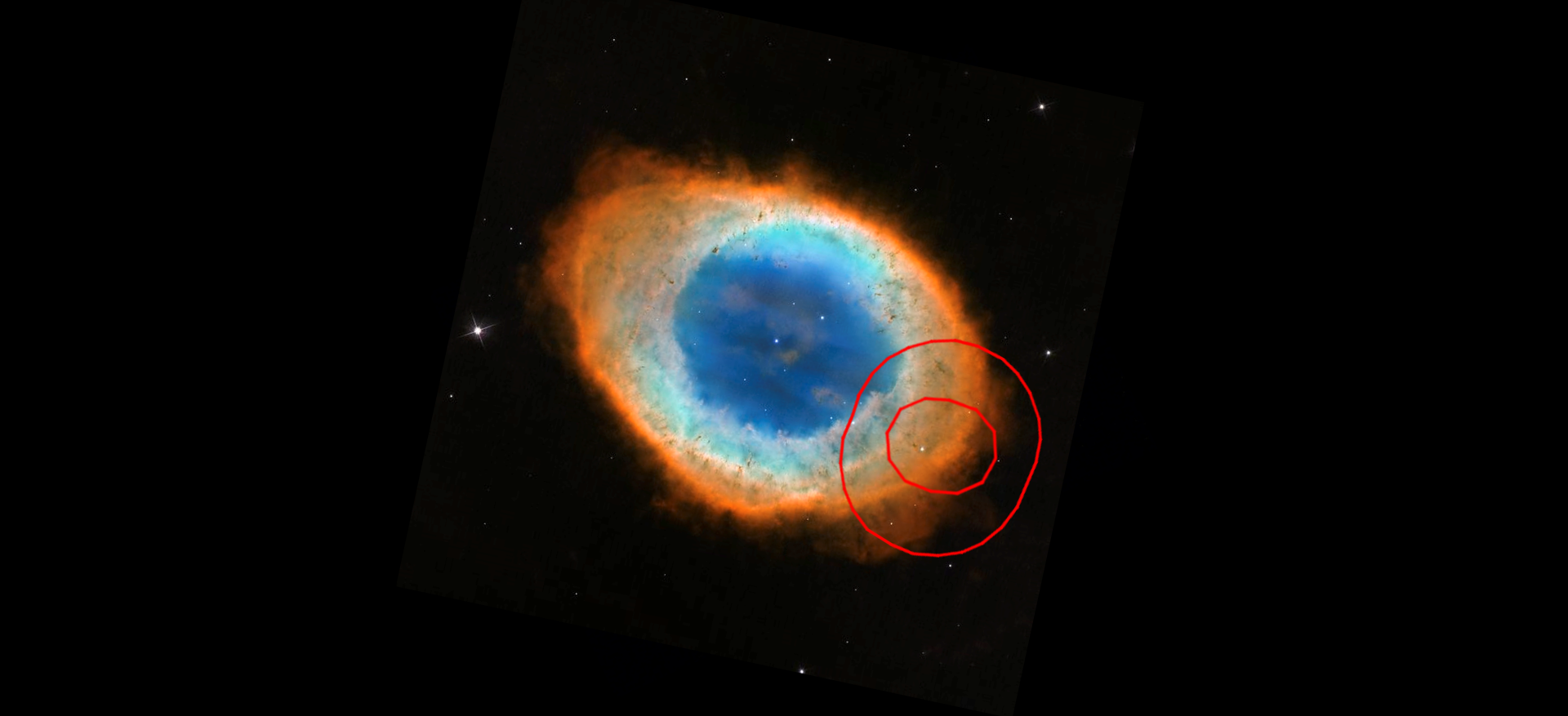}
     \caption{NASA/ESA \textit{Hubble Space Telescope} image of NGC 6720, a composite of R filters 6580 $\AA$ and 6730 $\AA$ (red), G filters 6560 $\AA$ and 6450 $\AA$ and V filters 5020 $\AA$ (green), B filters 4690 $\AA$ and 4870 $\AA$ (blue).}\label{fig:NGC6720_optico}
   \end{minipage}\hfill
   \begin{minipage}{0.45\textwidth}
     \centering
     \includegraphics[width=\linewidth]{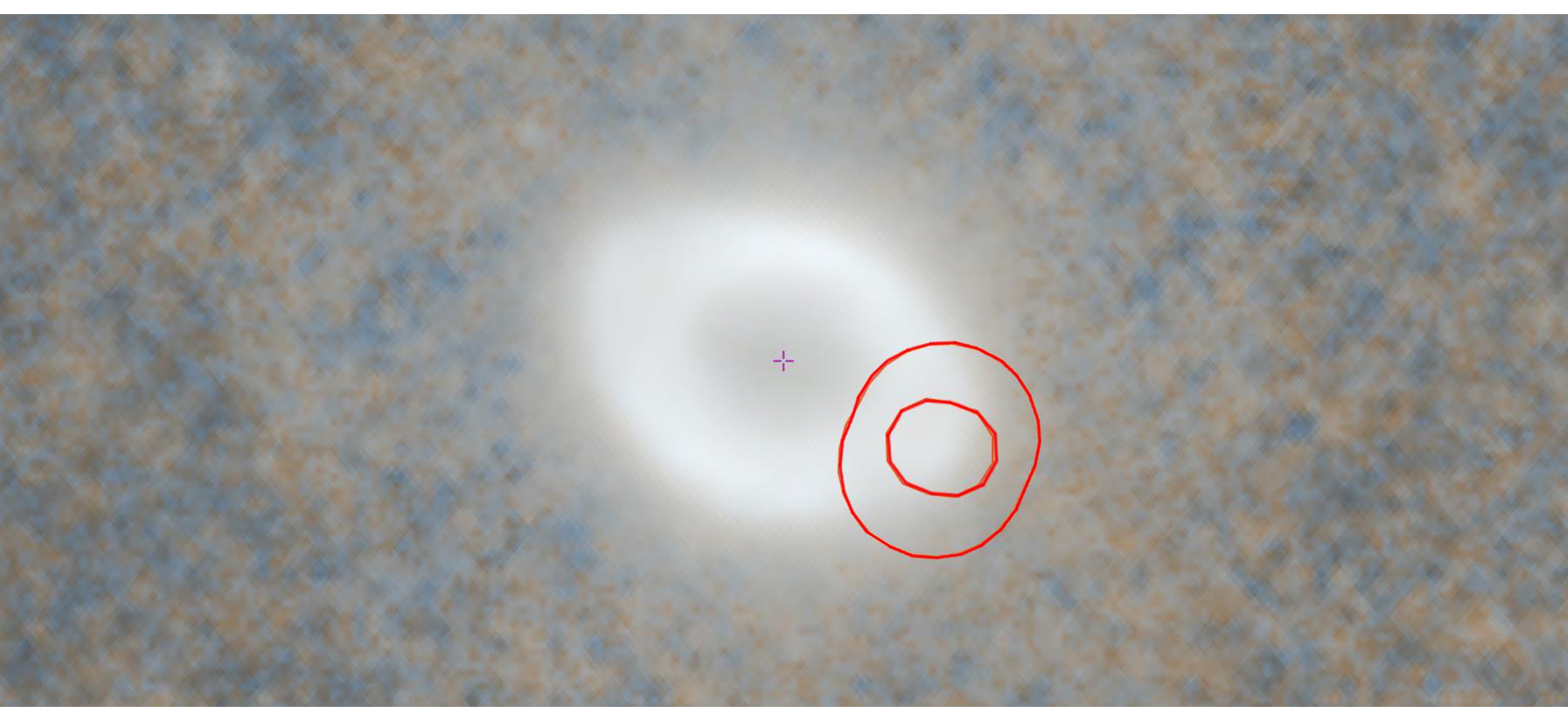}
     \caption{FIR image of NGC 6720 obtained by Herschel PACS RGB 70,160 microns.}\label{fig:NGC6720_FIR}
   \end{minipage}
   
 \begin{minipage}{0.45\textwidth}
     \centering
     \includegraphics[width=\linewidth]{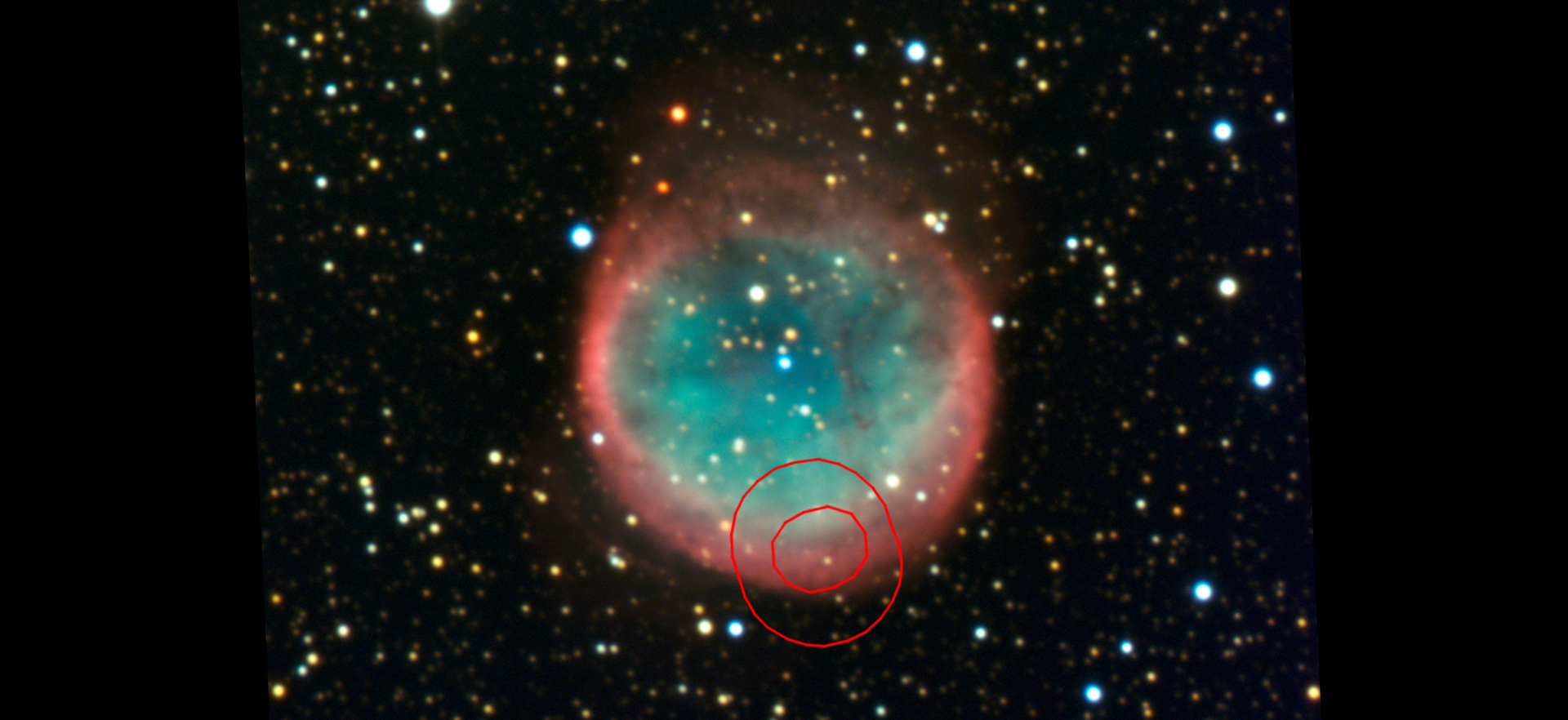}
     \caption{ESO image of NGC 6781, a composite of R and H$\alpha$ (red), V band and [OIII] (green), B band (blue).}\label{fig:NGC6781_optico}
   \end{minipage}\hfill
   \begin{minipage}{0.45\textwidth}
     \centering
     \includegraphics[width=\linewidth]{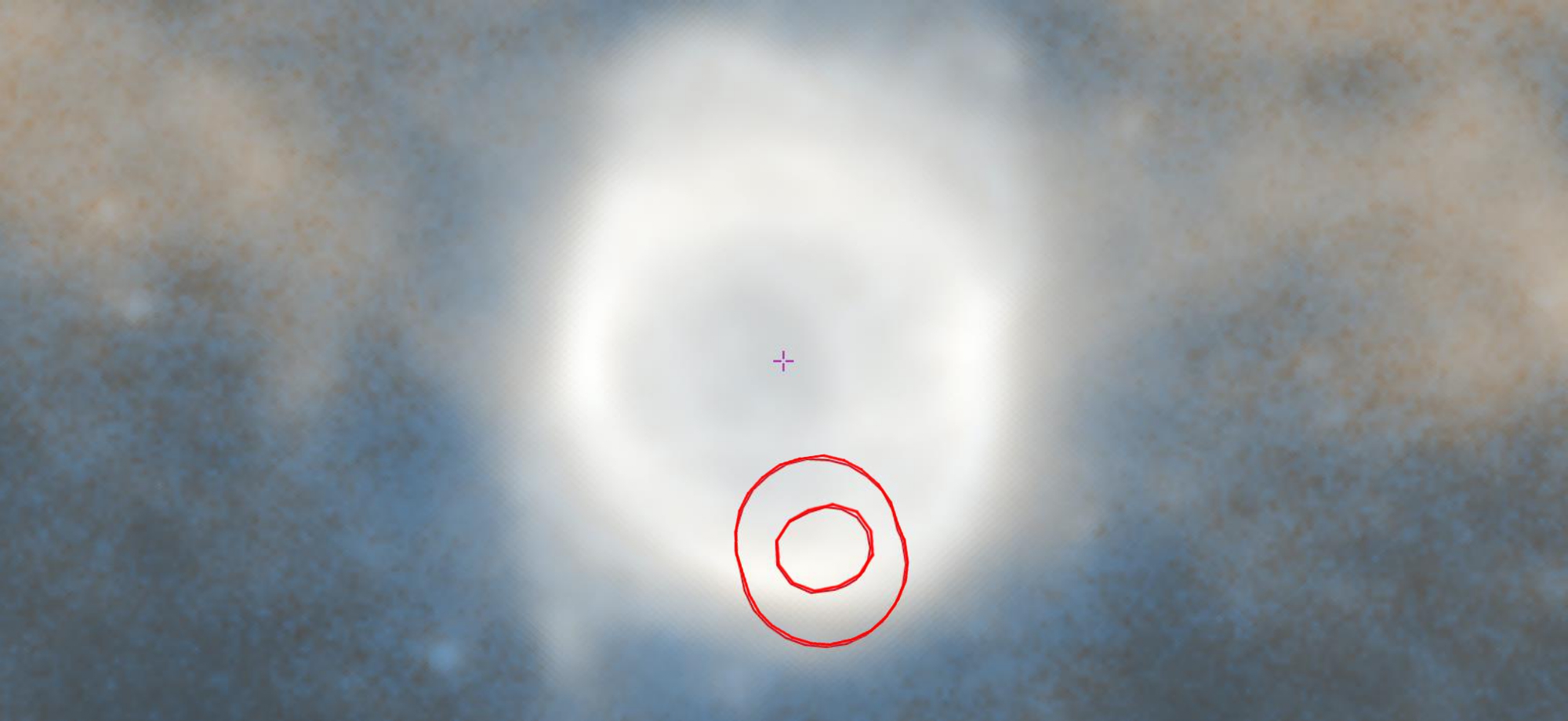}
     \caption{FIR image of NGC 6781 obtained by Herschel PACS RGB 70,160 microns.}\label{fig:NGC6781_FIR}
   \end{minipage}
   
 \begin{minipage}{0.45\textwidth}
     \centering
     \includegraphics[width=\linewidth]{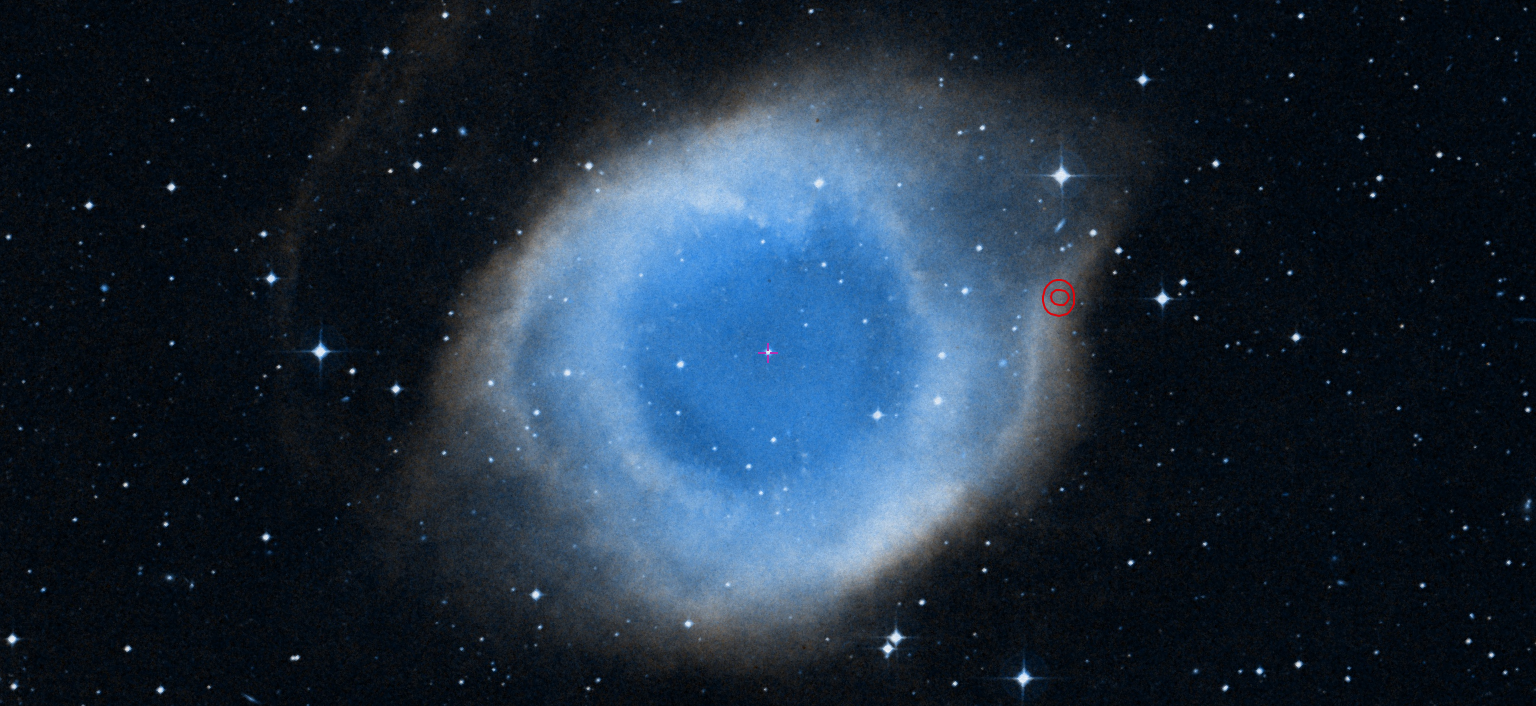}
     \caption{Image of NGC 7293 obtained from DSS2 colour.}\label{fig:NGC7293_optico}
   \end{minipage}\hfill
   \begin{minipage}{0.45\textwidth}
     \centering
     \includegraphics[width=\linewidth]{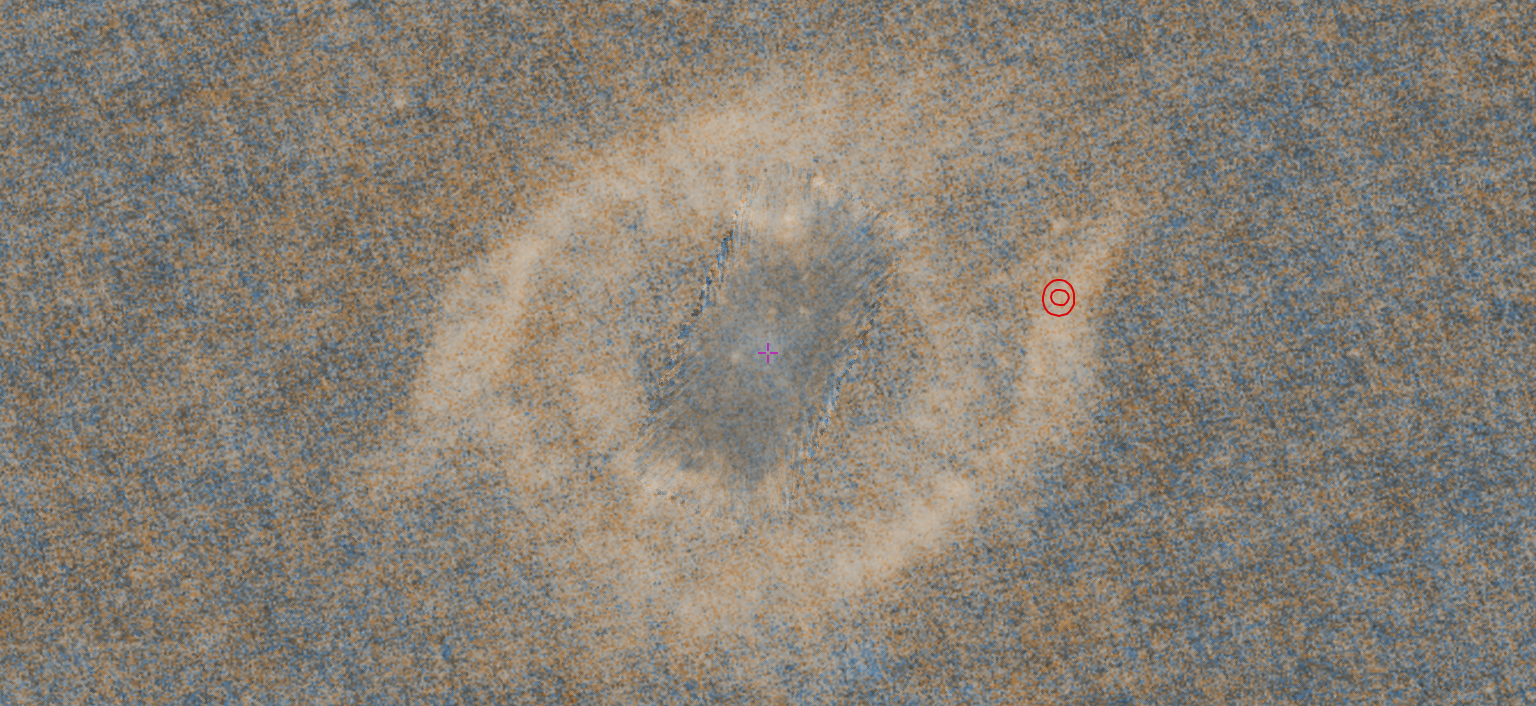}
     \caption{FIR image of NGC 7293 obtained by Herschel PACS RGB 70,160 microns.}\label{fig:NGC7293_FIR}
   \end{minipage}
\end{figure*}

\end{document}